\documentclass[twocolumn,superscriptaddress,aps,preprintnumbers,amsmath,amssymb,prl,footinbib]{revtex4-2}

\usepackage[titletoc,toc,title]{appendix}
\usepackage[english]{babel}
\usepackage[hidelinks]{hyperref}
\usepackage[normalem]{ulem}
\usepackage{adjustbox}
\usepackage{amsfonts}
\usepackage{bm}
\usepackage{enumitem}
\usepackage{epsfig}
\usepackage{cancel}
\usepackage{centernot}
\usepackage{color}
\usepackage{comment}
\usepackage{contour}
\usepackage{flushend}
\usepackage{footmisc}
\usepackage{graphics}
\usepackage{graphicx}
\usepackage{mathrsfs}
\usepackage{mdframed}
\usepackage{pifont}
\usepackage{relsize}
\usepackage{shadowtext}
\usepackage{slashed}
\usepackage{soul}
\usepackage{tcolorbox}
\usepackage{textgreek}
\usepackage{placeins}

\usepackage[T1]{fontenc}
\usepackage{ragged2e} 

\usepackage{titlesec}
\usepackage{titletoc}
\usepackage{verbatim}
\usepackage{xcolor}
\usepackage{xfrac}

\contourlength{0.2em}

\definecolor{myred}{RGB}{123, 28, 20}
\definecolor{zima_blue}{HTML}{1393C1}
\hypersetup{setpagesize=false,bookmarksnumbered=true,bookmarksopen=true,colorlinks=true,linkcolor=zima_blue,urlcolor=zima_blue,citecolor=zima_blue,linktocpage=false}

\usepackage{fontawesome5}
\definecolor{linkcolor}{rgb}{0.7752941176470588, 0.22078431372549023, 0.2262745098039215}
\definecolor{linkcolor}{HTML}{1393C1}

\newcommand{\deltaPT}{{\texttt{deltaPT}}}
\newcommand{\deltaPTvTwo}{\texttt{deltaPT\,2.0}}

\newcommand{\be}{\begin{equation}} 
\newcommand{\ee}{\end{equation}}
\newcommand{\bea}{\begin{equation}\begin{aligned}} 
\newcommand{\eea}{\end{aligned}\end{equation}}
\newcommand{\td}{{\rm d}}

\newif\ifarxiv
 \arxivtrue    

\begin{document}

\title{Curvature Perturbations from First-Order Phase Transitions:
\texorpdfstring{\\}{}Implications to Black Holes and Gravitational Waves}

\author{Gabriele Franciolini}
\email{gabriele.franciolini@cern.ch}
\affiliation{CERN, Theoretical Physics Department, Esplanade des Particules 1, Geneva 1211, Switzerland}
\author{Yann Gouttenoire}
\email{yann.gouttenoire@gmail.com}
\affiliation{School of Physics and Astronomy, Tel-Aviv University, Tel-Aviv 69978, Israel}
\affiliation{PRISMA+ Cluster of Excellence $\&$ MITP, Johannes Gutenberg University, 55099 Mainz, Germany}
\author{Ryusuke Jinno}
\email{jinno@phys.sci.kobe-u.ac.jp}
\affiliation{Department of Physics, Graduate School of Science, Kobe University, 1-1 Rokkodai, Kobe, Hyogo 657-8501, Japan}

\begin{abstract}


{Understanding whether primordial black holes form during strong first-order phase transition (FOPT) is a crucial open question in cosmology.} We address this using a fully covariant formalism to study cosmological perturbations, highlighting previously overlooked gauge dependencies. We show that non-covariant treatments can overestimate primordial black holes and scalar-induced gravitational waves. Once gauge dependencies are accounted for, both signals are strongly suppressed, {with direct implications for the FOPT interpretation of the Pulsar Timing Array signal.}

\vspace*{5pt} \noindent \textit{GitHub}: The {\tt deltaPT\,2.0} code to compute cosmological perturbations generated during strongly supercooled first-order phase transitions \href{https://github.com/YannGou/deltaPT2.0}{\faGithub\;YannGou/deltaPT\,2.0}.

\vspace*{5pt}
\noindent
\textit{Published version:~\href{https://doi.org/10.1103/tfcx-kzqx}{Phys.Rev.Lett. 136 (2026) 17, 171404}}
\end{abstract}

\preprint{CERN-TH-2025-044, KOBE-COSMO-25-05, MITP-25-016}

\maketitle

\textit{\textbf{ Introduction.}}
{The Universe in the primordial epochs} is expected to possess a higher degree of symmetry than today~\cite{Kirzhnits:1972ut}.  As the Universe cools, these symmetries may be broken through first-order phase transitions (FOPTs), a generic feature of many early-Universe scenarios, during which bubbles of the new vacuum nucleate and expand, converting the old vacuum energy into heat and bulk fluid motion (see Ref.~\cite{Gouttenoire:2022gwi} for a review).
If the energy stored in the bubbles constitutes a large fraction of the total universe energy, {FOPTs have been shown to} leave sizable imprints on the spacetime metric, in terms of gravitational waves~\cite{Caprini:2015zlo,Caprini:2019egz}, primordial black holes (PBHs)~\cite{Kodama:1982sf,Hsu:1990fg,Liu:2021svg,Hashino:2021qoq,Jung:2021mku,Kawana:2022olo,Lewicki:2023ioy,Gouttenoire:2023naa,Baldes:2023rqv,Gouttenoire:2023bqy,Salvio:2023ynn,Gouttenoire:2023pxh,Jinno:2023vnr,Banerjee:2023qya,Flores:2024lng,Lewicki:2024ghw,Lewicki:2024sfw,Ai:2024cka,Conaci:2024tlc,Kanemura:2024pae,Banerjee:2024cwv,Hashino:2025fse,Murai:2025hse,Zou:2025sow,Ghoshal:2025dmi,Byrnes:2025tji}, and curvature perturbations~\cite{Sasaki:1982fi,Liu:2022lvz,Giombi:2023jqq,Elor:2023xbz,Lewicki:2024ghw,Buckley:2024nen,Cai:2024nln,Jinno:2024nwb}. {These curvature perturbations can source scalar-induced gravitational waves (SIGWs), studied in this context in~\cite{Lewicki:2024ghw} and more generally in~\cite{Tomita:1975kj,Matarrese:1992rp,Matarrese:1993zf,Matarrese:1997ay,Acquaviva:2002ud,Mollerach:2003nq,Carbone:2004iv,Ananda:2006af,Baumann:2007zm,Espinosa:2018eve,Kohri:2018awv,Domenech:2021ztg}.}

Since cosmological perturbations generally exhibit gauge dependence~\cite{Bardeen:1980kt}, reliable predictions for these late-time observables require a covariant treatment of perturbations and their evolution. {Such a treatment is required to assess whether FOPTs can produce observable PBHs and SIGWs, a question of central importance when evaluating the FOPT interpretation of the Pulsar Timing Array signal~\cite{Gouttenoire:2023bqy,Ellis:2023oxs,Lewicki:2024ghw}.}

In this $\textit{letter}$, we derive the statistical properties of cosmological perturbations produced during a strong first-order phase transition, with a particular emphasis on the gauge choice and the calculation of gauge-invariant quantities.
We evaluate the abundance of PBHs and the spectrum of SIGWs using the most appropriate quantities for each: the density contrast in the comoving gauge $\delta^{(C)}$ for the former, to match the convention used to report the threshold for collapse in numerical simulations, while the gauge-invariant curvature perturbation $\mathcal{R}$ for the latter.

\textit{\textbf{ Covariant linear perturbations from FOPTs.}} We model the nucleation rate by
\begin{equation}
\Gamma = H_n^4 \,e^{\beta (t-t_n)},
\end{equation}
where $H_n$ is the Hubble factor ($H\equiv \dot a(t) / a $, with $a$ being the scale factor) at $t=t_n$, defined as the nucleation time.
Adopting a relativistic bag equation of state, the energy and pressure density decompose as
\begin{equation}
\label{eq:rho_p_avg}
\rho=\rho_R+\rho_V,\qquad p=\rho_R/3-\rho_V,
\end{equation}
where $R$ and $V$ stand for radiation and vacuum components, respectively. 
We consider the supercooled limit for which the latent heat $\Delta V$ is larger than the radiation energy density just before nucleation $\alpha \equiv \Delta V/\rho_{R} \gg 1$. At the background level, the radiation and vacuum energy density obeys the continuity equation
\begin{align}
\label{eq:bkg_continuity}
\overline{\rho_{\rm R}}'+4\mathcal{H}\overline{\rho_{\rm R}}=-\overline{\rho_{\rm V}}',\qquad \textrm{with}\quad \overline{\rho_{\rm V}} = \overline{F}\Delta V,
\end{align}
and the Friedmann equation $3M_{\rm pl}^2H^2=\overline{\rho_{\rm R}}+\overline{\rho_{\rm V}}$. The prime denotes the derivative with respect to the conformal time $\eta$ related to the cosmic time $t$ by $
\eta(t_2) =\eta(t_1) +\int^{t_2}_{t_1} d\tilde{t}/a(\tilde{t})$, and we defined $\mathcal{H}=aH$.
The remaining fraction of false vacuum $\overline{F}(t) $ is given by 
\begin{equation}
\overline{F}(t) = \exp\left[-\frac{4\pi}{3}\!\! \int_{-\infty}^t \!\!\!\td t_n\, \Gamma(t_n) a(t_n)^3 \left(\int_{t_n}^t\!\! \frac{d\tilde{t}}{a(\tilde{t})}\right)^{\!3}\right],
\end{equation}
where we assumed bubble walls to expand at the speed of light $v_w \simeq 1$.

Due to the stochastic nature of bubble nucleation, the stress-energy tensor $T^\mu_{\,\,\,\,\nu}$ acquires fluctuations, assumed to be statistically isotropic, $T^0_{\,\,\,\,0}=\overline{\rho}+\delta \rho$, $T^i_{\,\,\,\,j}=-(\overline{p}+\delta p)\delta^i_j+\Sigma^i_{~j}$, and $T^i_{\,\,\,\,0}=(\overline{\rho}+\overline{p})v^i$ where $v^i=dx^i/dt$ is the coordinate velocity. In the following, we use $\delta \equiv \delta \rho/\overline{\rho}$. 
At the linear level, the 10 degrees of freedom of the symmetric $4\times4$ space-time metric $g_{\mu\nu}$ decomposes into 4 scalars, 4 vectors and 2 tensors. Keeping only the scalars ($A,B,C,E$), the spacetime metric reads
\begin{multline}
    \mathrm{d}s^2 
    = 
    a^2 \Bigl[
    \bigl(1 + 2A\bigr)\,\mathrm{d}\eta^2 
    - 2\,\partial_i B\,\mathrm{d}x^i\,\mathrm{d}\eta  \\
-\big((1-2C)\delta_{ij}+2\partial_{\langle i}\partial_{j\rangle}E\big)dx^i dx^j\Bigr],
\end{multline}
where $\partial_{\langle i}\partial_{j\rangle} E \equiv (\partial_i\partial_j - \tfrac{1}{3}\delta_{ij}\nabla^2)E$. In the absence of anisotropic stress, scalar metric perturbations contain only one physical degree of freedom. The apparent redundancy in their description is known as ``gauge dependence''~\cite{Bardeen:1980kt}. To extract physical predictions, one must fix the coordinate freedom by choosing a gauge. The assumption of isotropy, $\Sigma^i_{~j} = 0$, imposes one constraint, leaving two additional scalar gauge degrees of freedom to fix.  A common choice is the spatially-flat gauge (F), defined by $C=E=0$, {in which we denote the density contrast by $\delta^{(F)}$.} Linearized Einstein equations in Fourier space lead to~\cite{SuppMat}
\color{black}
\begin{equation}
\label{eq:SF_continuity_prelude}
\tilde{\delta}^{(F)'}_\mathbf{k}+3\mathcal{H}(c_s^2-\omega)\tilde{\delta}^{(F)}_\mathbf{k}=(1+\omega) k^2\tilde{\mathcal{V}}_\mathbf{k}-3\mathcal{H}\tilde{\delta}_{p,\rm nad,\mathbf{k}},
\end{equation}
where $\tilde{X}_{\mathbf{k}}(t)$ denotes the Fourier-transform of $X(\mathbf{x},t)$. 
The quantities $\omega \equiv \overline{p}/\overline{\rho}$ and $c_s^2\equiv {\overline{p}}'/{\overline{\rho}}'$ are the equation of state and speed of sound, respectively.
The quantity $\delta_{p,\rm nad}$ is the non-adiabatic pressure perturbation
\begin{align}
\delta_{p,\rm nad} \equiv \frac{\delta p_{\rm nad}}{\overline{\rho}},\quad \textrm{with}\quad \delta p_{\rm nad} \equiv \delta p^{(F)} - c_s^2 \delta \rho^{(F)},\label{eq:comoving_definition_deltapnad}
\end{align}
which is gauge-invariant in spite of the superscript $(F)$ on the right hand-side. Using Eq.~\eqref{eq:rho_p_avg}, Eq.~\eqref{eq:comoving_definition_deltapnad} becomes
\begin{equation}
\label{eq:delta_nad_FOPT_main}
\delta p_{\rm nad} = \frac{1-3c_s^2}{3}\overline{\rho}\,\delta^{(F)}-\frac{4}{3}\Delta V\delta F^{(F)},
\end{equation} 
where $\delta F^{(F)} = F^{(F)}-\overline{F}$ with $F^{(F)}\equiv \rho_V^{(F)}/\Delta V$.
The quantity $\mathcal{V}$ in Eq.~\eqref{eq:SF_continuity_prelude} is the gauge-invariant scalar velocity $\mathcal{V}\equiv v+E'$ with $\partial_i(v)=\delta_{ij}v^j$. Einstein equation along $0i$ gives
\begin{equation}
\label{eq:GI_velocity}
\mathcal{V}= -\frac{2}{3(1+\omega)} \frac{\mathcal{H}\Phi+\Phi'}{\mathcal{H}^2},
\end{equation}
where $\Phi$ is the gauge-invariant Newtonian potential~\cite{Bardeen:1980kt}. A combination of Einstein equations along $00$ and $ij$ gives
\begin{equation}
\label{eq:Newtonian_potential}
\tilde{\Phi}_{\mathbf{k}}'' + 3 (1 + c_s^2)\mathcal{H} \tilde{\Phi}_{\mathbf{k}}' 
+[3 (c_s^2-\omega)\mathcal{H}^2+c_s^2 k^2]\tilde{\Phi}_{\mathbf{k}} 
= \frac{3}{2}\mathcal{H}^2\tilde{\delta}_{p,\rm nad,\mathbf{k}}.
\end{equation}
The initial conditions for the system Eqs.~\eqref{eq:SF_continuity_prelude} and \eqref{eq:Newtonian_potential} are 
\begin{equation}
\tilde{\delta}^{(F)}_{\mathbf{k}}(0)=0,\qquad \tilde{\Phi}_{\mathbf{k}}(0)=0,\qquad \tilde{\Phi}_{\mathbf{k}}'(0)=0,
\end{equation}
where we have neglected the primordial curvature perturbations (i.e.~generated during inflation), which we assume to be small at these scales. 
For more details on cosmological perturbation theory, we refer the reader to the Supplemental Material (SuM) or Refs.~\cite{Kodama:1984ziu,Mukhanov:1990me,Mukhanov:2005sc,Baumann:2022mni,Peter:2013avv}.

In the spatially-flat gauge, constant-time hypersurfaces are unperturbed 
$ds^2\big|_{\eta={\rm const.}}=-a^2dx^2$.
This is particularly convenient since by plugging Eq.~\eqref{eq:delta_nad_FOPT_main} into Eq.~\eqref{eq:SF_continuity_prelude}, we get a perturbation equation
\begin{align}
\delta{\tilde{\rho}}_{R,{\mathbf{k}}}^{(F)'}+4\mathcal{H}\delta{\tilde{\rho}}_{R,{\mathbf{k}}}^{(F)}=-\delta{\tilde{\rho}}_{\rm V,{\mathbf{k}}}^{(F)'}+\frac{4}{3}\overline{\rho_R}\, k^2 \tilde{\mathcal{V}}_{\mathbf{k}},\label{eq:SF_continuity}
\end{align}
which coincides with the background equation in Eq.~\eqref{eq:bkg_continuity} in the super-Hubble limit $k\ll \mathcal{H}$. 
Hence, in the spatially-flat gauge and in the super-Hubble limit, different patches evolve as distinct FLRW universes \cite{Wands:2000dp}, 
and we can identify
\begin{equation}
\label{eq:F_k_F}
\tilde{F}^{(F)}_{\mathbf{k}} \underset{k\ll\mathcal{H}}{\simeq} \tilde{F}^{\rm (FLRW)}_{\mathbf{k}},
\end{equation}
where $F^{\rm (FLRW)}$ is the false vacuum fraction in a flat FLRW universe. Notice that such equality does not hold in other gauges. The code \deltaPT~developed in Ref.~\cite{Lewicki:2024ghw} calculates the false vacuum fraction $F^{\rm (FLRW)}_{\rm avg}(R,t)$ averaged over a ball of radius $R$, in a flat FLRW universe. In this work, we rely on the approximation~\footnote{We leave a proper de-convolution from the top-hat window function $W(r,R)=\Theta(R-r)/V$ followed by inverse-Fourier transformation for future work.
}
\begin{equation}
\label{eq:F_k_F_avg}
\tilde{F}^{\rm (FLRW)}_{\mathbf{k}}(t) \simeq V F^{\rm (FLRW)}_{\rm avg}(R=k^{-1},t),
\end{equation}
with $V=4\pi R^3/3$.
The factor $V$ ensures that Fourier transforms carry the usual volume dimension.
Due to Eq.~\eqref{eq:F_k_F_avg}, it follows that $\tilde{X}_{\mathbf{k}}(t)\simeq V X_{\rm avg}(R=k^{-1},t)$ for any perturbations $X$ derived in this work. 
This approximation should become exact in the super-Hubble limit $k\ll \mathcal{H}$ when gradient terms are negligible. We now denote perturbations by $X\equiv \tilde{X}_{\mathbf{k}}/V\simeq X_{\rm avg}(R=k^{-1})$.

\begin{figure}[t]
\centering
\includegraphics[width=0.48\textwidth]{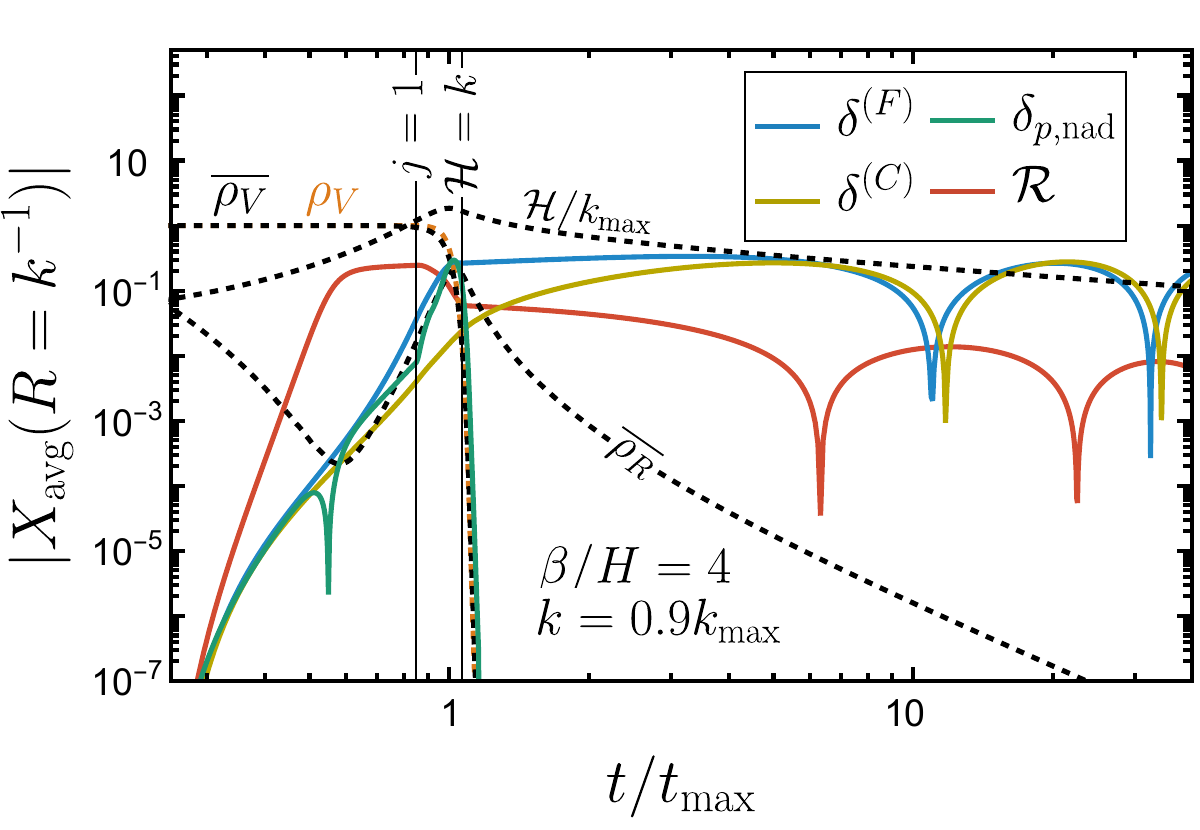}
\caption{\label{fig:delta_Phi_numerical}
Time evolution of the comoving curvature perturbation $\mathcal{R}$, and the density contrast in the spatially-flat $\delta^{(F)}$ and comoving $\delta^{(C)}$ gauge during a FOPT. We show perturbations $X$ averaged over a comoving volume $4\pi k^{-3}/3$.
We also report the non-adiabatic pressure $\delta p_{\rm nad}$ peaking close to Hubble crossing epoch $\mathcal{H}=k$. For presentation purposes we assume $\beta / H = 4$ and show the wavenumber $k = 0.9 k_{\rm max}$, where we identify $k_{\rm max}$ as the maximal ${\cal H}$. $j=1$ indicates the time of first bubble nucleation, which is slightly delayed with respect to average (orange vs black dashed line).}
\end{figure}

\begin{figure}[ht!]
\centering
\includegraphics[width=0.48\textwidth]{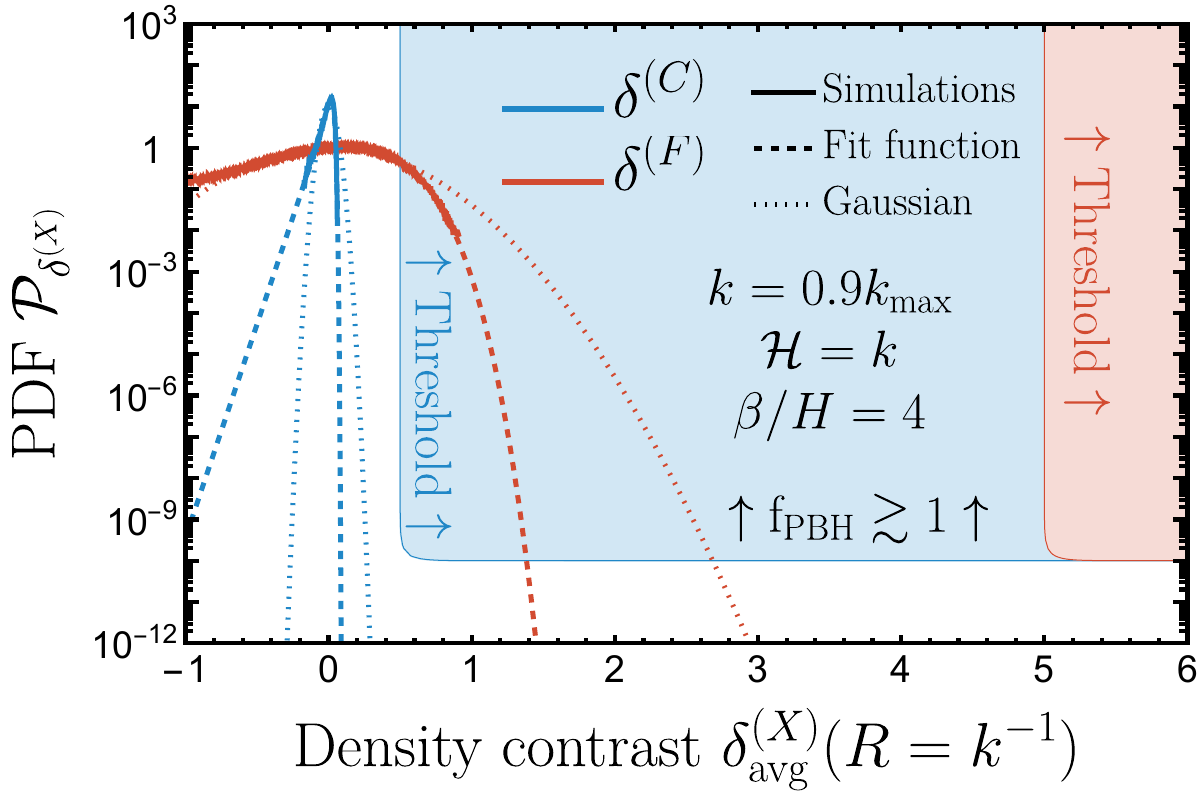}
\caption{\label{fig:proba_distrib_P_deltaC}
Probability distribution of the density contrast in the comoving (blue) and spatially-flat (red) gauges, obtained from {solving cosmological perturbations equations for $N_{\rm sim}=10^6$ distinct random nucleation histories inside a patch of comoving volume $4\pi k^{-3}/3$.} The dashed lines show the best fits using the distribution in Eq.~\eqref{eq:fit_function}. The dotted lines shows the best fits using a Gaussian distribution.
We choose $k=0.9k_{\rm max}$ where $k_{\rm max}$ is the scale entering at percolation. 
The colored boxes indicate the collapse condition in the comoving (blue) and spatially-flat (red) gauges. We cut the box from below at the indicative amplitude for a stellar mass PBH. }
\end{figure}

\begin{figure}[ht]
\centering
\includegraphics[width=0.48\textwidth]{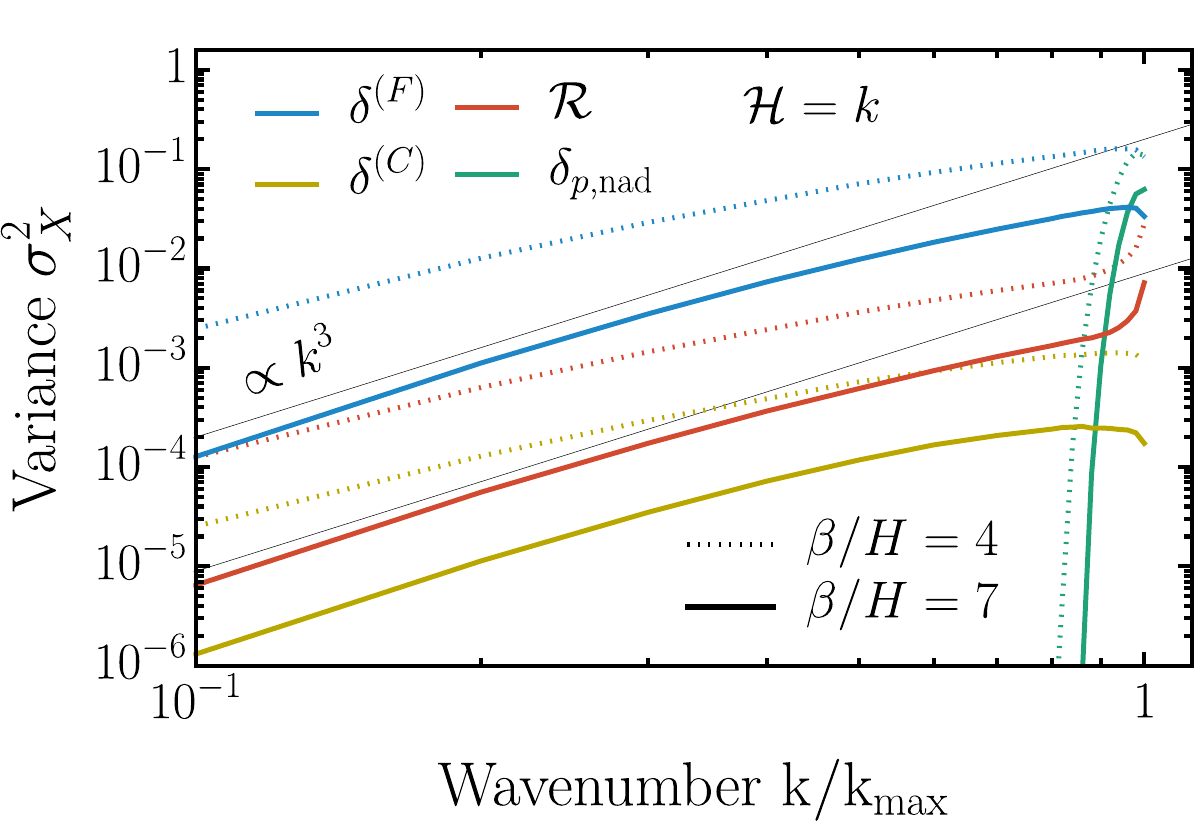}
\caption{\label{fig:PS_allquantities}
Variance of the perturbations $X$ averaged over a ball of comoving radius $k^{-1}$, $\sigma_{X}^2\equiv \left< X_{\rm avg}^2(R=k^{-1},t_H)\right>$, evaluated at Hubble crossing epoch $t_H$. We show the comoving curvature perturbation $\mathcal{R}$ (red), the density contrast in the spatially-flat $\delta^{(F)}$ (blue) and comoving gauge $\delta^{(C)}$ (yellow), and the non-adiabatic pressure perturbation $\delta_{p,{\rm nad}}$ (green).
}
\end{figure}

In practice, the evolution of perturbations can be derived as follows: from Eqs.~\eqref{eq:F_k_F}, \eqref{eq:F_k_F_avg} and \deltaPT, we calculate $\delta p_{\rm nad}$ in Eq.~\eqref{eq:delta_nad_FOPT_main}, then we solve the closed system of equations in \eqref{eq:SF_continuity_prelude} and \eqref{eq:Newtonian_potential} to determine $\Phi$ and $\delta^{(F)}$, see SuM for more details. Another common choice of coordinate is the comoving gauge (C), in which the fluid velocity perturbation vanishes ($v + B = 0,~ E=0$).
The density contrast in the comoving gauge can be derived using the transformation rules
\begin{align}
&\delta^{(C)}=\delta^{(F)}+ 3(1+\omega)\mathcal{R},\label{eq:delta_C_F}
\end{align}
where $\mathcal{R}$ is the comoving curvature perturbation given by the gauge-invariant expression~\cite{Bardeen:1980kt}
\begin{align}
\mathcal{R} ~\equiv ~\Phi - \mathcal{H}\mathcal{V} ~=~ \frac{5+3\omega}{3+3\omega} \Phi +\frac{2\Phi'}{3(1+\omega)\mathcal{H}}.
\label{eq:mathcalR_def_main}
\end{align}
{The transformation rule in Eq.~\eqref{eq:delta_C_F}  follows from the textbook relationship $\zeta =\mathcal{R}-\delta^{(C)}/3(1+\omega)$~\cite{Baumann:2022mni} where $\zeta~\equiv~ -\delta^{(F)}/3(1+\omega)$ is the curvature perturbation on uniform-density hypersurfaces, see SuM for details.}  {Building on this framework, our public GitHub release \deltaPTvTwo~\cite{deltaPTv2} extends \deltaPT~\cite{Lewicki:2024ghw} to compute the covariant cosmological perturbations analyzed in this work.}
We show the time evolution of $\delta^{(F)}$, $\delta^{(C)}$, $\mathcal{R}$ and $\delta p_{\rm nad}$ in Fig.~\ref{fig:delta_Phi_numerical}, for a single realization of the bubble nucleation dynamics within a ball of radius $R=k^{-1}$ operated with \deltaPTvTwo~\cite{deltaPTv2} . {We derive the perturbations statistics by making $N_{\rm sim}=10^6$ realizations of random bubble nucleation histories and storing the Hubble-crossing values of the perturbations in a histogram~\footnote{{In each patch, we randomly sample the nucleation times $t_j$ and positions $d_j$ of the first $j_c=50$ bubbles and assume a deterministic evolution for subsequent nucleation dynamics~\cite{Lewicki:2024ghw}. The probability distributions for $t_j$ and $d_j$ within a patch of radius $R$ are derived analytically in App.~A of Ref.~\cite{Lewicki:2024ghw} and are summarized in the SuM.}
}.}
In Fig.~\ref{fig:proba_distrib_P_deltaC}, we show the resulting probability density function (PDF) of the density contrast. The variance of the density contrast in the spatially-flat and comoving gauges, as well as the comoving curvature perturbation are shown in Fig.~\ref{fig:PS_allquantities}. 

\textit{\textbf{ Primordial Black Hole formation.}}
Once the density contrast in a Hubble patch exceeds a threshold, it collapses into a PBH~\cite{Zeldovich:1967lct,Hawking:1971ei,Carr:1974nx,Carr:1975qj,Chapline:1975ojl}. Numerical relativity simulations~\cite{Escriva:2021aeh} define the collapse threshold in terms of the compaction function evaluated at Hubble crossing $t_H$~\cite{Shibata:1999zs,Harada:2023ffo}. At this epoch, it reduces to the density contrast averaged over a Hubble-scale region, $\delta_{\rm avg}(R=k^{-1},t_H)$~\cite{Musco:2018rwt,Escriva:2019phb}, which in our notation corresponds to $\delta(k\simeq\mathcal{H})$.
As we have seen, the latter is gauge-dependent, see also Ref.~\cite{Harada:2015yda}. Recent simulations have relied on the comoving gauge, see e.g.~\cite{Musco:2018rwt}. This gauge is advantageous since the $00$ component of Einstein's equations reduces to Poisson’s equation
\begin{equation}
\label{eq:Poisson_eq}
    \Delta \Phi = \frac{3}{2}\mathcal{H}^2\delta^{(C)}.
\end{equation}
 In this gauge, a patch collapses into a PBH if~\cite{Musco:2018rwt,Escriva:2019phb,Escriva:2021aeh}
\begin{equation}
\label{eq:PBH_threshold}
\delta^{(C)}(k\simeq\mathcal{H}) \gtrsim \delta_c^{(C)} \in [0.40,\,0.67],
\end{equation}
with the exact value depending on the curvature {real space} profile \cite{Musco:2020jjb}. These simulations assume adiabatic evolution from super-Hubble scales~\cite{Musco:2018rwt,Escriva:2019phb,Escriva:2021aeh}. In our scenario, {non-adiabatic effects are present} but decay rapidly after percolation (see Fig.~\ref{fig:delta_Phi_numerical}), and we therefore adopt the standard collapse criterion evaluated at $k=\mathcal{H}$.

We proceed to estimate the abundance of primordial black holes. Denoting $P(\delta^{(C)})$ as the PDF for $\delta^{(C)}$ at $k={\cal H}$, the PBH mass fraction at formation reads
\begin{align}
\label{eq:beta_k_M_1stOPT}
\beta_k(M) &= \int_{\delta^{(C)}_c}^{\infty} 
{\rm d}\delta^{(C)} \,
\left (\frac{M(\delta^{(C)})}{M_k} \right )
P(\delta^{(C)}) .
\end{align} 
In the first line of Eq.~\eqref{eq:beta_k_M_1stOPT}, the PBH mass $M$ is related to $\delta^{(C)}$ through the critical scaling law \cite{Choptuik:1992jv,Evans:1994pj,Niemeyer:1997mt,Musco:2012au}
\begin{equation}
\label{eq:choptuik_law}
M(\delta^{(C)}) = {\cal K} M_k (\delta^{(C)} - \delta_c^{(C)})^\gamma,
\end{equation}
with $\gamma = 0.36$, ${\cal K} \sim 3$, and 
$M_k = 4\pi M_{\rm pl}^2/H_k$ 
represents the mass within the Hubble sphere when scale $k$ re-enters, $H_k=k/a$.
Using that PBHs redshift like matter, we obtain the DM abundance composed of PBH $f_{\rm PBH}$
\begin{equation}
\frac{df_{\rm PBH}}{d\ln(M)} \simeq \int d\ln{k}\,\left(\frac{\beta_k(M)}{3 \times 10^{-10}}\right)\left(\frac{T_k}{\rm GeV}\right),
\end{equation}
where $T_k$ is the temperature at Hubble crossing.
Approximating the spectrum to a nearly monochromatic distribution close to $k \sim k_{\rm max}$, one parametrically finds
\begin{equation}
f_{\rm PBH} ~\sim~ 
\left (\frac{P(\delta_c^{(C)})}{10^{-10}}\right )\,
\left (\frac{T_{k}}{\rm GeV} \right ),
\end{equation}
where the Hubble crossing temperature is related to the Hubble mass as
\begin{equation}
M_k(T_k) 
= 4.8\times 10^{-2} M_\odot 
\left(\frac{106.75}{g_*} \right)^{1/2} \left( \frac{\rm GeV}{T_k} \right)^{2},
\end{equation}
and $g_*$ is the number of relativistic degrees of freedom. 

As shown in blue in Fig.~\ref{fig:proba_distrib_P_deltaC}, the distribution 
$P(\delta^{(C)})$, here plotted for $k=0.9k_{\rm max}$, is negatively skewed with an extended tail in the negative range and an exponential suppression of large overdensities, reflecting the exponentially-low probability of forming late-blooming patches~\cite{Gouttenoire:2023naa}. As in Refs.~\cite{Tomberg:2023kli,Lewicki:2024ghw}, the fit (with $\epsilon, \sigma > 0$) 
\begin{equation}
\label{eq:fit_function}
P(\delta) \propto \exp\!\left[\frac{\epsilon}{2}(\delta-\mu) - \frac{2}{\epsilon^2\sigma^2}\left(1 - e^{\frac{\epsilon}{2}(\delta-\mu)}\right)^2\right]
\end{equation}
accurately describes the data (see SuM).

Due to the negative non-Gaussianity (NG), even for a completion rate as low as $\beta/H=4$, 
$P({\delta^{(C)}})$ drops sharply around $\delta^{(C)}\sim 0.1$ and is strongly suppressed above the optimistic PBH threshold $\delta^{(C)}_c\gtrsim 0.4$ in Eq.~\eqref{eq:PBH_threshold}. We found similar results for $\beta/H =2$, just above the no-completion boundary. We conclude that within the framework considered here, slow FOPTs do not appear to yield PBHs in observable amounts. The PDF for $\delta^{(F)}$ in the spatially-flat gauge, shown in red in Fig.~\ref{fig:proba_distrib_P_deltaC}, is broader but still heavily skewed, and with the collapse threshold shifted to{
\begin{equation}
\label{eq:delta_F_delta_C_link}
    \delta^{(F)}_c\simeq 10\,\delta^{(C)}_c
\end{equation}
following from Eqs.~\eqref{eq:delta_C_F} and \eqref{eq:Poisson_eq} with $k=\mathcal{H}$, $\omega=1/3$, and $\Phi'\simeq 0$~\footnote{As known from textbooks, e.g.~\cite{Baumann:2022mni}, and re-derived in the SuM, the approximation $\Phi'\simeq0$ holds for superhorizon modes only in the absence of non-adiabatic pressure perturbations, $\delta_{p,\mathrm{nad}}=0$. In our scenario, bubble dynamics generate $\delta_{p,\mathrm{nad}}\neq0$, leading to departures from $\Phi'\simeq0$ on superhorizon scales. Nevertheless, $\Phi'\simeq0$ remains a good approximation near horizon crossing, $k\simeq \mathcal{H}$, justifying its use in evaluating the PBH collapse threshold. The time evolution of $\Phi(t)$ is shown in the SuM.
},} the conclusion remains unchanged. Note that our analysis neglects non-linear corrections to the density-curvature relation, which would slightly reduce the variance of $\delta^{(C)}$ further \cite{DeLuca:2019qsy,Young:2019yug}.

\textit{\textbf{ Induced Gravitational Waves.}} We now focus on the GW spectrum from FOPT, seizing the relevance of SIGWs from curvature perturbations. 
The GW abundance today can be written as
\begin{equation}
\Omega_{\rm GW}(k)=\frac{1.7\times10^{-5}}{g_*^{1/3}(T_k)}\Big[\Omega_{\rm PGW}(k,T_k)+\Omega_{\rm SIGW}(k,T_k)\Big],
\end{equation}
where $\Omega_{\rm PGW}$ and $\Omega_{\rm SIGW}$ are the spectrum of primary GWs (PGWs) and SIGWs. 
Primary GW are produced from bubble collision and relativistic shells. Assuming that the latter remain thin and conserve their energy after collision, their GW spectrum is approximated by the broken power-law bulk flow formula~\cite{Jinno:2017fby,Konstandin:2017sat,Baldes:2024wuz}
\begin{equation}
\Omega_{\rm PGW} \simeq 0.06 \left( \frac{H}{\beta}\right)^2
\frac{(a+b) f^a f_\star^b}{\left(af^{a+b}+b f_\star^{a+b} \right)}S_H(f),
\end{equation}
with $f_\star=0.8(\beta/H)(\mathcal{H}/2\pi)$, $a=0.9$, $b=2.1$, and $S_H(f)$ imposing a $f^3$ behavior for $f<\mathcal{H}/2\pi$.
SIGWs are given by~\cite{Tomita:1975kj,Matarrese:1992rp,Matarrese:1993zf,Matarrese:1997ay,Acquaviva:2002ud,Mollerach:2003nq,Carbone:2004iv,Ananda:2006af,Baumann:2007zm,Domenech:2021ztg,Espinosa:2018eve,Kohri:2018awv}
\begin{equation}
\Omega_{\rm SIGW}(k,\eta)= \frac{1}{24}\left(\frac{k}{{\cal H}(\eta)}\right)^2\overline{{\cal P}_h(k,\eta)},
\end{equation}
where the power spectrum induced at second-order from curvature perturbations is
\begin{align}
\overline{ {\cal P}_h(k,\eta)} &
(2\pi)^3\delta_D^{(3)}(\mathbf{k}+\mathbf{k}')
= \frac{16 k^3}{2 \pi^2}
\sum_s
\int\frac{{\rm d}^3\mathbf{p}}{(2\pi)^{3}}
\int\frac{{\rm d}^3\mathbf{q}}{(2\pi)^{3}}
\nonumber \\
&
\times 
\overline{J_s(\mathbf{k},\mathbf{p},\eta) 
J_s(\mathbf{k}',\mathbf{q},\eta) }
\langle 
\tilde{\mathcal{R}}_{\mathbf{p}}
\tilde{\mathcal{R}}_{\mathbf{k}-\mathbf{p}}
\tilde{\mathcal{R}}_{\mathbf{q}}
\tilde{\mathcal{R}}_{\mathbf{k'}-\mathbf{q}}
\rangle\,,
\label{eq:Phfull}
\end{align}
having defined
\begin{equation}
J_s(\mathbf{k},\mathbf{p},\eta) 
\equiv Q_{s}(\mathbf{k},\mathbf{p})
I(\mathbf{k},\mathbf{p},\eta) 
\end{equation}
\begin{figure}[ht!]
\centering
\includegraphics[width=0.48\textwidth]{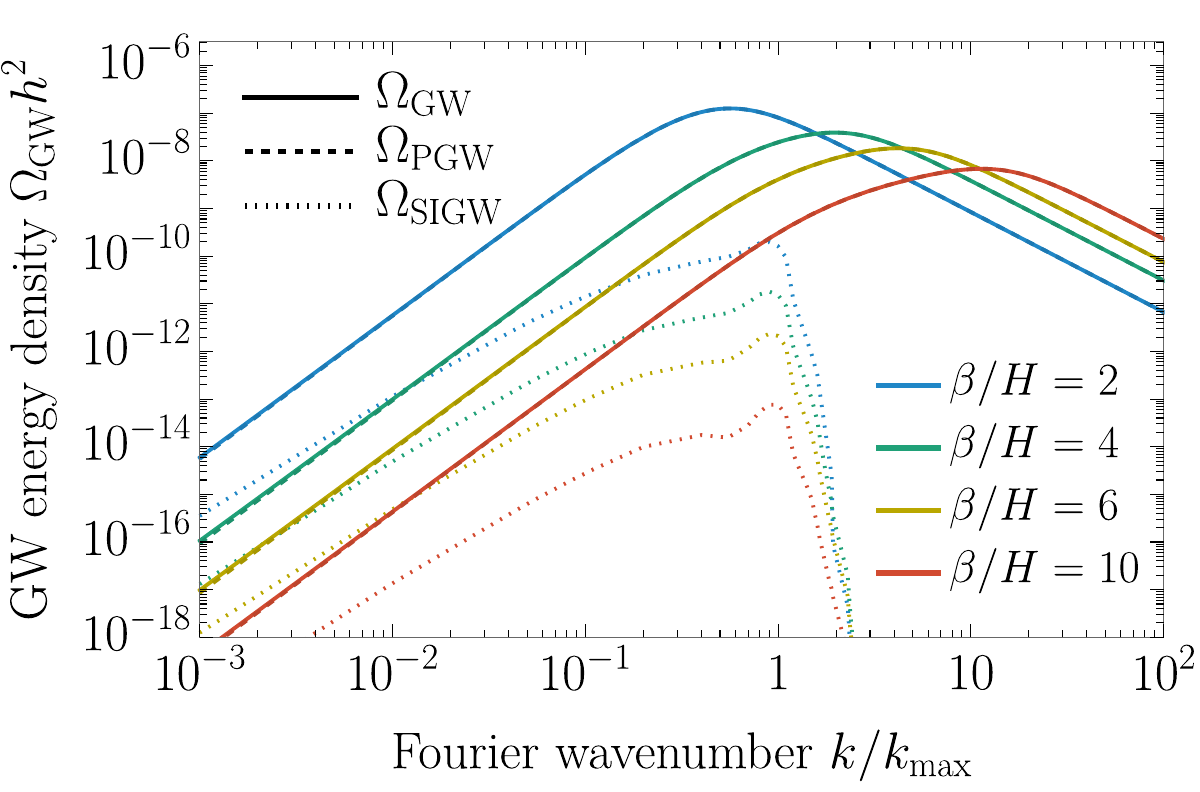}
\caption{\label{fig:2ndGW}
SIGW contribution to the GWs from FOPT. The solid line reports the total $\Omega_{\rm GW}$, while the dotted line the subdominant SIGWs, which includes the (negligible) NG corrections. 
}
\end{figure}
as the product of the polarization tensor contracted with the loop momenta $Q_{s}(\mathbf{k},\mathbf{p}) = e_{s}^{ij}(\hat{\mathbf{k}})p_ip_j$, and the kernel function $ I(\mathbf{k},\mathbf{p},\eta)$ being the convolution of source and Green function, see e.g.~Ref.~\cite{LISACosmologyWorkingGroup:2025vdz} and references therein for the explicit expressions. 
The overline denotes a time average over oscillations~\cite{Maggiore:1999vm} and $s$ labels the two GW polarizations. We take the Gaussian limit in which only disconnected contractions contribute to the four-point function in \eqref{eq:Phfull}.
We extract the curvature power spectrum,
$\langle \tilde{\mathcal{R}}(\mathbf{k})\tilde{\mathcal{R}}(\mathbf{k}')\rangle = (2\pi)^3\delta_D^{(3)}(\mathbf{k}+\mathbf{k}')(2\pi^2/k^3)\mathcal{P}_{\mathcal{R}}(k)$, from the variance of the volume-averaged curvature perturbation $\sigma_{\mathcal{R}}^2$ shown in Fig.~\ref{fig:PS_allquantities} by inverting the formula
\begin{align}
\label{eq:variance_power_spectrum}
\sigma_{\mathcal{R}}^2\equiv \left< \mathcal{R}_{\rm avg}^2\right>
= \int \frac{{\rm d}k}{k}\,| \tilde{W}(k,R)|^2 \mathcal{P}_{\mathcal{R}}(k),
\end{align}
where $\tilde{W}(k,R)$ is the Fourier-transform of the top-hat window function. Using $\mathcal{P}_\mathcal{R}(k)\propto k^3$ we obtain $\mathcal{P}_{\mathcal{R}}(k=R^{-1}) \simeq 2\sigma_\mathcal{R}^2/3\pi$. 
Fig.~\ref{fig:2ndGW} shows the resulting GW spectra for various $\beta/H$. 
We conclude that SIGWs are never strong enough to form a secondary peak in the total spectrum and at most affect the far-IR tail due to the logarithmic scaling 
$\Omega_{\rm SIGW} (k \ll k_{\star}) \, {\propto}\, k^3(1 + \tilde A \ln^2(k/\tilde k))$~\cite{Atal:2021jyo}. {In the SuM, we show that non-Gaussianities, quantified by the skewness $\langle\mathcal{R}^3\rangle$~\cite{Luo:1992er,Komatsu:2001rj,Nakama:2016gzw,Garcia-Bellido:2017aan,Unal:2018yaa,Cai:2018dig,Cai:2019amo,Ragavendra:2020sop,Yuan:2020iwf,Adshead:2021hnm,Davies:2021loj,Abe:2022xur,Garcia-Saenz:2022tzu,Garcia-Saenz:2023zue,Li:2023xtl,Yuan:2023ofl,Perna:2024ehx,Gouttenoire:2025jxe}, have negligible contributions to the SIGW signal.}

{\textit{\textbf{ Discussion and implications of our results.}}}
In this work we have employed a covariant formalism to study cosmological perturbations originating from a FOPT, with implications for PBHs and SIGWs.
Upon modeling bubble dynamics on a FLRW background, previous works have implicitly chosen the spatially-flat gauge (F). However, the values of the density contrast at the PBH formation threshold $\delta^{(C)}_c\in[0.4,0.67]$ and its relation to the comoving curvature perturbation at Hubble crossing $\mathcal{R}\simeq -9\delta^{(C)}/4 $~\footnote{{The relation $\mathcal{R}\simeq -9\delta^{(C)}/4$ follows from Eqs.~\eqref{eq:mathcalR_def_main} and \eqref{eq:Poisson_eq} for $k=\mathcal{H}$, $\omega=1/3$, and $\Phi'\simeq 0$.}} -- widely quoted in the PBH literature -- are valid in the comoving gauge (C).
At Hubble crossing, the density contrast in the spatially-flat and comoving gauges are related by $\delta^{(F)} \simeq 10\delta^{(C)}$, {see Eq.~\eqref{eq:delta_F_delta_C_link}.} 
Misidentifying these two gauges leads to an underestimation of the PBH collapse threshold by a factor of 10, leading to a dramatic overestimation of the PBH abundance.
The same gauge confusion overestimates the comoving curvature perturbation $\mathcal{R}$ by a factor $10$ and therefore overestimates SIGWs by a factor of $10^4$.
Additionally, we identified a factor of $2/3\pi$ between the variance of the spatially averaged $\mathcal{R}$ and its power spectrum, {see below Eq.~\eqref{eq:variance_power_spectrum},} further suppressing SIGWs by a factor of $20$, {leading to lower SIGWs amplitude with respect to Ref.~\cite{Lewicki:2024ghw} by a factor $2\times 10^5$.}

{Let us now discuss the implication of the fact that PBH and SIGW production from FOPTs appear to be far smaller than previously thought. First, this implies it would be harder to use PBHs to search for evidence of strong FOPTs having taken place in the primordial universe. 
Secondly, this result has important implications for the FOPT interpretation of the PTA signal: the loudest region of the posterior is not necessarily excluded by PBH overproduction constraints, particularly those from LIGO-Virgo-KAGRA~\cite{Gouttenoire:2023bqy,Ellis:2023oxs,Lewicki:2024ghw}, and FOPTs with a reheating temperature above a GeV might not explain the PTA signal~\cite{Lewicki:2024ghw}.
}

{\textit{\textbf{Limitations and Future Directions}}}
We have neglected gravitational effects on nucleation rates~\cite{Coleman:1980aw} and on the expansion of bubbles~\cite{Giombi:2023jqq,Jinno:2024nwb}, as well as the gradient energy of bubble walls~\cite{Lewicki:2023ioy,Flores:2024lng,Hashino:2025fse}.
The actual distribution of radiation converted from the vacuum energy is also important in understanding perturbations on small scales $k \gtrsim \mathcal{H}$~\cite{Jinno:2019jhi,Lewicki:2022pdb}.
{While asphericities have a small impact on the threshold for PBH collapse in the standard scenario \cite{Yoo:2020lmg}, it is not yet clear whether the same conclusion applies in the present case.}
Future work should incorporate these effects and perform full numerical simulations tracking non-linear curvature perturbations and non-adiabatic pressure effects to refine PBH formation criteria.

\textit{\textbf{ Acknowledgements.}}---%
YG thanks Caner Unal for his valuable contributions during early stages of this project, and Sokratis Trifinopoulos and Miguel Vanvlasselaer for interesting discussions. We thank Pedro Schwaller for spotting a typo in Fig. 4 and Karsten Jedamzik for drawing our attention to typographical issues in Eqs.~6 and~8.
YG acknowledges support by the Cluster of Excellence ``PRISMA+'' funded by the German Research Foundation (DFG) within the German Excellence Strategy (Project No. 390831469), and by a fellowship awarded by the Azrieli Foundation.
The work of RJ is supported by JSPS KAKENHI Grant Numbers 23K17687, 23K19048, and 24K07013.

\appendix
\onecolumngrid

\fontsize{11}{13}\selectfont

\makeatletter
\long\def\@makecaption#1#2{%
  \vskip\abovecaptionskip
  \parbox{\hsize}{%
    \fontsize{10.5}{12.5}\selectfont 
    \justifying
    #1.\ #2%
  }%
  \vskip\belowcaptionskip
}
\makeatother




\newpage
\vspace{1cm}

\ifarxiv
\begin{center}
\textbf{\it\Large Supplemental Material}
\end{center}
\renewcommand{\tocname}{\it
\vspace{-1.5cm}
}

\else
\begin{center}
\textbf{\large Supplemental Material for ``Curvature Perturbations from First-Order Phase Transitions:
Implications to Black Holes and Gravitational Waves''}
\end{center}

\noindent
\renewcommand{\tocname}{\it
\vspace{-1cm}
Gabriele Franciolini, Yann Gouttenoire, Ryusuke Jinno
}
\fi

\titleformat{\section}
{\normalfont\fontsize{12}{14}\bfseries  \centering }{\thesection.}{1em}{}
\titleformat{\subsection}
{\normalfont\fontsize{12}{14}\bfseries \centering}{\thesubsection.}{1em}{}
\titleformat{\subsubsection}
{\normalfont\fontsize{12}{14}\bfseries \centering}{\thesubsubsection)}{1em}{}

\titleformat{\paragraph}
{\normalfont\fontsize{12}{14}\bfseries  }{\thesection:}{1em}{}

 {
 \hypersetup{linkcolor=black}
 \tableofcontents
 }

\section{Background equations}

We consider a perfect fluid in a homogeneous, isotropic universe whose stress-energy tensor can be written as
\begin{equation}
    \overline{T}^\mu_{\ \nu} = (\overline{\rho} + \overline{p})\overline{u}^\mu \overline{u}_\nu - \overline{p}\delta^\mu_\nu,
\end{equation}
where $\overline{u}_\mu = a\delta^0_\mu$ and $\overline{u}^\mu = a^{-1}\delta_0^\mu$. The Einstein field equations in this background lead to the usual Friedmann and continuity equations:
\begin{equation}
    H^2 = \frac{8\pi G}{3}\overline{\rho}, \quad
    \dot{H} = -\frac{3(1+\omega)}{2}H^2, \quad
    \dot{\overline{\rho}} =-3(1+\omega)H\overline{\rho},
\end{equation}
where $H \equiv \dot{a}/a$. 
Expressing time in terms of the conformal variable $\eta$, with $\mathrm{d}t/\mathrm{d}\eta = a(\eta)$, the above relations become:
\begin{equation}
\label{eq:Friedmann_comoving}
    \mathcal{H}^2 = \frac{8\pi G}{3}\overline{\rho}a^2, \quad
    \mathcal{H}' =-\frac{1 + 3\omega}{2}\mathcal{H}^2, \quad
    \overline{\rho}' =- 3(1+\omega)\mathcal{H}\overline{\rho} ,
\end{equation}
where $\mathcal{H} \equiv a'/a$ is the comoving Hubble factor and $\omega$ is the equation of state of the universe
\begin{equation}
\label{eq:EoS}
\omega~\equiv~ \overline{p} /\overline{\rho}.
\end{equation}
Integration of the continuity and Friedmann equation for constant $\omega$ give
\begin{equation}
    \overline{\rho} \propto a^{-3(1+\omega)},\qquad \qquad
    a \propto t^{\,\frac{2}{3(1+\omega)}},\qquad \textrm{and}\qquad
    a \propto \eta^{\,\frac{2}{1+3\omega}}.
\end{equation}
Useful relationships between dot and prime derivatives include:
\begin{equation}
    H = \frac{\mathcal{H}}{a}, \quad
    \dot{H} = \frac{\mathcal{H}'}{a^2} - \frac{\mathcal{H}^2}{a^2}, \quad
    \dot{F} = \frac{F'}{a}, \quad
    \ddot{F} = \frac{F''}{a^2} - H\frac{F'}{a},
\end{equation}
where $F$ is an arbitrary function.

\section{Cosmological Scalar Perturbation Theory}

We briefly review the theory of linear cosmological scalar perturbations and the associated gauge-invariant variables used throughout this work.
This framework provides the basis for our analysis of perturbations generated during first-order phase transitions.

\subsection{Gauge-invariant linear perturbations}

\textbf{Linear metric perturbations ---}
The line element with linear perturbations around a flat FLRW metric can be written as:
\begin{equation}
    \mathrm{d}s^2 
    = 
    a^2 \Bigl[
    -\bigl(1 + 2A\bigr)\,\mathrm{d}\eta^2 
    + 2\,B_i\,\mathrm{d}x^i\,\mathrm{d}\eta 
    + \bigl(\delta_{ij} + h_{ij}\bigr)\,\mathrm{d}x^i\,\mathrm{d}x^j
    \Bigr].
\end{equation}
We follow the notations of Ref.~\cite{Peter:2013avv,Baumann:2022mni}.
The quantity $B_i$ can be split into a gradient plus a divergence-free component:
\begin{equation}
    B_i = \partial_i B + \hat{B}_i,
\end{equation}
and the symmetric perturbation $h_{ij}$ can be further decomposed into scalar, vector, and tensor contributions:
\begin{equation}
    h_{ij} = 2C\delta_{ij} 
    + 2\partial_{\langle i}\partial_{j\rangle}E
    + 2\partial_{(i}\hat{E}_{j)} 
    + \hat{E}_{ij}
\end{equation}
where $\partial_{\langle i}\partial_{j\rangle} E \equiv (\partial_i\partial_j - \tfrac{1}{3}\delta_{ij}\nabla^2)E$ and $\partial_{(i} \hat{E}_{j)}
= \frac{1}{2} \left( \partial_i \hat{E}_j + \partial_j \hat{E}_i \right)$. The “hatted” functions ($\hat{B}_i$, $\hat{E}_i$, $\hat{E}_{ij}$) are divergence-free, while $\hat{E}_{ij}$ is also traceless. Overall, the ten metric degrees of freedom decompose into four scalars ($A,B,C,E$), four vector components ($\hat{B}_i, \hat{E}_i$), and two tensor modes ($\hat{E}_{ij}$).
Popular gauge choices are, e.g.~\cite{Kodama:1984ziu}:
\begin{enumerate}
    \item \textit{Spatially-Flat gauge} ($C=E=0$)  
    \begin{equation}
    \label{eq:SF_gauge}
        \mathrm{d}s^2 
        = 
        a^2 \bigl[
        -(1 + 2A)\,\mathrm{d}\eta^2 
        + 2\,B_i\,\mathrm{d}x^i\,\mathrm{d}\eta 
        + \mathrm{d}\mathbf{x}^2
        \bigr].
    \end{equation}

    \item \textit{Newtonian gauge} ($B=E=0$)  
    \begin{equation}
    \label{eq:Newtonian_gauge}
        \mathrm{d}s^2 
        =
        a^2 \bigl[
        -(1 + 2A)\,\mathrm{d}\eta^2 
        + (1 + 2C)\,\mathrm{d}\mathbf{x}^2
        \bigr].
    \end{equation}

    \item \textit{Comoving gauge} ($v + B = 0,~ E=0$)  
    \begin{equation}
    \label{eq:Comoving_gauge}
        \mathrm{d}s^2 
        =
        a^2\bigl[
        -(1 + 2A)\,\mathrm{d}\eta^2 
        + 2\,B_i\,\mathrm{d}x^i\,\mathrm{d}\eta 
        + (1 + 2C)\,\mathrm{d}\mathbf{x}^2
        \bigr].
    \end{equation}
\end{enumerate}

\textbf{Linear matter perturbations ---}
Starting from
$T^\mu{}_\nu=(\rho+p)u^\mu u_\nu+p\,\delta^\mu{}_\nu+\Sigma^\mu{}_\nu$,
the linear perturbation reads~\cite{Peter:2013avv,Baumann:2022mni}
\begin{equation}
\delta T^\mu{}_\nu
=(\delta\rho+\delta p)\,\overline{u}^\mu\overline{u}_\nu
+(\overline{\rho}+\overline{p})\big(\delta u^\mu\overline{u}_\nu+\overline{u}^\mu\delta u_\nu\big)
-\delta p\,\delta^\mu{}_\nu-\Sigma^\mu{}_\nu,
\end{equation}
where $\Sigma^\mu{}_\nu$ is the first order anisotropic stress.
The unit norm $u^\mu u_\mu=\overline{u}^\mu \overline{u}_\mu=-1$ implies
\begin{equation}
\delta g_{\mu\nu}\overline{u}^\mu \overline{u}^\nu + 2\overline{u}_\mu\delta u^\mu=0.
\end{equation}
Using $\overline{u}^\mu=a^{-1}\delta^\mu_0$ and $\delta g_{00}=-2a^2A$, we get $\delta u^0=-Aa^{-1}$.
Parametrizing the spatial velocity by $\delta u^i=v^i/a$, the 4-velocity reads
\begin{equation}
u^\mu=a^{-1}(1-A,\,v^i),\qquad
u_\mu=a(-(1+A),\,v_i+B_i),
\end{equation}
where we used $u_\alpha=g_{\alpha\beta}u^\beta$ and the convention $v_i=\delta_{ij}v^j$. By definition, the anisotropic stress is purely spatial in the fluid rest
frame and therefore satisfies $\Sigma^\mu{}_\nu u^\nu=0$ and
$\Sigma^\mu{}_\mu=0$.
We deduce that the perturbed stress-energy components are:
\begin{align}
    &T^0_{\ 0} = -(\overline{\rho} + \delta\rho), &T^0_{\ i} = (\overline{\rho} + \overline{p})(v_i + B_i),  \\
   & T^i_{\ 0} = -(\overline{\rho} + \overline{p})\,v^i\equiv q^i, &T^i_{\ j} = (\overline{p} + \delta p)\delta^i_j + \Sigma^i_{\ j}.
\end{align}
We introduce the scalar part of the 3-velocity and of the 3-momentum~\footnote{We follow the convention of \cite{Peter:2013avv,Baumann:2022mni} where $v_i=\delta_{ij}v^j$. Instead Refs.~\cite{Mukhanov:1990me,Mukhanov:2005sc} define $v_i=-a^2v^i$.}
\begin{equation}
    \partial_i( v) \equiv  v_i,\qquad \textrm{and}\qquad  q\equiv (\overline{\rho}+\overline{p})v.
\end{equation}
We define the density contrast $\delta \equiv \delta \rho/\rho$.

\textbf{Gauge-invariant quantities ---}
Under an infinitesimal coordinate transformation generated by scalar functions $T$ and $L$
\begin{equation}
\label{eq:coordinate_transformation}
x^\mu(q)\rightarrow \tilde{x}^\mu(q)\equiv x^\mu(q)+\xi^\mu(q),\quad \textrm{where}\quad \xi^0\equiv T,\quad\textrm{and}\quad \xi^i\equiv \partial^i L,
\end{equation}
where $q$ denotes a physical spacetime point, the scalar metric perturbations transform as~\cite{Baumann:2022mni}
\begin{equation}
\label{eq:metric_shift}
    A\rightarrow A-T'-\mathcal{H}T,\quad B\rightarrow B+T-L',\quad C\rightarrow C-\mathcal{H}T-\frac{1}{3}\nabla^2 L,\quad E\rightarrow E-L.
\end{equation}
These transformation laws motivate the definition of two gauge-invariant combinations, the Newtonian (or Bardeen) potentials~\cite{Bardeen:1980kt, Baumann:2022mni}:
\begin{align}
\label{eq:Bardeen_Potentials}
    \Psi \equiv A + \mathcal{H}(B - E') + (B - E')', \qquad \Phi \equiv -C + \frac{1}{3}\nabla^2 E - \mathcal{H}(B - E').
\end{align}
Under the same coordinate transformation in Eq.~\eqref{eq:coordinate_transformation}, matter perturbations transform as~\cite{Baumann:2022mni}
\begin{equation}
\label{eq:matter_shift}
    \delta \rightarrow \delta-\frac{\overline{\rho}'}{\overline{\rho}} T,\quad \delta p\rightarrow \delta p-\overline{p}'T,\quad v\to v+L',\quad \Sigma_{i j}\rightarrow \Sigma_{ij}.
\end{equation}
This in turn motivates the introduction of the following gauge-invariant matter variables~\cite{Peter:2013avv}
\begin{equation}
\label{eq:density_constrast_defs}
   \mathcal{V} \equiv v + E',\quad \delta^{(N)} \equiv \delta + \frac{\overline{\rho}'}{\overline{\rho}}(B - E'),\quad \delta^{(C)}\equiv \delta  + \frac{\overline{\rho}'}{\overline{\rho}}(v+B) ,\quad \delta^{(F)}\equiv \delta+\frac{\overline{\rho}'}{\mathcal{H}\overline{\rho}}\left( -C+\frac{1}{3}\nabla^2E\right).
\end{equation}
The quantities $ \delta^{(N)}$, $ \delta^{(C)}$, and  $\delta^{(F)}$ reduce to the contrast density in the Newtonian, comoving and spatially-flat gauges~\footnote{Note that Ref.~\cite{Peter:2013avv} introduces $\delta^{(F)}\equiv \delta-(\overline{\rho}'/\overline{\rho})(C/\mathcal{H})$ instead of $\delta^{(F)}\equiv \delta-(\overline{\rho}'/\overline{\rho})(C/\mathcal{H})+(\overline{\rho}'/\overline{\rho})(\nabla^2E/3\mathcal{H})$ which is however not invariant under the transformation rules in Eq.~\eqref{eq:metric_shift}.}, defined in Eqs.~\eqref{eq:SF_gauge}, \eqref{eq:Newtonian_gauge} and \eqref{eq:Comoving_gauge}, respectively. The quantity $ \mathcal{V}$ reduce to the coordinate velocity $v$ in those three gauges.
Similarly, we define the gauge-invariant pressure perturbations,
\begin{equation}
    \delta p^{(N)} \equiv \delta p + \overline{p}'(B - E'),\quad \delta p^{(C)}\equiv \delta p  + \overline{p}'(v+B) ,\quad \delta p^{(F)} \equiv \delta p+\frac{\overline{p}'}{\mathcal{H}}\left( -C+\frac{1}{3}\nabla^2E\right).
\end{equation}

\textbf{Linearized Einstein equations ---}
In the absence of anisotropic stress ($\Sigma^i_{\ j}=0$) the trace-free part of Einstein equations along $ij$ leads to
\begin{equation}
\label{eq:no_anisotropy}
    \Psi = \Phi.
\end{equation}
Their expressions in Eqs.~\eqref{eq:Bardeen_Potentials} expressed in the gauges defined in Eqs.~\eqref{eq:SF_gauge}, \eqref{eq:Newtonian_gauge} and \eqref{eq:Comoving_gauge}, are 
\begin{equation}
\label{eq:Psi_Phi_different_gauges}
    \Psi =
    \begin{cases}
        A, & \\
        A + \mathcal{H}B + B', &  \\
        A + \mathcal{H}B + B', & .
    \end{cases}\qquad  \Phi =
    \begin{cases}
        -C, & \text{(Newtonian gauge)}, \\
        -\mathcal{H}B, & \text{(spatially-flat gauge)}, \\
        -C - \mathcal{H}B, & \text{(Comoving gauge)}.
    \end{cases}
\end{equation}
The condition in Eq.~\eqref{eq:no_anisotropy} imposes specific relationships among $A$, $B$, and $C$ in different gauges
\begin{equation}
\label{eq:Psi_equal_Phi}
    \Psi=\Phi\quad \Rightarrow \quad\begin{cases}
    A=-C,\qquad \qquad \qquad \qquad \qquad\qquad\textrm{(Newtonian gauge)},\\
    A+2\mathcal{H}B+B'=0,\quad \qquad \qquad \qquad\textrm{(Spatially-flat gauge)},\\
  A+C+2\mathcal{H}B+B'=0,\qquad \qquad \qquad \textrm{(Comoving gauge)}.
    \end{cases}
\end{equation} 
The linearised Einstein equations, $00$ component, spatial integration of $0j$ component and trace of $ij$ components, give in the Newtonian gauge, e.g.~\cite{Mukhanov:1990me,Baumann:2022mni}:
\begin{align}
    \nabla^2\Phi - 3\mathcal{H}(\Phi' + \mathcal{H}\Phi) &= \frac{3}{2}\mathcal{H}^2\delta^{(N)},\label{eq:Einstein_eq_0_0}\\
    \Phi' + \mathcal{H}\Phi &= -\frac{3}{2}\mathcal{H}^2 (1+\omega)\mathcal{V},\label{eq:Einstein_eq_0_i}\\
    \Phi'' + 3\mathcal{H}\Phi' -3\omega\mathcal{H}^2\Phi &= \frac{3}{2}\mathcal{H}^2\frac{\delta p^{(N)}}{\overline{\rho}}.\label{eq:Einstein_eq_i_j}
\end{align}
 Combining Eqs.~\eqref{eq:Einstein_eq_0_0} and \eqref{eq:Einstein_eq_i_j}, we obtain the evolution equation for the gauge-invariant Newtonian potential
\begin{equation}
\label{eq:Newtonian_potential_evolution_app}
\Phi'' + 3 (1 + c_s^2)\mathcal{H} \Phi' + (3 (c_s^2-\omega)\mathcal{H}^2-c_s^2 \Delta)\Phi =\frac{3}{2}\mathcal{H}^2 \frac{\delta p_{\rm nad}}{\overline{\rho}},
\end{equation}
where we used $\mathcal{H}' = -(1 + 3\omega)\mathcal{H}^2/2$ given by Eq.~\eqref{eq:Friedmann_comoving} and where $\delta p_{\rm nad}$ is the non-adiabatic (or entropic) part of the density perturbation
\begin{equation}
\label{eq:delta_nad_app_def}
    \delta p_{\rm nad}\equiv \delta p-c_s^2\delta \rho,
\end{equation} 
where $c_s^2$ is the (adiabatic) speed of sound of the universe. It is a background quantity defined by
\begin{equation}
\label{eq:SoS}
c_s^2 \equiv \frac{\delta p}{\delta \rho}\Big|_{\delta p_{\rm nad}=0} = \frac{\overline{p}'}{\overline{\rho}'} = \omega -\frac{\omega'}{3(1+\omega)\mathcal{H}}.
\end{equation}
The adiabatic condition $\delta p_{\rm nad}=0$ implies that perturbations correspond to a local time shift of the background, $\eta\to\eta+T(\eta,\mathbf{x})$~\cite{Baumann:2022mni}. To first order, this gives $\delta\rho=\overline{\rho}'\,T$ and $\delta p=\overline{p}'\,T$, yielding the second equality in Eq.~\eqref{eq:SoS}. Another interpretation is that fluctuations in pressure arise solely from fluctuations in density, hence the local equation of state $p(\rho)$ is uniform and $\delta p/\delta \rho = d\overline{p}/d\overline{\rho}$~\cite{Bardeen:1983qw}. The last equality in Eq.~\eqref{eq:SoS} follows from Eqs.~\eqref{eq:EoS} and \eqref{eq:Friedmann_comoving}.
Combining Eq.~\eqref{eq:Einstein_eq_0_0} and \eqref{eq:Einstein_eq_0_i}, we get the relativistic Poisson equation
\begin{equation}
\label{eq:Einstein_eq_00_0i_Poisson}
    \nabla^2\Phi ~=~ \frac{3}{2}\mathcal{H}^2\delta^{(C)},
\end{equation}
where the comoving density contrast $\delta^{(C)}$ is given by the gauge-invariant expressions:
\begin{align}
\label{eq:Delta_def}
    \delta^{(C)}= \delta^{(N)} +2\Phi+\frac{2\Phi'}{\mathcal{H}} =\delta^{(N)}  + \frac{\overline{\rho}'}{\overline{\rho}}\mathcal{V}=\delta  + \frac{\overline{\rho}'}{\overline{\rho}}(v+B) ,
\end{align}
with the three equalities being derived from Eqs.~\eqref{eq:Einstein_eq_0_0}, \eqref{eq:Einstein_eq_0_i}, \eqref{eq:density_constrast_defs}, respectively.
Using Eqs.~\eqref{eq:Bardeen_Potentials},~\eqref{eq:density_constrast_defs} and \eqref{eq:Einstein_eq_0_i}, we deduce the transformation rules between the density contrast in the different gauges
\begin{align}
    &\delta^{(C)}-\delta^{(N)}=+2\Phi+\frac{2\Phi'}{\mathcal{H}}, \label{eq:delta_C_delta_N}\\
    &\delta^{(F)}-\delta^{(N)}=-3(1+\omega)\Phi, \label{eq:delta_F_delta_N}\\
    &\delta^{(C)}-\delta^{(F)}= (5+3\omega)\Phi +\frac{2\Phi'}{\mathcal{H}}.\label{eq:delta_C_delta_F}
\end{align}
We can check that those transformation rules are in agreement with the literature, e.g. \cite[Eqs.~5.87]{Peter:2013avv}.
The continuity equation in the different gauges are given by
\begin{align}
&\delta^{(N)'}+3\mathcal{H}(c_s^2-\omega)\delta^{(N)}=-(1+\omega)(\Delta \mathcal{V}-3\Phi')-3\mathcal{H}\frac{\delta p_{\rm nad}}{\overline{\rho}},\label{eq:delta_N_eq}\\
&\delta^{(F)'}+3\mathcal{H}(c_s^2-\omega)\delta^{(F)}=-(1+\omega)\Delta \mathcal{V}-3\mathcal{H}\frac{\delta p_{\rm nad}}{\overline{\rho}},\label{eq:delta_F_eq}\\
&\delta^{(C)'}-3\omega\mathcal{H}\delta^{(C)}=-(1+\omega)\Delta \mathcal{V}.\label{eq:delta_C_eq}
\end{align}
The first equation in Eq.~\eqref{eq:delta_N_eq} can be calculated from a combination of Eq.~\eqref{eq:Einstein_eq_0_0} and its derivative. The second equation in Eq.~\eqref{eq:delta_F_eq} can be deduced from the first using Eq.~\eqref{eq:delta_F_delta_N}. The third equation in Eq.~\eqref{eq:delta_C_eq} can be derived from taking the derivative of Eq.~\eqref{eq:Einstein_eq_00_0i_Poisson}, using Eq.~\eqref{eq:Einstein_eq_0_i} and finally using Eq.~\eqref{eq:Einstein_eq_00_0i_Poisson} again. The first two equations are also given in \cite[Eqs.~5.125 and 5.128]{Peter:2013avv}.

\subsection{Gauge-invariant curvature perturbations}
We can introduce two additional gauge-invariant quantities to express curvature perturbations. At first the comoving curvature perturbation $\mathcal{R}$~\cite{Bardeen:1980kt}
\begin{align}
\mathcal{R} ~\equiv ~ -C+\frac{1}{3}\nabla^2E-\mathcal{H}(v+B) ~=~\Phi - \mathcal{H}\mathcal{V} ~=~ \frac{5+3\omega}{3+3\omega} \Phi +\frac{2\Phi'}{3(1+\omega)\mathcal{H}},
\label{eq:mathcalR_def}
\end{align}
where the third equality follows from Eq.~\eqref{eq:Einstein_eq_0_i}.
The curvature perturbation on uniform density hypersurface $\zeta$~\cite{Bardeen:1983qw,Wands:2000dp,Lyth:2004gb}:
\begin{equation}
\label{eq:zeta_def}
\zeta ~\equiv ~ -C+\frac{1}{3}\nabla^2E+\mathcal{H}\frac{\overline{\rho}}{\overline{\rho}'}\delta ~=~ \Phi-\frac{1}{3}\frac{\delta^{(N)}}{1+\omega} = -\frac{1}{3}\frac{\delta^{(F)}}{1+\omega} ~=~ \mathcal{R}-\frac{\delta^{(C)}}{3(1+\omega)},
\end{equation}
where the third and fourth equalities follow from Eqs.~\eqref{eq:delta_F_delta_N} and \eqref{eq:delta_C_delta_F}. Those two gauge-invariant quantities, $\mathcal{R}$  and $\zeta$,  are particularly convenient due to their laws of conservation. Using $\mathcal{H}' = -(1+3\omega)\mathcal{H}/2$, $\omega'=-3\mathcal{H}(1+\omega)(c_s^2-\omega)$, Eq.~\eqref{eq:Einstein_eq_0_i} and Eq.~\eqref{eq:Newtonian_potential_evolution_app}, we calculate
\begin{align}
\label{eq:R_prime_Phi_prime}
\mathcal{R}' ~=~  \mathcal{H} \frac{\delta p_{\rm nad}}{\overline{\rho}+\overline{p}} - \frac{2c_s^2}{3(1+\omega)}\frac{k^2}{\mathcal{H}}\Phi,\qquad \textrm{and} \qquad\zeta' ~= ~ \mathcal{H} \frac{\delta p_{\rm nad}}{\overline{\rho}+\overline{p}} - \frac{k^2 \mathcal{V}}{3}.
\end{align}
The Fourier index $k$ is suppressed for clarity.
In the absence of large-scale non-adiabatic pressure fluctuation $\delta p_{\rm nad}=0$, the curvature perturbations are conserved $\mathcal{R}'\simeq \zeta'\simeq 0$ on super-horizon scales $k\ll \mathcal{H}$. Plugging $\delta p_{\rm nad}=0$ and $c_s^2\simeq \omega$ into Eq.~\eqref{eq:Newtonian_potential_evolution_app}, we deduce that the Newtonian potential is also conserved  $\Phi'\simeq 0$ for adiabatic evolution and super-horizon scales.
For constant Newtonian potential $\Phi'\simeq 0$, Eqs.~\eqref{eq:Einstein_eq_00_0i_Poisson}, \eqref{eq:delta_C_delta_N}, \eqref{eq:delta_F_delta_N}, \eqref{eq:delta_C_delta_F}, and \eqref{eq:mathcalR_def}  lead to
\begin{equation}
\label{eq:R_relations}
-\mathcal{R}\simeq -\left( \frac{5+3\omega}{3+3\omega} \right)\Phi \simeq  \frac{5+3\omega}{2(1+\omega)}\left(\frac{\mathcal{H}}{k}\right)^2\delta^{(C)}\simeq\frac{5+3\omega}{1+\omega}\frac{\left(\mathcal{H}/k\right)^2\delta^{(F)}}{2+3(5+3\omega)(\mathcal{H}/k)^2}\simeq\frac{5+3\omega}{2(1+\omega)}\frac{\left(\mathcal{H}/k\right)^2\delta^{(N)}}{1+3(\mathcal{H}/k)^2}.
\end{equation}
Using Eqs.~\eqref{eq:zeta_def}, we deduce 
\begin{equation}
\label{eq:zeta_relations}
-\zeta  \simeq  \frac{(2+3(5+3\omega)(\mathcal{H}/k)^2)}{6(1+\omega)}\delta^{(C)}\simeq \frac{\delta^{(F)}}{3(1+\omega)}\simeq\frac{(2+3(\mathcal{H}/k)^2(5+3\omega)}{6(1+\omega)(1+3(\mathcal{H}/k)^2)}\delta^{(N)},
\end{equation}
as well as 
\begin{equation}
\label{eq:R_zeta_relations}
    \zeta \simeq  \frac{(2+3(5+3\omega)(\mathcal{H}/k)^2)}{3(5+3\omega)(\mathcal{H}/k)^2}\mathcal{R}.
\end{equation}
We recover the well-known result that $\zeta \simeq \mathcal{R}$ in the super-horizon limit $k\ll \mathcal{H}$ in the absence of large-scale non-adiabatic pressure perturbation.
Eq.~\eqref{eq:R_relations}, \eqref{eq:zeta_relations}, and \eqref{eq:R_zeta_relations} become exact in the limit $\Phi' \to 0$.
Evaluating those expression at horizon crossing $\mathcal{H}=k$ and assuming a radiation-dominated universe $\omega=1/3$, we obtain
\begin{align}
\label{eq:R_relations_rad}
 -\mathcal{R}~\simeq~ -\frac{9}{10}\zeta~\simeq~ -\frac{3}{2}\Phi~ \simeq  ~ \frac{9\delta^{(F)}}{40}~\simeq~ \frac{9}{16}\delta^{(N)}~ \simeq ~ \frac{9}{4}\delta^{(C)} .
\end{align}
Fig.~\ref{fig:Variance_betaOH4_perturbations} (right) shows that the adiabatic and horizon-entry approximations in Eq.~\eqref{eq:R_relations_rad} remain accurate during
a first-order phase transition, except for modes entering the horizon near
percolation, $k \sim k_{\rm max}$, due to violation of the non-adiabaticity condition $\delta p_{\rm nad}\neq 0$.

\section{Application to a First-Order Phase Transition}
\label{app:Bubble_growth_theory}

In the previous section, we reviewed the theory of cosmological perturbations. This framework can be applied to any specific scenario once the relevant input quantities are specified, namely the equation-of-state parameter $\omega$, the sound speed $c_s^2$, the non-adiabatic pressure perturbation $\delta p_{\rm nad}$, and the initial conditions.
We now apply this formalism to a first-order phase transition. In Sec.~\ref{subsec:deltapnad}, we derive the expressions for the aforementioned input quantities. The non-adiabatic pressure perturbation $\delta p_{\rm nad}$ relies on the false vacuum fraction $F$ which we show how to compute with $\deltaPT$ in Sec.~\ref{subsec:falsevacuum}. Finally, the cosmological perturbations are numerically calculated in Sec.~\ref{subsec:cosmo_pert}.

\subsection{The non-adiabatic pressure fluctuation}
\label{subsec:deltapnad}

During a FOPT, the energy and pressure densities are 
\begin{equation}
\label{eq:rho_p_bar_app}
\rho=\rho_R+\rho_V,\qquad \textrm{and}\qquad p=\rho_R/3-\rho_V,\qquad \textrm{with}\qquad \rho_V(t) = \Delta V F(t),
\end{equation} 
where $\Delta V$ is the vacuum energy difference accross the phase transition and $F$ is the volume fraction still remaining in the old vacuum.
We deduce the equation of state $\omega=\overline{p}/\overline{\rho}$ and speed of sound $c_s^2=\overline{p}'/\overline{\rho}'$  of the universe
\begin{equation}
\label{eq:omega_cs_app}
\omega = \frac{\overline{\rho}_R/3-\overline{\rho_V}}{\overline{\rho_R}+\overline{\rho_V}},\qquad \textrm{and}\qquad  c_s^2 = \frac{1}{3}\left(1+\frac{\overline{\rho_V}'}{\mathcal{H}\overline{\rho_{R}}} \right).
\end{equation}
Neglecting prior seeds like primordial curvature perturbations produced during inflation and responsible for galaxy formation, the growth of the linear perturbations --- whether it is the gravitational potential $\Phi$, the density contrast $\delta^{(X)}$ in a given gauge $(X)$ or the gauge-invariant velocity $\mathcal{V}$ --- are sourced by the non-adiabatic perturbation $\delta p_{\rm nad} =\delta p^{(X)}-c_s^2\delta\rho^{(X)}$. Using Eqs.~\eqref{eq:rho_p_bar_app} and \eqref{eq:omega_cs_app}, we can write
\begin{equation}
\label{eq:delta_nad_FOPT}
\delta p_{\rm nad} = \frac{1-3c_s^2}{3}\overline{\rho}\,\delta^{(X)}-\frac{4}{3}\Delta V(F_k^{(X)}-\overline{F}),
\end{equation}
where $F_k^{(X)}=\rho_{\rm V}^{(X)}/\Delta V$ is the false vacuum fraction in a finite comoving volume $V(k)=4\pi k^{-3}/3$ and $\overline{F}$ is the background value. 
The spatially-flat gauge (F) is particularly convenient,
since constant-time hypersurfaces are flat, cf.~Eq.~\eqref{eq:SF_gauge}. As discussed in the main text, this suggests that the false
vacuum fraction in the spatially-flat gauge can be modeled by assuming an FLRW
background, which we adopt here as a working approximation,
\begin{equation}
\label{eq:F_k_F_k_FLRW}
F_k^{(F)} \simeq F_k^{(\mathrm{FLRW})},
\end{equation}
where $F_k^{(\mathrm{FLRW})}$ is evaluated using \deltaPT~\cite{Lewicki:2024ghw}.
Note that the expression for $F_k^{(X)}$ in the other gauges can determined by recalling that the non-adiabatic pressure perturbation $\delta p_{\rm nad}$ is gauge-independent, see e.g. Eq.~\eqref{eq:R_prime_Phi_prime}. In order to leave Eq.~\eqref{eq:delta_nad_FOPT} invariant under a gauge transformation $X\to Y$, the false vacuum fraction $F_k^{(X)}$ must transform as
\begin{equation}
F_k^{(X)}- F_k^{(Y)} = \frac{1-3c_s^2}{4}\frac{\overline{\rho}}{\Delta V}\left(\delta^{(X)}-\delta^{(Y)}\right).
\end{equation}

\subsection{The false vacuum fraction}
\label{subsec:falsevacuum}

The false vacuum fraction in a finite comoving volume $V(k)=4\pi k^{-3}/3$ in a FLRW background $F_k^{(\rm FLRW)}$, introduced in Eq.~\eqref{eq:F_k_F_k_FLRW}, can be calculated with the public code \deltaPT~\cite{Lewicki:2024ghw}{, now upgraded to \deltaPTvTwo~\cite{deltaPTv2}.}
We now review the calculation.
The false vacuum fraction is decomposed into two components: one that arises from the first $j_c$ bubbles and another from the remaining bubbles~\cite{Lewicki:2024ghw}
\begin{equation}
\label{eq:F_FLRW_def_app}
F_k^{(\rm FLRW)}(t) = F_k^{(j\leq j_c)}(t)F_k^{(j> j_c)}(t).
\end{equation} 
The contribution from the first $j_c$ bubbles is computed stochastically by drawing,
from their respective probability distributions, the times $\{t_j\}_{j\le j_c}$ at
which the $j^{\rm th}$ bubble first nucleates inside or reaches the volume $V(k)$,
together with the corresponding distances $\{d_j\}_{j\le j_c}$ from the center of
$V(k)$. The probability distribution for $t_j$ is a Poisson distribution~\cite{Lewicki:2024ghw}
\begin{equation} \label{eq:poisson}
p_{t,j}(t;k) = \frac{\td \bar{N}_k(t)}{\td t} \frac{\bar{N}_k(t)^{j-1}}{\Gamma(j)} e^{-\bar{N}_k(t)} ,\qquad \textrm{with}\quad \bar{N}_k(t) = \frac{4\pi}{3} \int_{-\infty}^t \td t_n, \Gamma(t_n) a(t_n)^3 \left[ k^{-1} + R(t;t_n) \right]^3.
\end{equation}
The quantity $\bar{N}_k(t)$ is the average number of bubbles with non-vanishing overlap with the volume $a^3 V(k)$ at time $t$. We recall that $\Gamma(t)=H_0^4\,\textrm{exp}(\beta t)$ is the nucleation rate and $H_0$ the Hubble rate in $t=0$.
A bubble that reaches or nucleates within the volume $V(k)$ at time $t_j$ must have formed at a distance $d_j < k^{-1} + R(t;t_n)$, where $t_n$ is the time of nucleation. The probability distribution for this distance, $d_j$, increases with $d_j^2$ up to the maximum radius~\cite{Lewicki:2024ghw}
\begin{equation} \label{eq:pd}
p_d(d_j;t,k) = \frac{4\pi d_j^2}{\bar{N}_k(t)} \int_{-\infty}^t \td t_n \Gamma(t_n) a(t_n)^3 \theta(k^{-1} + R(t;t_n) - d_j) .
\end{equation}
The contribution of the first $j_c$ bubbles is~\cite{Lewicki:2024ghw} 
\begin{equation} \label{eq:Fk_first}
F_k^{(j\leq j_c)}(t) \simeq \prod_{j=1}^{j_c} \left[1-f(t;t_j,d_j,k)\right] ,\qquad \textrm{with} \quad f(t;t_j,d_j,k) = \frac{V_{\rm int}(d_j,\max[0,d_j-k^{-1}]+R(t;t_j),k^{-1})}{V(k)}.
\end{equation}
The quantity $f(t;t_j,d_j,k)$ is the fraction of the volume $V(k)$ occupied by the $j$th bubble.
The quantity $V_{\rm int}$ is the volume of the overlap between two spheres, with radii $R$ and $r$ and separated by a distance $d$~\cite{Lewicki:2024ghw}
\begin{equation}
V_{\rm int}(d,R,r) =
\begin{cases}
\frac{4\pi}{3} d^3 \xi\left(\frac{R}{d},\frac{r}{d}\right) & \text{if } |R-r| \leq d \leq R+r, \\
\frac{4\pi}{3} \min[r,R]^3 & \text{if } d < |R-r|, \\
0 & \text{if } d > R+r,
\end{cases}
\end{equation}
where $\xi(x,y)=(1-x-y)^2(1+2(x+y)-3(x-y)^2)/16$.
The contribution from the remaining bubbles is approximated as an averaged quantity~\cite{Lewicki:2024ghw}
\begin{align}
\label{eq:F_k_second_1}
F_k^{(j> j_c)}(t) \simeq \textrm{exp}\left[-\sum^{+\infty}_{j=j_c+1}\int dt_jdd_j p_{t,j}(t_j;k)p_d(d_j;t_j,k)f(t;t_j,d_j,k) \right].
\end{align}
Approximating the sum over $j$ by an integral and the Poisson distribution in Eq.~\eqref{eq:poisson} by a delta function $p_{t,j}(t;k)=\dot{\overline{N}}_k(t)\delta_{D}\left(\overline{N}_k(t)-j\right)$, the expression in Eq.~\eqref{eq:F_k_second_1} simplifies to~\cite{Lewicki:2024ghw}
\begin{align}
\label{eq:F_k_second_2}
F_k^{(j> j_c)}(t)\simeq \textrm{exp}\left[-\int_{-\infty}^t d\tilde{t}d\tilde{d} ~\Theta\left(\overline{N}_k(\tilde{t})-j_c\right)\frac{d\overline{N}_k(\tilde{t})}{d\tilde{t}}p_d(\tilde{d};\tilde{t},k)f(t;\tilde{t},\tilde{d},k) \right].
\end{align}

\begin{figure}[ht!]
\centering
\includegraphics[width=0.49\textwidth]{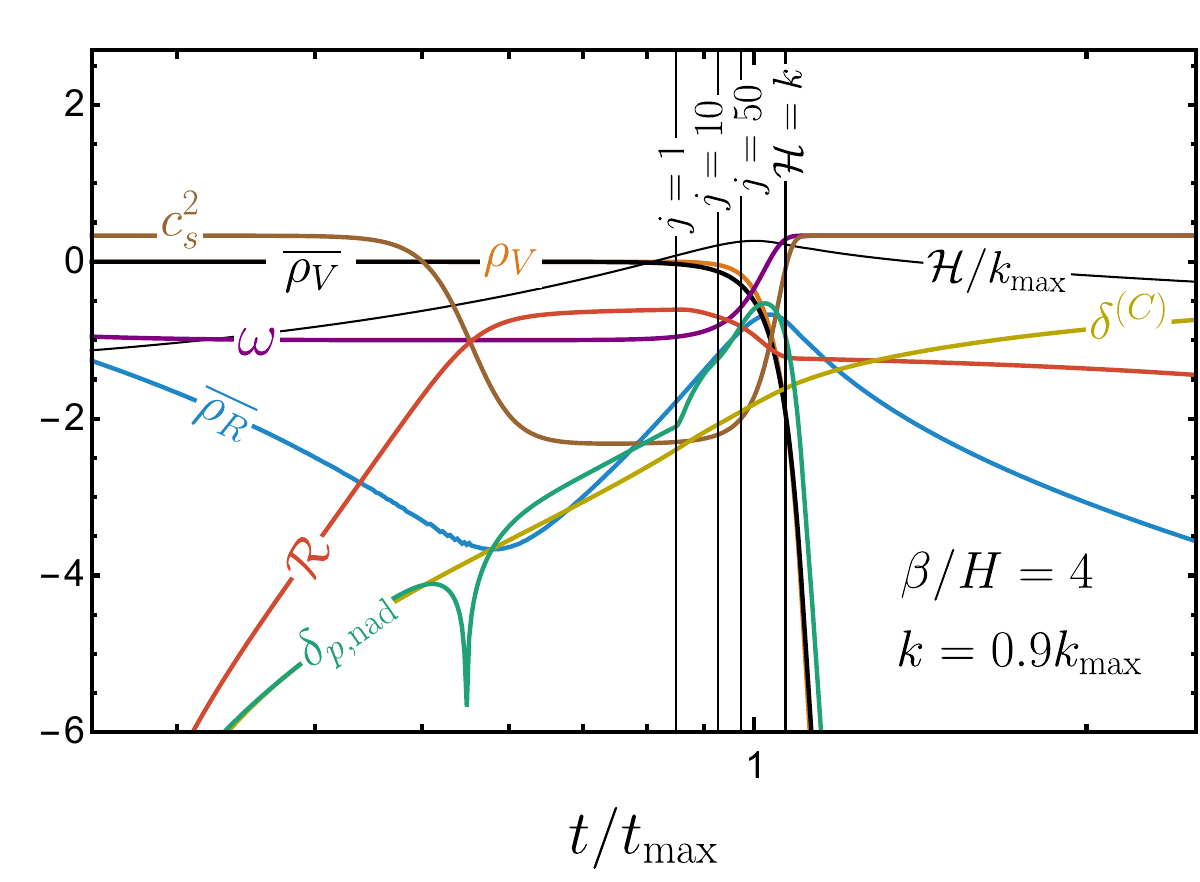}
\includegraphics[width=0.49\textwidth]{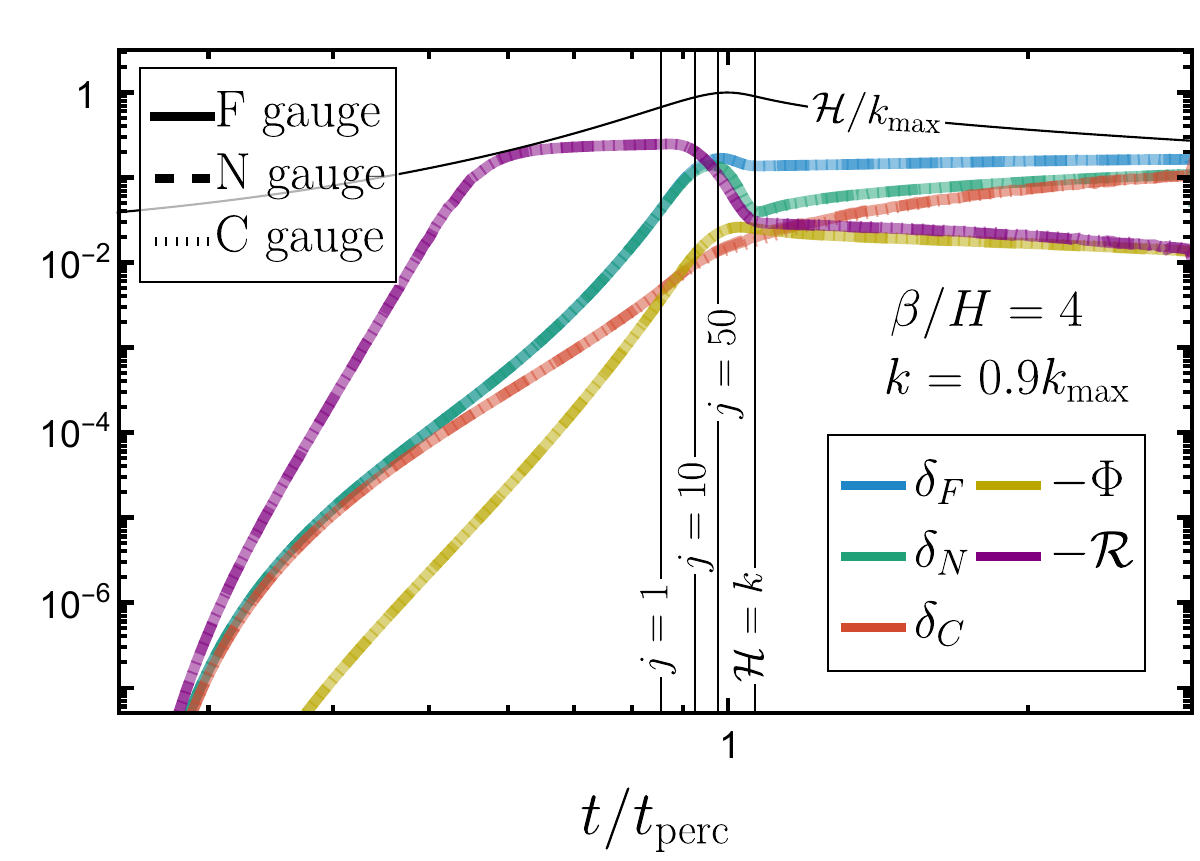}
\caption{  \label{fig:plot_delta_Phi_gauge_inv} 
\textbf{Left:} Background quantities  $\rho_{\rm R}$, $\rho_{\rm V}$,  $\omega$, $c_s^2$ and linear perturbations $\delta^{(C)}$, $\delta p_{\rm nad}$ and $\mathcal{R}$. The vertical lines indicate the nucleation times of the $j^{\rm th}$ bubble appearing within the spherical patch of comoving size $R=k^{-1}$, modeled by {\deltaPTvTwo~\cite{deltaPTv2}.} \textbf{Right:} 
The exact agreement between the solid, dashed, and dotted curves confirms the covariance of the system of equations and the internal consistency of the numerical implementation.   The solid, dashed and dotted lines employ Eqs.~\eqref{eq:delta_F_eq}, \eqref{eq:delta_N_eq}, and \eqref{eq:delta_C_eq} respectively.
In all cases, the Newtonian potential $\Phi$ is computed from Eq.~\eqref{eq:Newtonian_potential_evolution_app} and the non-adiabatic pressure perturbation $\delta p_{\rm nad}$ from Eq.~\eqref{eq:delta_nad_FOPT}.
The perturbation $\delta^{(X)}$ are reconstructed using Eqs.~\eqref{eq:delta_C_delta_N}, \eqref{eq:delta_F_delta_N}, and \eqref{eq:delta_C_delta_F}.}
\end{figure}
\begin{figure}[ht!]
\centering
\includegraphics[width=0.48\textwidth]{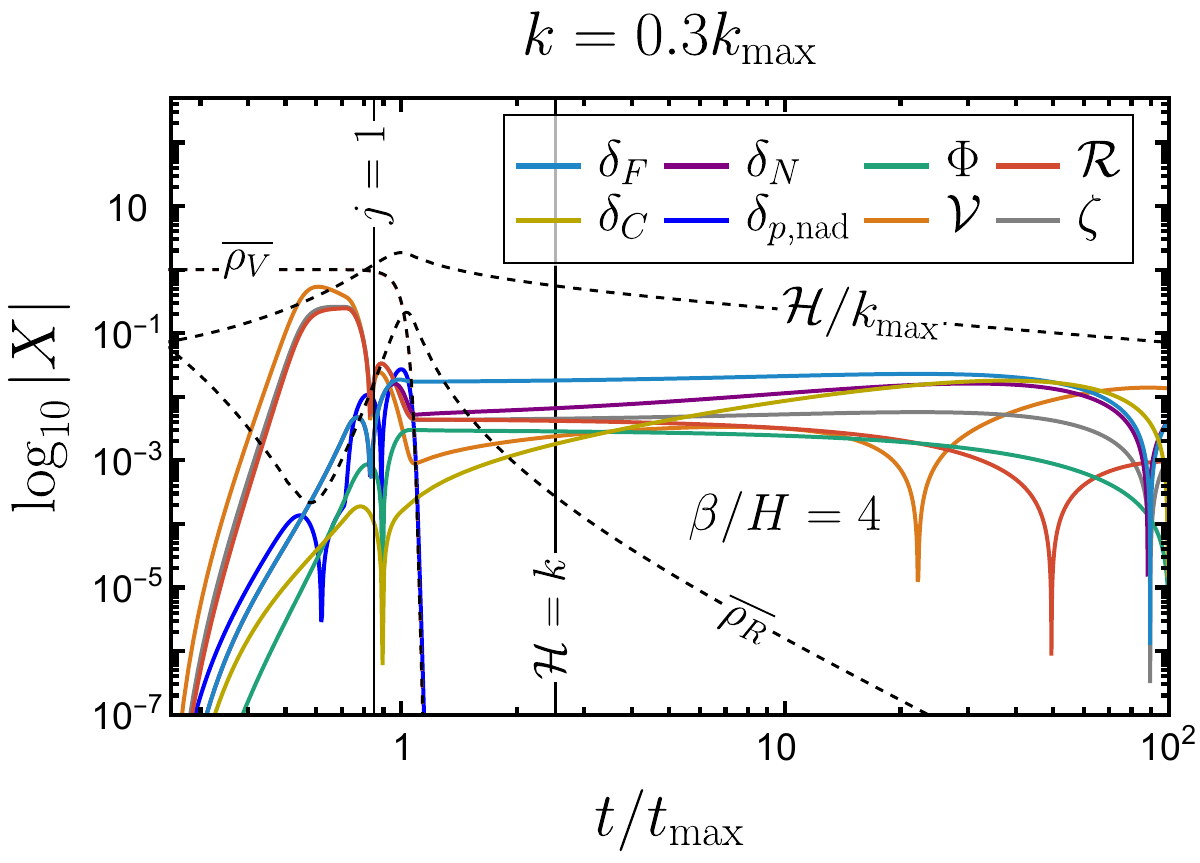}
\includegraphics[width=0.48\textwidth]{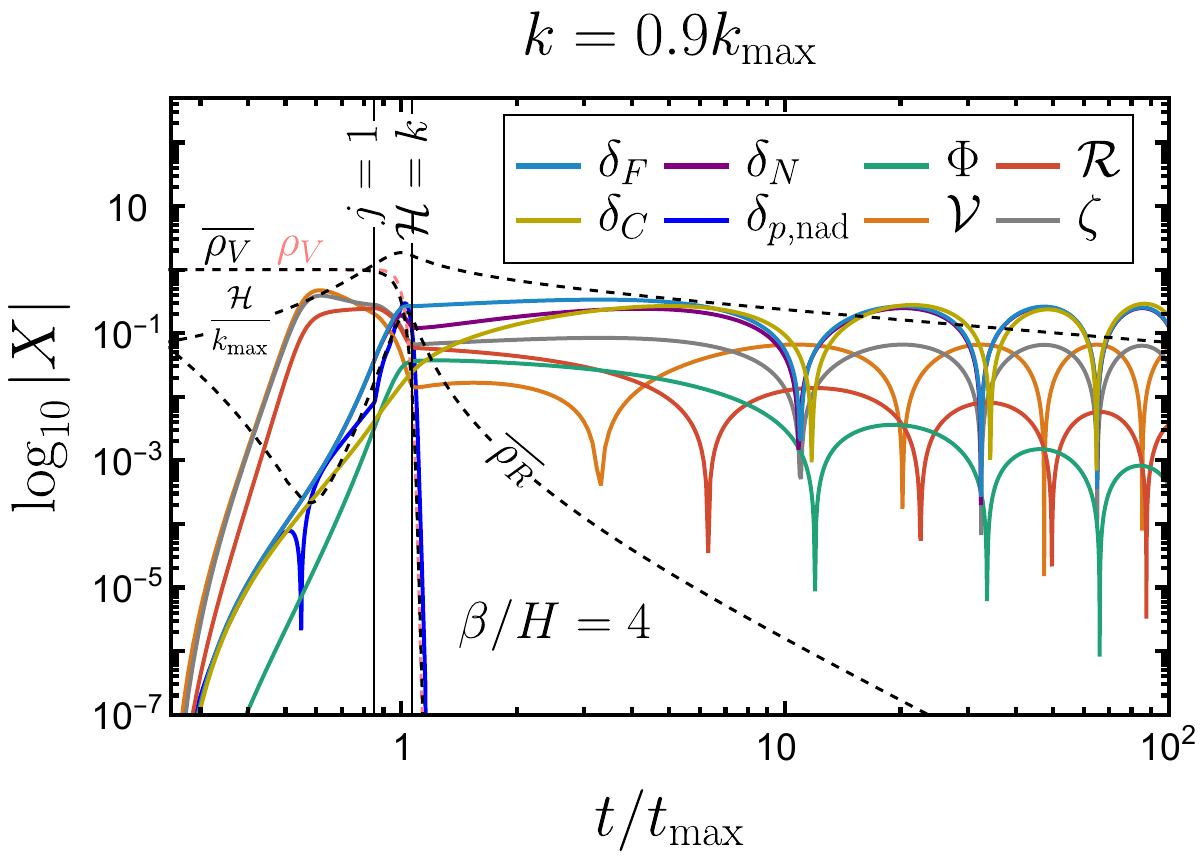}
\includegraphics[width=0.48\textwidth]{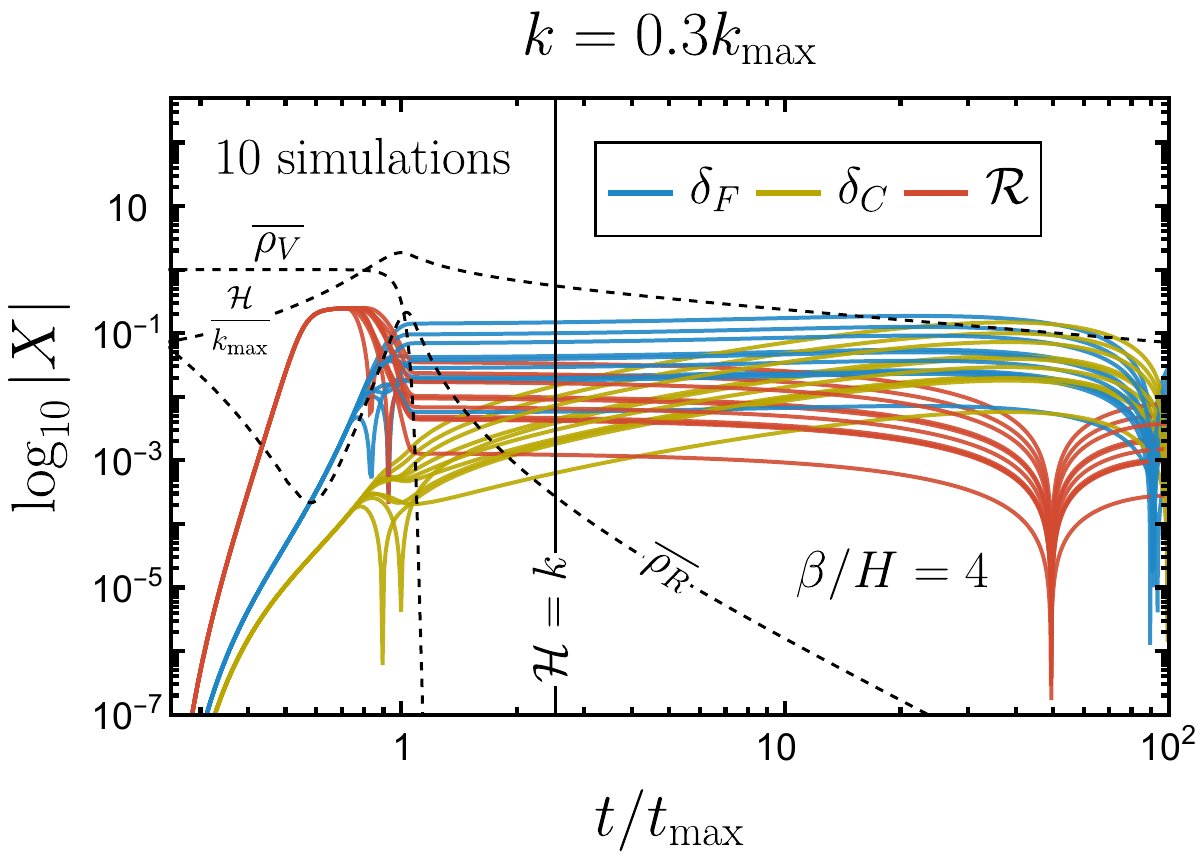}
\includegraphics[width=0.48\textwidth]{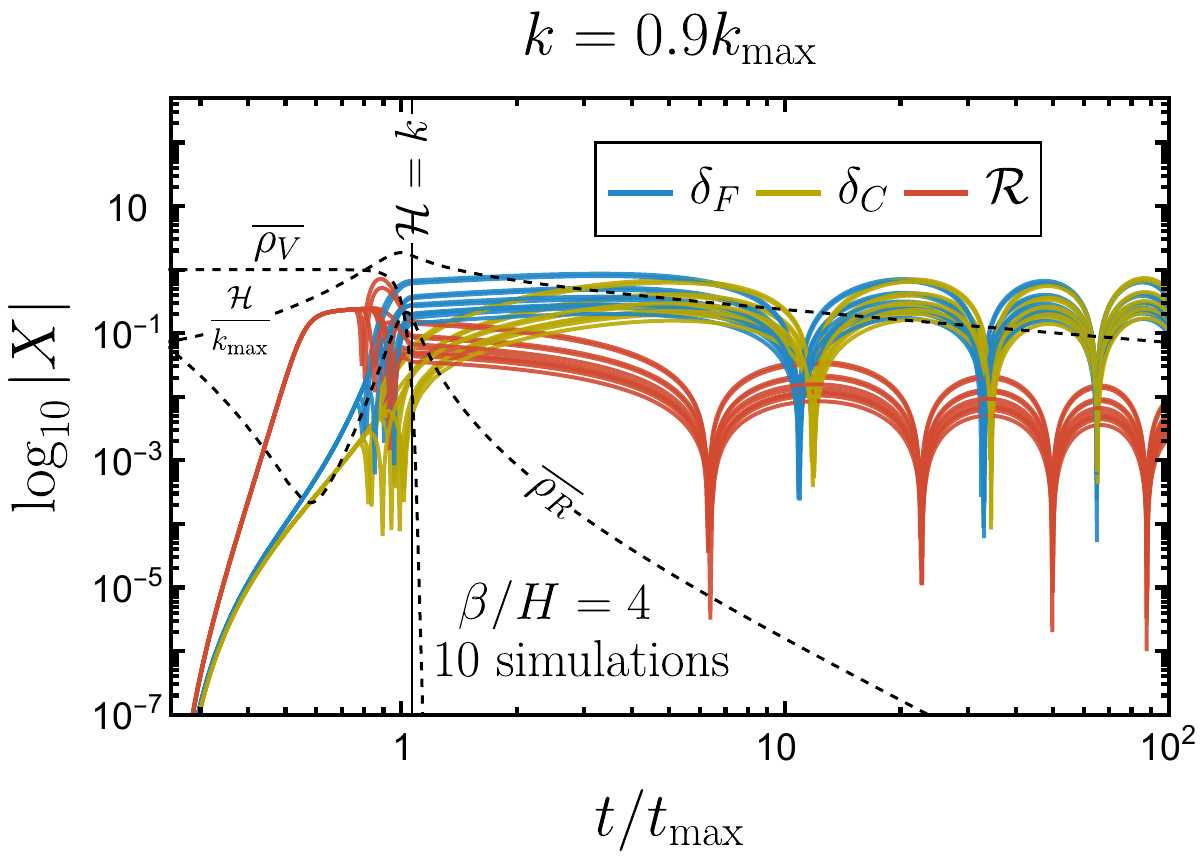}
\caption{  \label{fig:perturbations_allvariables}  
Evolution of cosmological linear perturbations during a first-order phase transition in a spherical patch of inverse comoving size $k=0.3k_{\rm max}$ (\textbf{left}) and $k=0.9k_{\rm max}$ (\textbf{right}). The top panels show a single representative simulation, while the bottom panels display an ensemble of $10$ simulations. In each case, the nucleation times and positions of the first $j_c=50$ bubbles are randomly sampled. }
\end{figure}

\subsection{Evolution of cosmological perturbations}
\label{subsec:cosmo_pert}

\begin{figure}[ht!]
\centering
\includegraphics[width=0.49\textwidth]{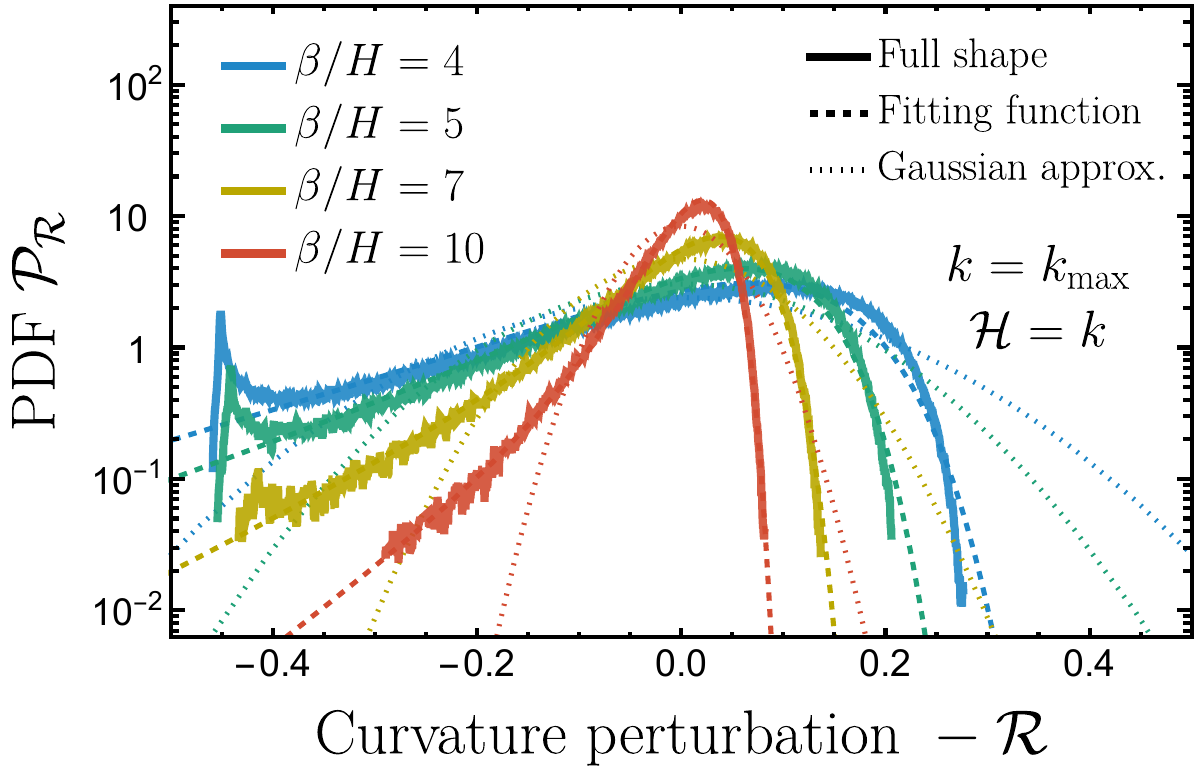}
\includegraphics[width=0.49\textwidth]{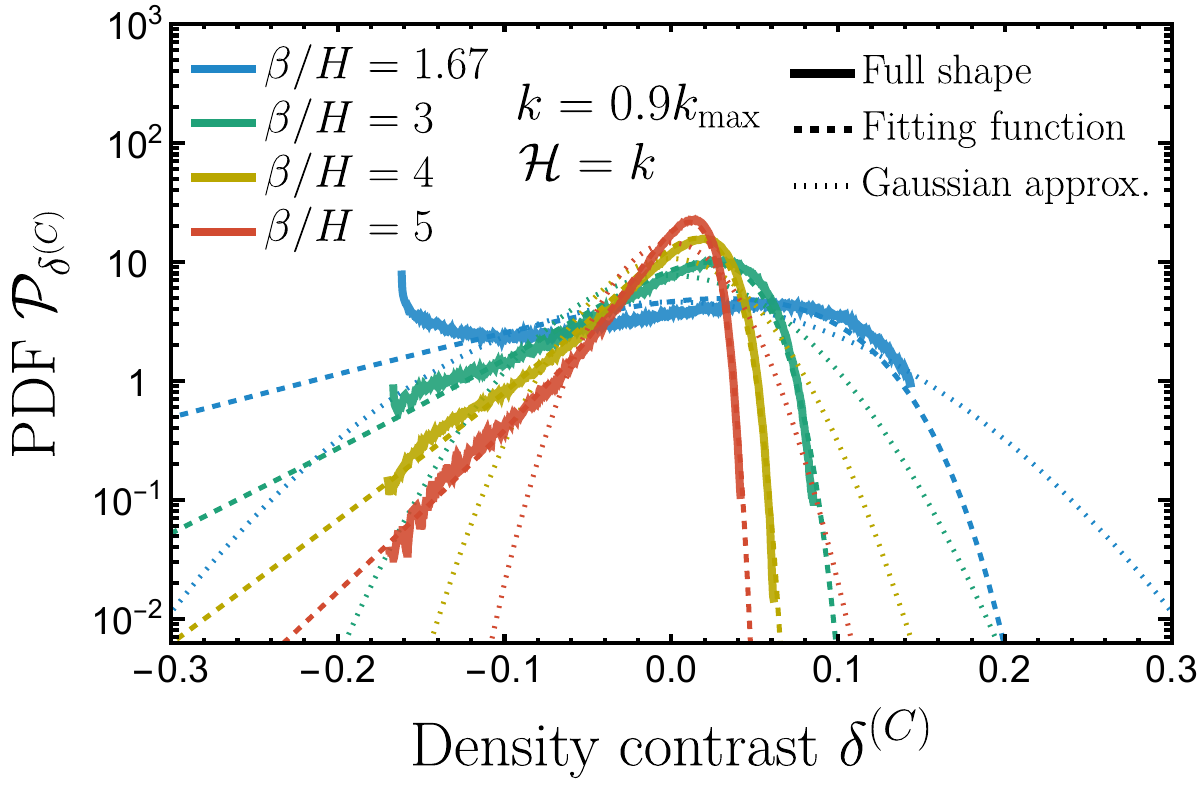}
\caption{  \label{fig:PDF_different_betaOH_deltaC}  
Probability density function for the curvature perturbations $\mathcal{R}$ (\textbf{left}) and density contrast in the comoving gauge $\delta^{(C)}$ (\textbf{right}).  }
\end{figure}

\begin{figure}[ht!]
\centering
\includegraphics[width=0.49\textwidth]{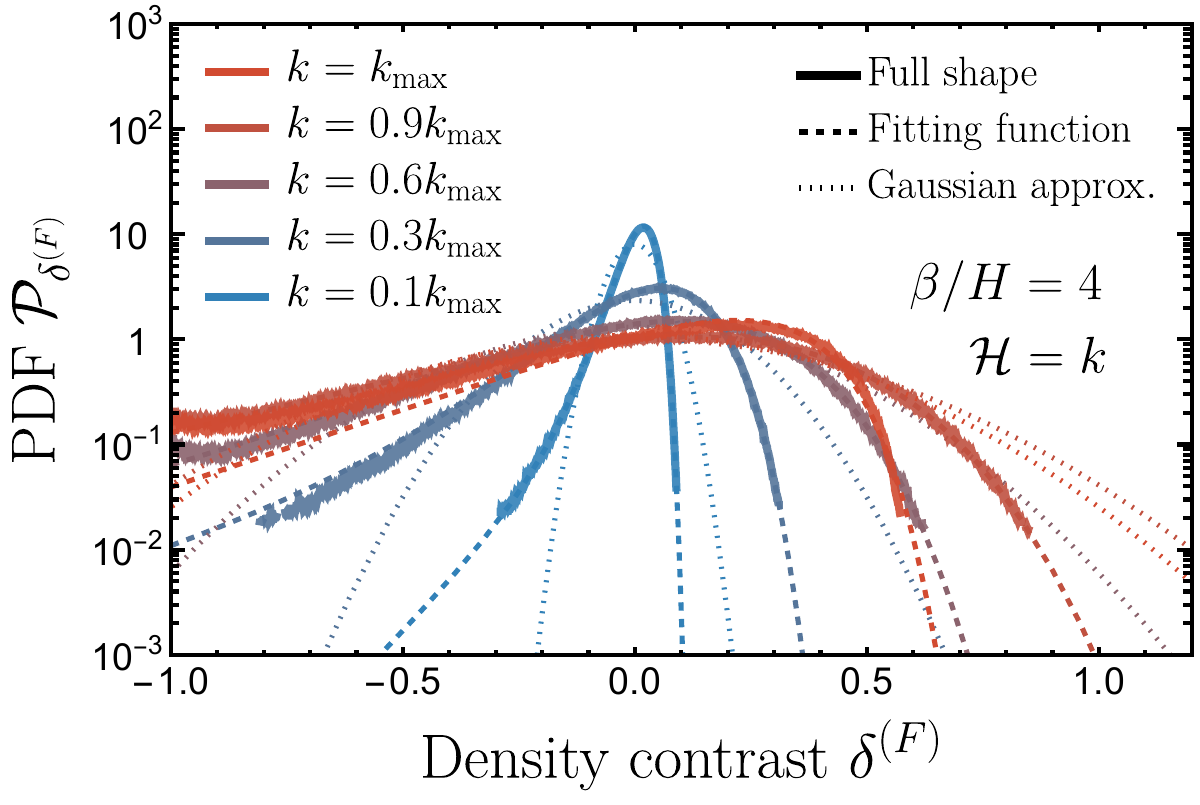}
\includegraphics[width=0.49\textwidth]{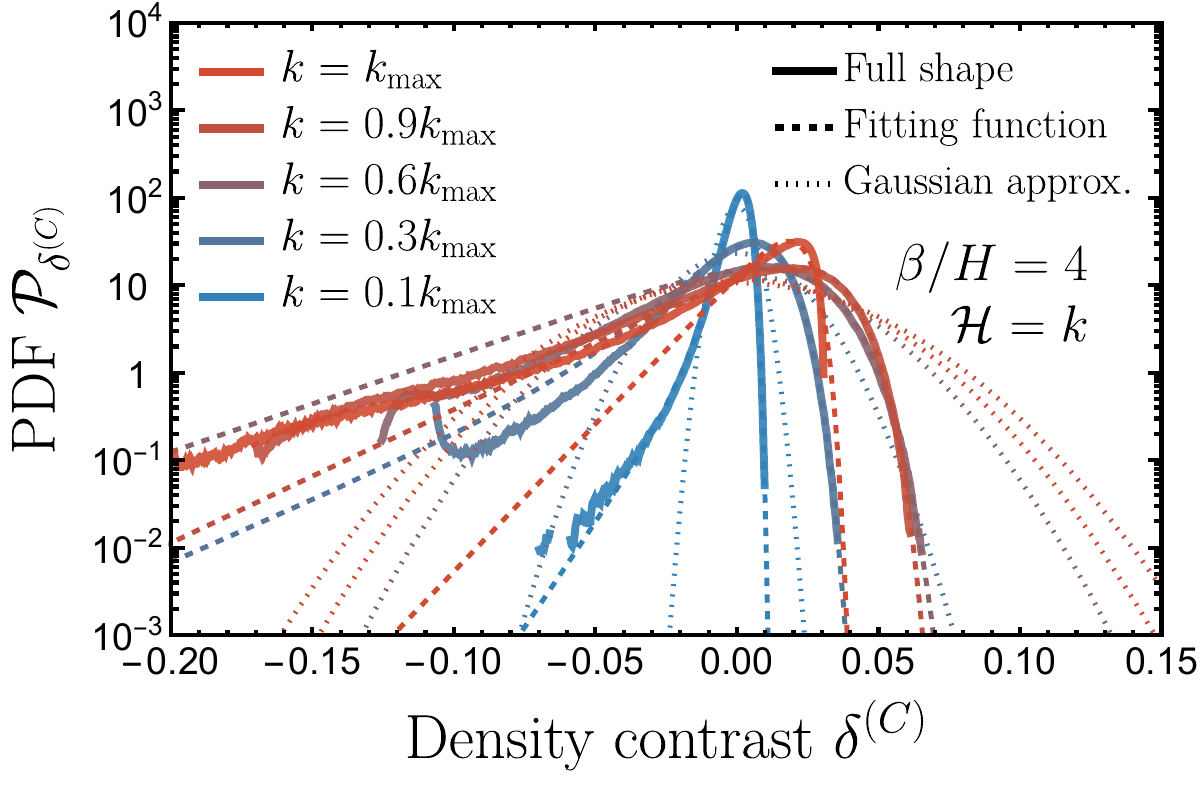}
\includegraphics[width=0.49\textwidth]{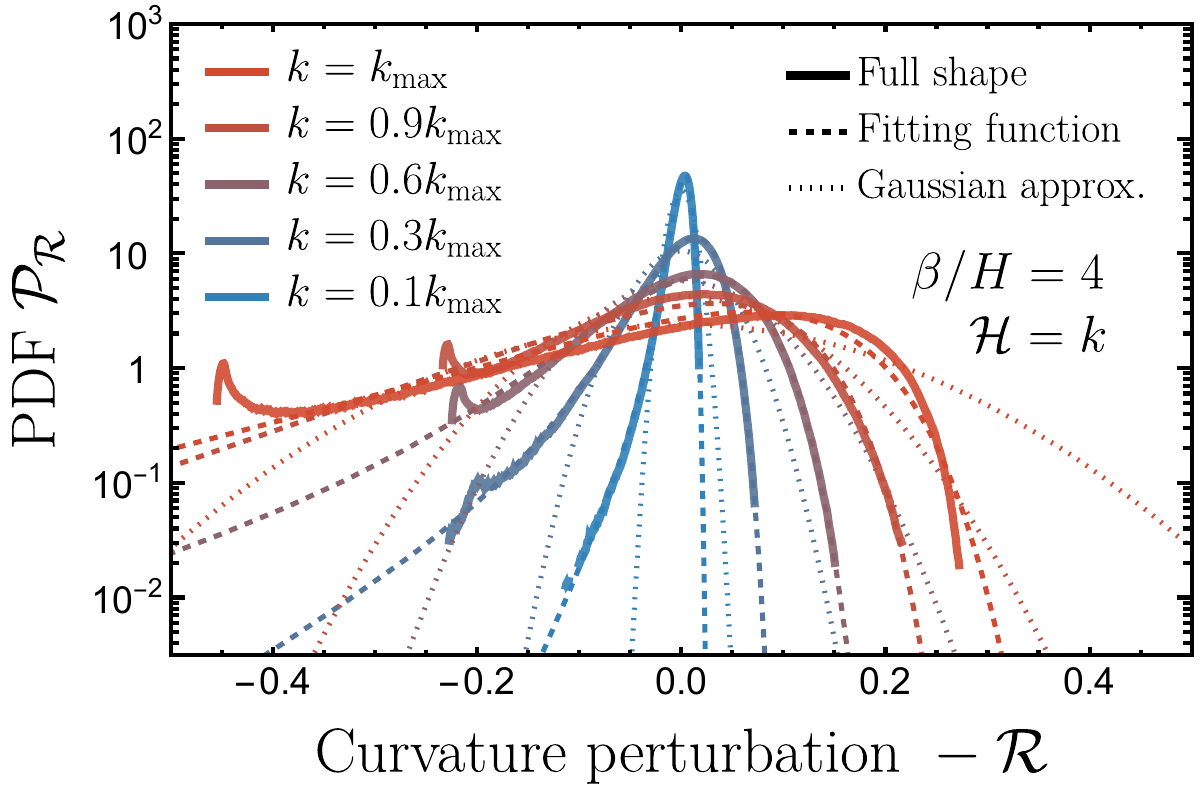}
\includegraphics[width=0.49\textwidth]{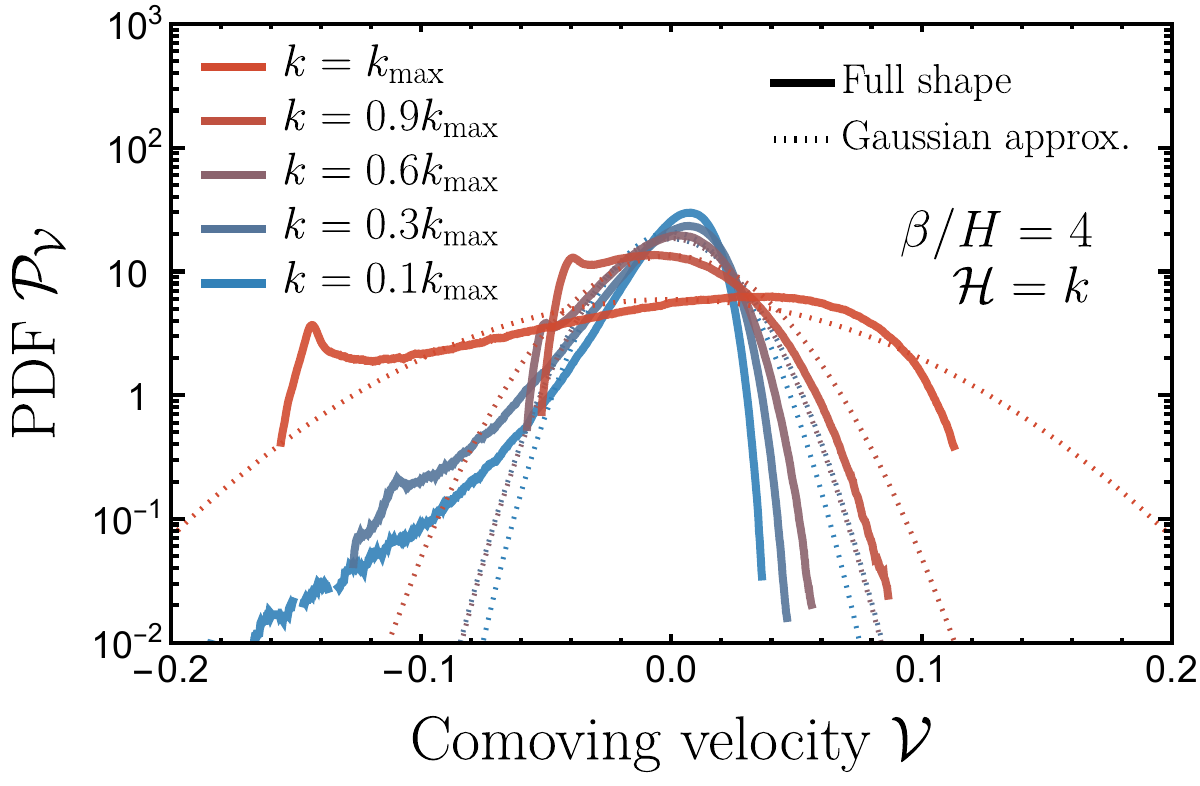}
\caption{  \label{fig:PDF_betaOH4_perturbations}  
Probability distribution for the density contrast in the spatially-flat gauge $\delta^{(F)}$ (\textbf{top-left}), comoving gauge $\delta^{(C)}$  (\textbf{top-right}), the comoving curvature perturbation $\mathcal{R}$  (\textbf{bottom-left}) and the comoving velocity $\mathcal{V}$ (\textbf{bottom-right}). }
\end{figure}

\begin{figure}[ht!]
\centering
\includegraphics[width=0.48\textwidth]{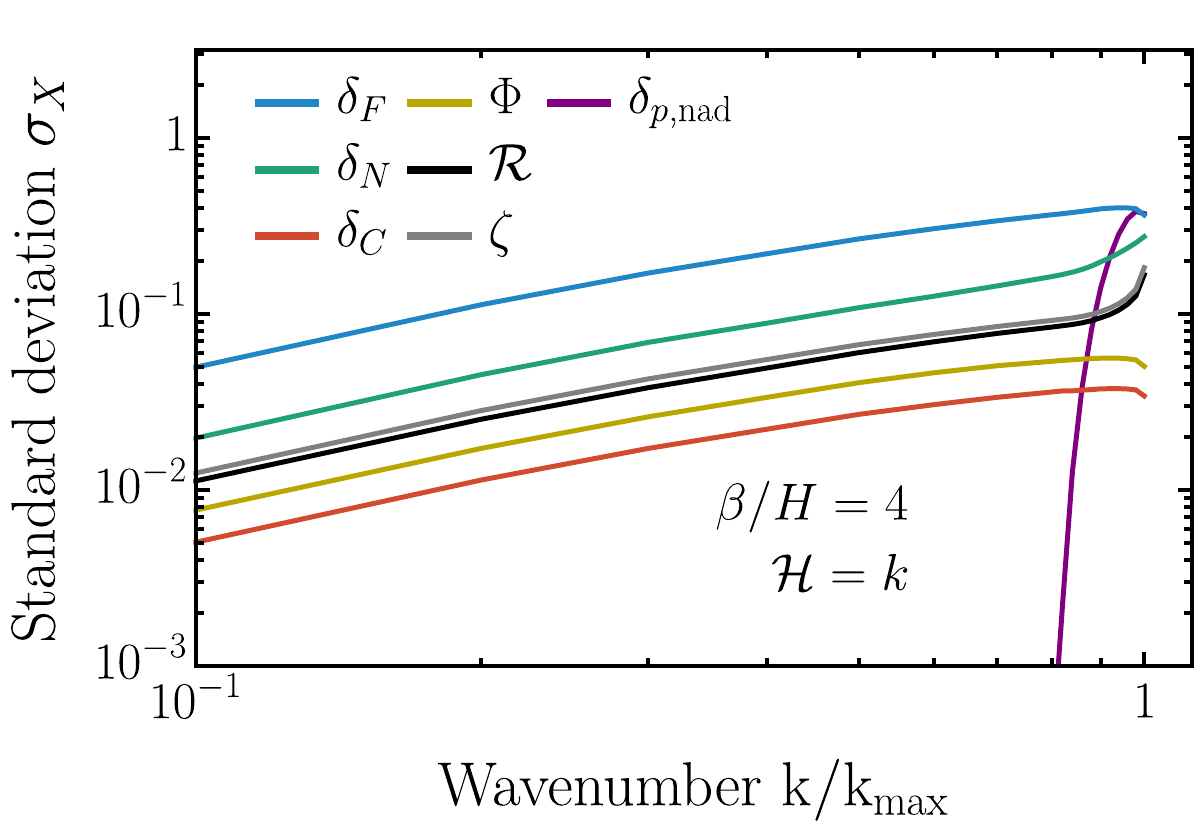}
\includegraphics[width=0.48\textwidth]{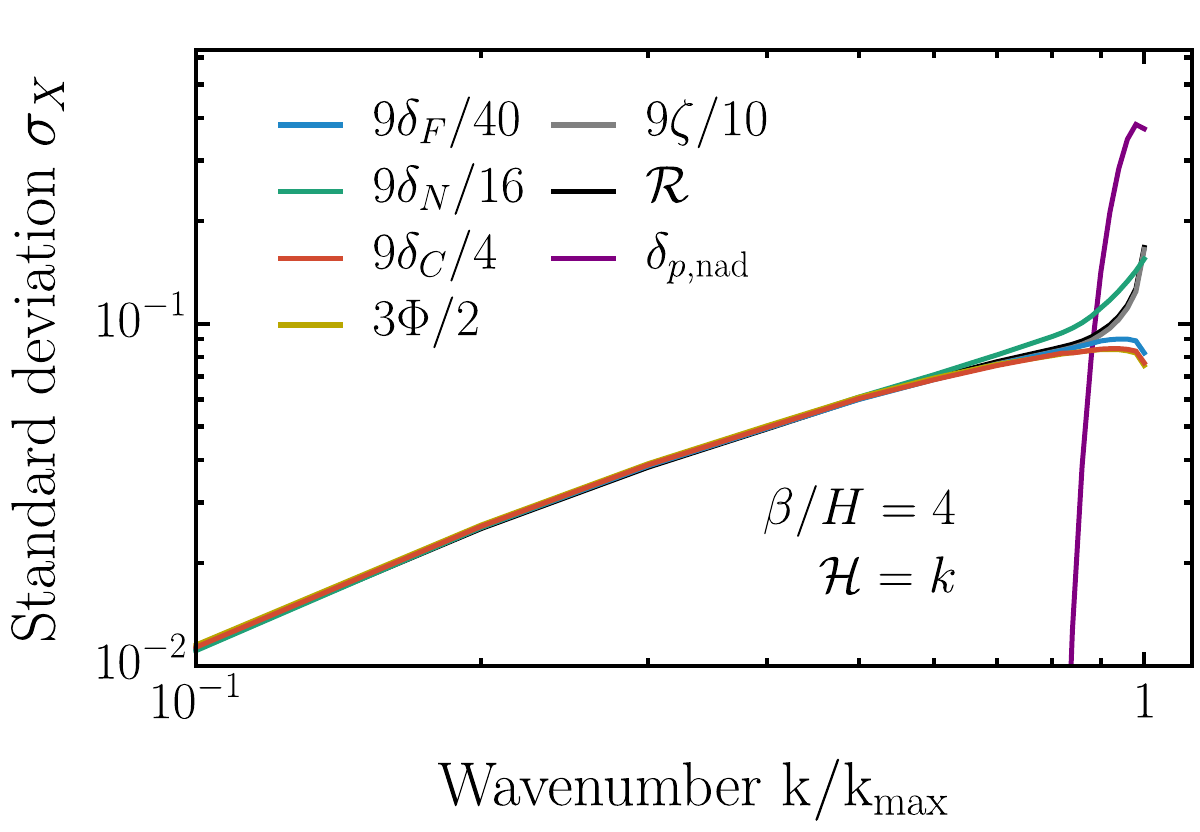}
\includegraphics[width=0.48\textwidth]{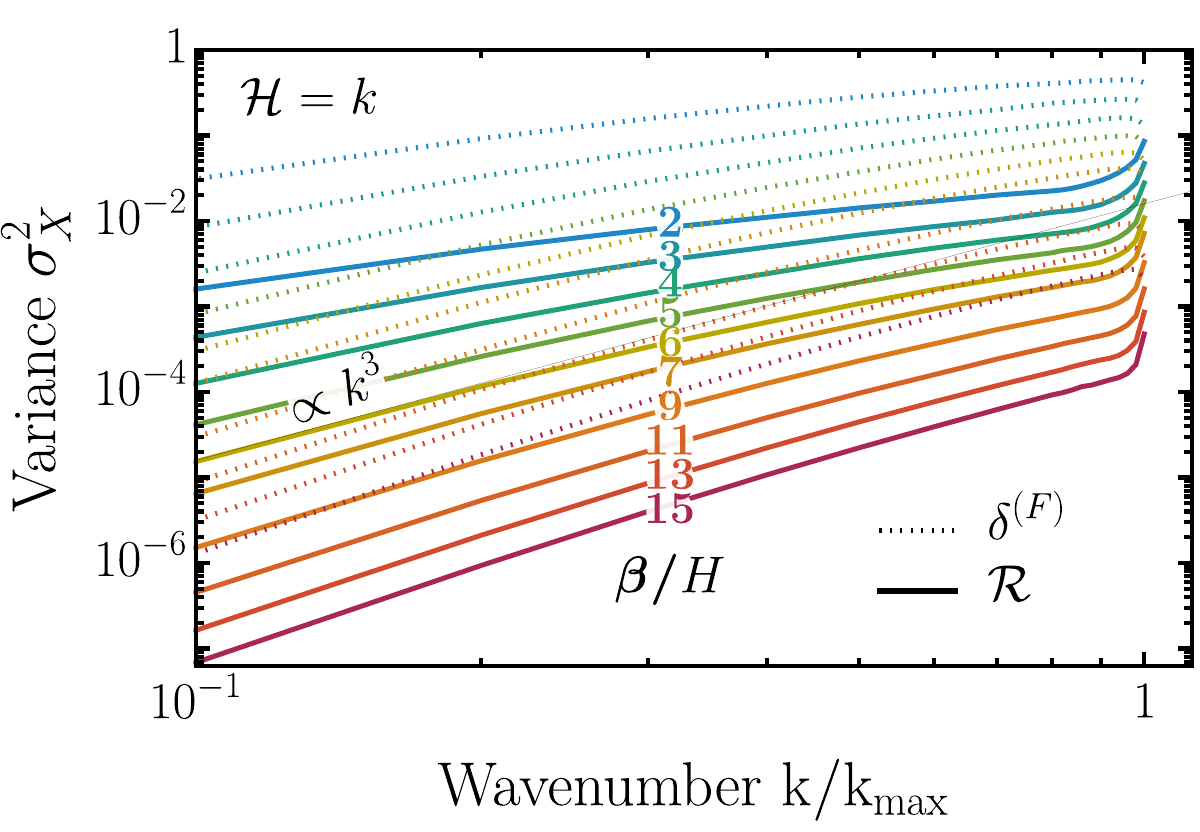}
\caption{  \label{fig:Variance_betaOH4_perturbations}
\textbf{Top-Left:} Standard deviation (square root of the variance) of the different cosmological variables at horizon entry $\mathcal{H}=k$. \textbf{Top-Right:} The adiabatic relationships in Eq.~\eqref{eq:R_relations_rad} are very well satisfied except for $k\sim k_{\rm max}$ due to the presence of large non-adiabatic pressure (\textbf{purple}). \textbf{Bottom:} Variance of the comoving curvature perturbation $\mathcal{R}$ and of the density contrast in the spatially-flat gauge $\delta^{(F)}$ for different values of $\beta/H$. }
\end{figure}

{In \deltaPTvTwo~\cite{deltaPTv2}, we extend
\deltaPT~\cite{Lewicki:2024ghw}} to add the calculation the gravitational potential $\Phi$ (cf. Eq.~\eqref{eq:Newtonian_potential_evolution_app}) as well as the calculation of the density contrast in the Newtonian gauge $\delta^{(N)}$ (cf. Eqs.~\eqref{eq:delta_N_eq}), spatially-flat gauge $\delta^{(F)}$ (cf. \eqref{eq:delta_F_eq}) and comoving gauge $\delta^{(C)}$ (cf. \eqref{eq:delta_C_eq}). The equation of state $\omega$ and speed of sound $c_s$ are calculated from Eq.~\eqref{eq:omega_cs_app} and the non-adiabatic curvature perturbation $\delta p_{\rm nad}$ is calculated from Eq.~\eqref{eq:delta_nad_FOPT}. We plot the time evolution of the different background and perturbed quantities in Fig.~\ref{fig:plot_delta_Phi_gauge_inv} and \ref{fig:perturbations_allvariables}. We run $N_{\rm sim}=10^6$ simulations, stores the values of perturbation at horizon-crossing $\mathcal{H}=k$ at each time, and draw their probability density function (PDF) in Fig.~\ref{fig:PDF_different_betaOH_deltaC} and \ref{fig:PDF_betaOH4_perturbations}. The variance of those PDFs are shown in Fig.~\ref{fig:Variance_betaOH4_perturbations}. The phenomenological consequences in terms of PBHs, and SIGWs are discussed in the next section.


\FloatBarrier

\section{Cosmological Implications}

Having derived the cosmological perturbations generated during FOPTs, we now turn to their implications for primordial black holes (PBHs) and scalar-induced gravitational waves (SIGWs).

\subsection{Primordial Black Holes}

This section summarizes the fitting procedure used to model the PDF of the comoving-gauge density contrast $P(\delta^{(C)})$, which controls the PBH abundance (see main text)
\begin{equation}
f_{\rm PBH} ~\sim~ 
\left (\frac{P(\delta_c^{(C)})}{10^{-10}}\right )\,
\left (\frac{T_{k}}{\rm GeV} \right ),
\end{equation}
through its exponentially sensitive high-density tail, i.e. at the PBH threshold value $\delta_c^{(C)} \in [0.40,\,0.67]$~\cite{Musco:2018rwt,Escriva:2019phb,Escriva:2021aeh}. The PDF is evaluated at horizon re-entry, $\delta^{(C)}(k\simeq\mathcal{H})$, where the standard collapse criterion applies.
The PDF of $\delta^{(C)}$ is shown for different values of $\beta/H$ in Fig.~\ref{fig:PDF_different_betaOH_deltaC} (right) and different values of $k$ in Fig.~\ref{fig:PDF_betaOH4_perturbations} (top-right).  We observe a strong suppression of large overdensities which strongly suppress the PBH abundance below observation level.
As in Ref.~\cite{Lewicki:2024ghw}, we find that the PDF is well described by the three-parameter fit introduced in
Ref.~\cite{Tomberg:2023kli} in the context of ultra-slow-roll inflation,
\begin{equation}
\label{eq:fit_function_app}
P(\delta) = P_0\exp\!\left[\frac{\epsilon}{2}(\delta-\mu)
- \frac{2}{\epsilon^2\sigma^2}
\left(1 - e^{\frac{\epsilon}{2}(\delta-\mu)}\right)^2\right],
\end{equation}
where $\mu$ sets the peak position, $\sigma$ the width, and $\epsilon>0$ controls the skewness. In the Gaussian limit $\epsilon\to0$, this expression reduces to a standard normal distribution. The best fit for those parameters is given in Table.~\ref{tab:sample}. 

{
\renewcommand{\arraystretch}{1.4}
\setlength{\tabcolsep}{15pt}
\begin{table}[h]
    \centering
    \begin{tabular}{|l|c|c|c|c|}
\hline
$\beta/H$ & $P_0$ & $\mu$ & $\epsilon$ & $\sigma$ 
\\
\hline
\hline
1 & 0.99 & -0.35 & 9.1 & 1.0
\\
\hline
2 & 0.66 & -0.14 & 21 & 0.52
\\
\hline
3 & 0.15 & -0.019 & 32 & 0.084
\\
\hline
4 & 0.084 & -0.0042 & 48 & 0.041
\\
\hline
5 & 0.056 & -2.4$\times10^{-4}$ & 61 & 0.025
\\
\hline
6 & 0.040 & 4.2$\times10^{-4}$ & 71 & 0.016
\\
\hline
7 & 0.031 & 4.6$\times10^{-4}$ & 85 & 0.012
\\
\hline
8 & 0.024 & 4.9$\times10^{-4}$ & $1.1\times10^{2}$ & 0.0091
\\
\hline
\end{tabular}
\caption{\label{tab:sample}
Fit of the PDF of the density contrast in the comoving gauge ${\cal P}_{\delta^{(C)}}$ for $k=0.9k_{\rm max}$.}
\end{table}
}

\begin{figure}[t]
\centering
\includegraphics[width=0.48\textwidth]{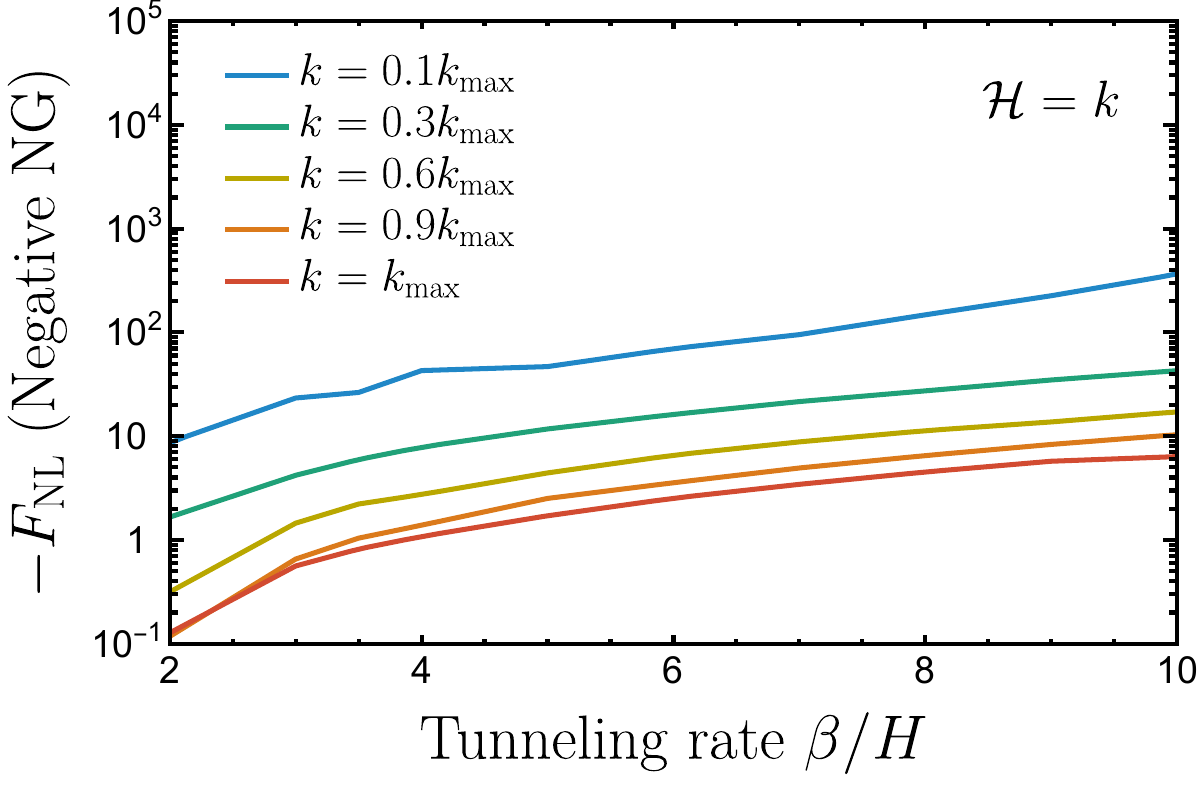}
\includegraphics[width=0.48\textwidth]{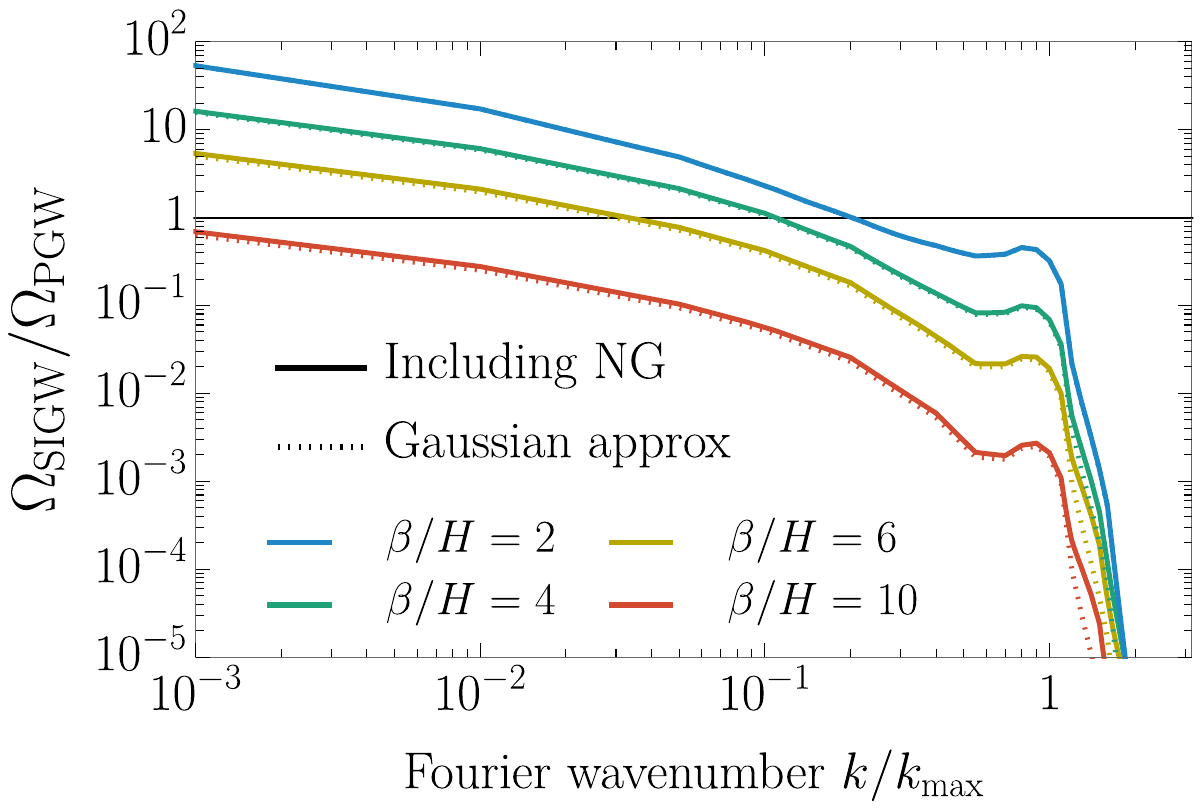}
\caption{
\label{fig:Negative_fNL}
Left: Effective NG parameter $F_{\rm NL}$ extracted from the skewness of the curvature perturbation $\mathcal{R}$ as a function of $\beta/H$, evaluated at $k\simeq0.9\,k_{\rm max}$.
Right: Ratio of the non-Gaussian contributions $\Omega_{\rm SIGW}^{(1)}+\Omega_{\rm SIGW}^{(2)}$ to the Gaussian SIGW spectrum $\Omega_{\rm SIGW}^{(0)}$.
For the parameter range relevant to first-order phase transitions, non-Gaussian corrections to the SIGW signal are subdominant.
}
\end{figure}

\subsection{Scalar-induced gravitational waves}

SIGWs generated at second order in cosmological perturbation theory are given by
\cite{Tomita:1975kj,Matarrese:1992rp,Matarrese:1993zf,Matarrese:1997ay,Acquaviva:2002ud,Mollerach:2003nq,Carbone:2004iv,Ananda:2006af,Baumann:2007zm,Espinosa:2018eve,Kohri:2018awv,Domenech:2021ztg}
\begin{equation}
\Omega_{\rm SIGW}(k,\eta)
= \frac{1}{24}
\left(\frac{k}{{\cal H}(\eta)}\right)^2
\overline{{\cal P}_h(k,\eta)} \,,
\label{eq:OmegaGW}
\end{equation}
where $\overline{{\cal P}_h}$ denotes the oscillation-averaged tensor power spectrum \cite{Maggiore:1999vm}.
The induced tensor spectrum is
\begin{align}
\overline{{\cal P}_h(k,\eta)}\,
(2\pi)^3 \delta_D^{(3)}(\mathbf{k}+\mathbf{k}')
&=
\frac{16 k^3}{2\pi^2}
\sum_s
\int \frac{d^3\mathbf{p}}{(2\pi)^3}
\int \frac{d^3\mathbf{q}}{(2\pi)^3}
\nonumber\\
&\hspace{-1.2cm}\times
\overline{
J_s(\mathbf{k},\mathbf{p},\eta)
J_s(\mathbf{k}',\mathbf{q},\eta)
}
\,
\big\langle
\mathcal{R}_{\mathbf{p}}
\mathcal{R}_{\mathbf{k}-\mathbf{p}}
\mathcal{R}_{\mathbf{q}}
\mathcal{R}_{\mathbf{k}'-\mathbf{q}}
\big\rangle \,,
\label{eq:Phfull_app}
\end{align}
where $s$ labels the two GW polarizations.
The kernel
\begin{equation}
J_s(\mathbf{k},\mathbf{p},\eta)
\equiv Q_s(\mathbf{k},\mathbf{p})\, I(\mathbf{k},\mathbf{p},\eta)
\end{equation}
is the product of the polarization contraction
$Q_s(\mathbf{k},\mathbf{p})=e^{ij}_s(\hat{\mathbf{k}})p_ip_j$
and the time-dependent convolution of the source with the Green function,
$I(\mathbf{k},\mathbf{p},\eta)$.
Explicit expressions can be found in
Refs.~\cite{Espinosa:2018eve,Kohri:2018awv,LISACosmologyWorkingGroup:2025vdz}.
For Gaussian curvature perturbations, only disconnected Wick contractions contribute to the four-point function in Eq.~\eqref{eq:Phfull_app}.
In the presence of primordial non-Gaussianities (NGs), additional disconnected contributions arise, together with genuinely connected terms proportional to the curvature trispectrum
\cite{Nakama:2016gzw,Garcia-Bellido:2017aan,Unal:2018yaa,Cai:2018dig,Cai:2019amo,Ragavendra:2020sop,Yuan:2020iwf,Adshead:2021hnm,Davies:2021loj,Abe:2022xur,Garcia-Saenz:2022tzu,Garcia-Saenz:2023zue,Li:2023xtl,Yuan:2023ofl,Perna:2024ehx,Gouttenoire:2025jxe}.

We quantify the level of NG by the skewness of the curvature perturbation,
\cite{Luo:1992er,Komatsu:2001rj}
\begin{equation}
F_{\rm NL}
\equiv
\frac{\langle \mathcal{R}^3 \rangle}
{6 \langle \mathcal{R}^2 \rangle^2} \,.
\end{equation}
As shown in Fig.~\ref{fig:Negative_fNL} (left), $F_{\rm NL}$ exhibits a scale dependence.
Since the SIGW signal is dominated by modes near the peak of the scalar power spectrum, we evaluate $F_{\rm NL}$ at
$k\simeq 0.9\,k_{\rm max}$.
For $\beta_H\equiv \beta/H \in [2,15]$, we find the empirical fit
\begin{equation}
\log_{10} F_{\rm NL}
\simeq
a \log_{10}^2(\beta_H/b)
+ c \log_{10}(\beta_H/b) \,,
\end{equation}
with $a\simeq -1.34$, $b\simeq 3.67$, and $c\simeq 2.95$.

The present-day SIGW spectrum can be decomposed as~\cite{Inomata:2019yww,DeLuca:2019ufz,Yuan:2019fwv,Domenech:2020xin} \begin{equation}\label{eq:Omegabar-total} \Omega_{\rm SIGW} =\Omega_{\rm SIGW}^{(0)} +\Omega_{\rm SIGW}^{(1)} +\Omega_{\rm SIGW} ^{(2)} \ , \end{equation} \begin{align} \Omega_{\rm SIGW}^{(0)} = \frac{1}{3} \int_1^\infty \td t \int_{-1}^1 \td s \frac{\overline{J^2(u,v)}}{(u v)^2} {\cal P}_\mathcal{R} (v k) {\cal P}_\mathcal{R} (u k) , \label{eq:P_h_ts} \end{align} where $u=(t + s)/2$, $v=(t - s)/2$ and the transfer function $\overline{J^2 (u,v)}$ depends on the cosmological background~\cite{Kohri:2018awv,Espinosa:2018eve}. In accordance with Refs.~\cite{Adshead:2021hnm, Li:2023qua, Zhu:2023faa, Ragavendra:2021qdu, Atal:2021jyo}, the higher-order terms in \eqref{eq:Omegabar-total} can be expressed as \begin{align} \Omega_{\rm SIGW}^{(1)} =& \frac{2F_{\rm NL}^2}{3} \prod_{i=1}^2 \biggl[\int_1^\infty \td t_i \int_{-1}^1 \td s_i\, u_iv_i\biggr] \Bigg\{ \frac{1}{2} \overline{J^2 (s_1,t_1)} \frac{ {\cal P}_\mathcal{R} (v_1 v_2 k) {\cal P}_\mathcal{R} (u_1 k) {\cal P}_\mathcal{R} (v_1 u_2 k)}{(u_1v_1u_2v_2)^3} \nonumber \\ +& \int_0^{2\pi} \frac{\td \varphi_{12}}{2\pi}\, \cos 2\varphi_{12} \overline{J (s_1,t_1) J (s_2,t_2)} \frac{{\cal P}_\mathcal{R} (v_2 k)}{v_2^3} \frac{{\cal P}_\mathcal{R} (w_{12} k)}{w_{12}^3} \bigg[ \frac{{\cal P}_\mathcal{R} (u_2 k)}{u_2^3} + \frac{{\cal P}_\mathcal{R} (u_1 k)}{u_1^3} \bigg] \Bigg\} \label{eq:Omega-C} \end{align} and \begin{align} \label{eq:Omega-N} \Omega_{\rm SIGW}^{(2)} &= \frac{F_{\rm NL}^4}{6} \prod_{i=1}^3 \biggl[\int_1^\infty \td t_i \int_{-1}^1 \td s_i\, u_iv_i\biggr] \Bigg\{\frac{1}{2}\overline{J^2 (s_1,t_1)} \frac{{\cal P}_\mathcal{R} (v_1 v_2 k) {\cal P}_\mathcal{R} (v_1 u_2 k) {\cal P}_\mathcal{R} (u_1 v_3 k) {\cal P}_\mathcal{R} (u_1 u_3 k)}{(u_1v_1u_2v_2u_3v_3)^3} \\ & + \int_0^{2\pi} \frac{\td \varphi_{12}}{2\pi}\frac{\td \varphi_{23}}{2\pi} \cos (2\varphi_{12} ) \overline{J (s_1,t_1) J (s_2,t_2)} \frac{{\cal P}_\mathcal{R}(u_3 k)}{u_3^3} \frac{{\cal P}_\mathcal{R} (w_{13} k)}{w_{13}^3} \frac{{\cal P}_\mathcal{R} (w_{23} k)}{w_{23}^3} \bigg[ \frac{{\cal P}_\mathcal{R} (v_3 k)}{v_3^3} + \frac{{\cal P}_\mathcal{R} (w_{123} k)}{w_{123}^3} \bigg]\Bigg\} \ , \nonumber \end{align} where we defined $s_i=u_i-v_i$, $t_i=u_i+v_i$, and \begin{subequations} \begin{eqnarray} y_{ij}&=&\frac{\cos\varphi_{ij}}{4}\sqrt{(t_i^2-1)(t_j^2-1)(1-s_i^2)(1-s_j^2)}+\frac{1}{4}(1-s_it_i)(1-s_jt_j)\ , \\ w_{ij}&=&\sqrt{v_i^2+v_j^2-y_{ij}}\, , \qquad w_{123}=\sqrt{v_1^2+v_2^2+v_3^2+y_{12}-y_{13}-y_{23}},\qquad \varphi_{13}=\varphi_{12}-\varphi_{23}.\ . \end{eqnarray} \end{subequations} The time-averaged integrated transfer functions $J(u,v)$ were derived in Refs.~\cite{Espinosa:2018eve,Kohri:2018awv,Atal:2021jyo,Adshead:2021hnm,Li:2023qua}, and are \begin{align}\label{eq:J-ave-12} & \overline{ J (s_i,t_i)J (s_j,t_j) } = \frac{9}{8}\frac{\left(t_i^2-1\right) \left(t_j^2-1\right) \left(1-s_i^2\right) \left(1-s_j^2\right) \left(t_i^2+s_i^2-6\right) \left(t_j^2+s_j^2-6\right) }{\left(t_i^2-s_i^2\right)^3 \left(t_j^2-s_j^2\right)^3} \nonumber\\ & \Bigg[ \left( \left(t_i^2+s_i^2-6\right) \ln \left| \frac{t_i^2-3}{3-s_i^2}\right| - 2\left(t_i^2-s_i^2\right) \right) \left( \left(t_j^2+s_j^2-6\right) \ln \left| \frac{t_j^2-3}{3-s_j^2}\right| - 2\left(t_j^2-s_j^2\right) \right) \nonumber\\ & + \pi^2\Theta\left(t_i-\sqrt{3}\right) \Theta\left(t_j-\sqrt{3}\right) \left(t_i^2+s_i^2-6\right) \left(t_j^2+s_j^2-6\right) \Bigg]\ . \end{align}
The contribution of scalar-induced gravitational waves to the total gravitational-wave signal from the first-order phase transition is shown in the main text to be numerically negligible,
$\Omega_{\rm SIGW} \ll \Omega_{\rm PGW}$.
As a consequence, the impact of primordial non-Gaussianities on the SIGW spectrum is also negligible,
$\Omega_{\rm SIGW}^{(1)},\,\Omega_{\rm SIGW}^{(2)} \ll \Omega_{\rm SIGW}^{(0)}$,
as illustrated in Fig.~\ref{fig:Negative_fNL} (right).

This behavior is expected on parametric grounds:
a suppressed SIGW contribution,
$\Omega_{\rm SIGW}/\Omega_{\rm PGW}\sim \langle \mathcal{R}^2\rangle$,
implies correspondingly small non-Gaussian corrections,
$\Omega_{\rm SIGW}^{(1)}/\Omega_{\rm SIGW}^{(0)}\sim F_{\rm NL}^2\langle \mathcal{R}^2\rangle$
and
$\Omega_{\rm SIGW}^{(2)}/\Omega_{\rm SIGW}^{(0)}\sim F_{\rm NL}^4\langle \mathcal{R}^2\rangle^2$.

\bibliographystyle{apsrev4-1}
\bibliography{biblio}

@misc{deltaPTv2,
  author       = {Franciolini, Gabriele and Gouttenoire, Yann and Jinno, Ryusuke},
  title        = {{\tt deltaPT\,2.0}: a C++ Code to compute cosmological perturbations generated during strongly supercooled first-order phase transitions \githubmaster{https://github.com/YannGou/deltaPT2.0}{\faGithub}},
  year         = 2025,
  url          = {https://github.com/YannGou/deltaPT2.0.git}
}

@article{Lewicki:2022pdb,
	archiveprefix = {arXiv},
	author = {Lewicki, Marek and Vaskonen, Ville},
	doi = {10.1140/epjc/s10052-023-11241-3},
	eprint = {2208.11697},
	journal = {Eur. Phys. J. C},
	number = {2},
	pages = {109},
	primaryclass = {astro-ph.CO},
	title = {{Gravitational waves from bubble collisions and fluid motion in strongly supercooled phase transitions}},
	volume = {83},
	year = {2023}}

@article{Yoo:2020lmg,
    author = "Yoo, Chul-Moon and Harada, Tomohiro and Okawa, Hirotada",
    title = "{Threshold of Primordial Black Hole Formation in Nonspherical Collapse}",
    eprint = "2004.01042",
    archivePrefix = "arXiv",
    primaryClass = "gr-qc",
    reportNumber = "RUP-20-12",
    doi = "10.1103/PhysRevD.102.043526",
    journal = "Phys. Rev. D",
    volume = "102",
    number = "4",
    pages = "043526",
    year = "2020",
    note = "[Erratum: Phys.Rev.D 107, 049901 (2023)]"
}

@article{Jinno:2019jhi,
	archiveprefix = {arXiv},
	author = {Jinno, Ryusuke and Seong, Hyeonseok and Takimoto, Masahiro and Um, Choong Min},
	doi = {10.1088/1475-7516/2019/10/033},
	eprint = {1905.00899},
	journal = {JCAP},
	pages = {033},
	primaryclass = {astro-ph.CO},
	reportnumber = {DESY 19-067, DESY-19-067, CTPU-PTC-19-12, KEK-TH-2125},
	title = {{Gravitational waves from first-order phase transitions: Ultra-supercooled transitions and the fate of relativistic shocks}},
	volume = {10},
	year = {2019}}

@article{Jinno:2024nwb,
	archiveprefix = {arXiv},
	author = {Jinno, Ryusuke and Kume, Jun'ya},
	doi = {10.1088/1475-7516/2025/02/057},
	eprint = {2408.10770},
	journal = {JCAP},
	pages = {057},
	primaryclass = {gr-qc},
	reportnumber = {RESCEU-12/24, KOBE-COSMO-24-02},
	title = {{Gravitational effects on fluid dynamics in cosmological first-order phase transitions}},
	volume = {02},
	year = {2025}}

@article{Atal:2021jyo,
	archiveprefix = {arXiv},
	author = {Atal, Vicente and Dom\`enech, Guillem},
	doi = {10.1088/1475-7516/2021/06/001},
	eprint = {2103.01056},
	journal = {JCAP},
	note = {[Erratum: JCAP 10, E01 (2023)]},
	pages = {001},
	primaryclass = {astro-ph.CO},
	title = {{Probing non-Gaussianities with the high frequency tail of induced gravitational waves}},
	volume = {06},
	year = {2021}}

@article{Acquaviva:2002ud,
	archiveprefix = {arXiv},
	author = {Acquaviva, Viviana and Bartolo, Nicola and Matarrese, Sabino and Riotto, Antonio},
	doi = {10.1016/S0550-3213(03)00550-9},
	eprint = {astro-ph/0209156},
	journal = {Nucl. Phys. B},
	pages = {119--148},
	reportnumber = {DFPD-A-02-21},
	title = {{Second order cosmological perturbations from inflation}},
	volume = {667},
	year = {2003}}

@article{Mollerach:2003nq,
	archiveprefix = {arXiv},
	author = {Mollerach, Silvia and Harari, Diego and Matarrese, Sabino},
	doi = {10.1103/PhysRevD.69.063002},
	eprint = {astro-ph/0310711},
	journal = {Phys. Rev. D},
	pages = {063002},
	title = {{CMB polarization from secondary vector and tensor modes}},
	volume = {69},
	year = {2004}}

@article{Baumann:2007zm,
	archiveprefix = {arXiv},
	author = {Baumann, Daniel and Steinhardt, Paul J. and Takahashi, Keitaro and Ichiki, Kiyotomo},
	doi = {10.1103/PhysRevD.76.084019},
	eprint = {hep-th/0703290},
	journal = {Phys. Rev. D},
	pages = {084019},
	title = {{Gravitational Wave Spectrum Induced by Primordial Scalar Perturbations}},
	volume = {76},
	year = {2007}}

@article{Ananda:2006af,
	archiveprefix = {arXiv},
	author = {Ananda, Kishore N. and Clarkson, Chris and Wands, David},
	doi = {10.1103/PhysRevD.75.123518},
	eprint = {gr-qc/0612013},
	journal = {Phys. Rev. D},
	pages = {123518},
	title = {{The Cosmological gravitational wave background from primordial density perturbations}},
	volume = {75},
	year = {2007}}

@article{Tomita:1975kj,
	author = {Tomita, Kenji},
	doi = {10.1143/PTP.54.730},
	journal = {Prog. Theor. Phys.},
	pages = {730},
	reportnumber = {RRK 75-3},
	title = {{Evolution of Irregularities in a Chaotic Early Universe}},
	volume = {54},
	year = {1975}}

@article{Kohri:2018awv,
	archiveprefix = {arXiv},
	author = {Kohri, Kazunori and Terada, Takahiro},
	doi = {10.1103/PhysRevD.97.123532},
	eprint = {1804.08577},
	journal = {Phys. Rev. D},
	number = {12},
	pages = {123532},
	primaryclass = {gr-qc},
	reportnumber = {KEK-TH-2046, KEK-COSMO-223},
	title = {{Semianalytic calculation of gravitational wave spectrum nonlinearly induced from primordial curvature perturbations}},
	volume = {97},
	year = {2018}}

@article{Unal:2018yaa,
	archiveprefix = {arXiv},
	author = {Unal, Caner},
	doi = {10.1103/PhysRevD.99.041301},
	eprint = {1811.09151},
	journal = {Phys. Rev. D},
	number = {4},
	pages = {041301},
	primaryclass = {astro-ph.CO},
	title = {{Imprints of Primordial Non-Gaussianity on Gravitational Wave Spectrum}},
	volume = {99},
	year = {2019}}

@article{Matarrese:1993zf,
	archiveprefix = {arXiv},
	author = {Matarrese, Sabino and Pantano, Ornella and Saez, Diego},
	doi = {10.1103/PhysRevLett.72.320},
	eprint = {astro-ph/9310036},
	journal = {Phys. Rev. Lett.},
	pages = {320--323},
	reportnumber = {DFPD-93-A-67},
	title = {{General relativistic dynamics of irrotational dust: Cosmological implications}},
	volume = {72},
	year = {1994}}

@article{Domenech:2021ztg,
	archiveprefix = {arXiv},
	author = {Dom\`enech, Guillem},
	doi = {10.3390/universe7110398},
	eprint = {2109.01398},
	journal = {Universe},
	number = {11},
	pages = {398},
	primaryclass = {gr-qc},
	title = {{Scalar Induced Gravitational Waves Review}},
	volume = {7},
	year = {2021}}

@article{Adshead:2021hnm,
	archiveprefix = {arXiv},
	author = {Adshead, Peter and Lozanov, Kaloian D. and Weiner, Zachary J.},
	doi = {10.1088/1475-7516/2021/10/080},
	eprint = {2105.01659},
	journal = {JCAP},
	pages = {080},
	primaryclass = {astro-ph.CO},
	title = {{Non-Gaussianity and the induced gravitational wave background}},
	volume = {10},
	year = {2021}}

@article{Cai:2024nln,
	archiveprefix = {arXiv},
	author = {Cai, Rong-Gen and Hao, Yu-Shi and Wang, Shao-Jiang},
	doi = {10.1007/s11433-024-2416-3},
	eprint = {2404.06506},
	journal = {Sci. China Phys. Mech. Astron.},
	number = {9},
	pages = {290411},
	primaryclass = {astro-ph.CO},
	title = {{Primordial black holes and curvature perturbations from false vacuum islands}},
	volume = {67},
	year = {2024}}

@article{Matarrese:1992rp,
	author = {Matarrese, Sabino and Pantano, Ornella and Saez, Diego},
	doi = {10.1103/PhysRevD.47.1311},
	journal = {Phys. Rev. D},
	pages = {1311--1323},
	reportnumber = {DFPD-92-A-39},
	title = {{A General relativistic approach to the nonlinear evolution of collisionless matter}},
	volume = {47},
	year = {1993}}

@article{Matarrese:1997ay,
	archiveprefix = {arXiv},
	author = {Matarrese, Sabino and Mollerach, Silvia and Bruni, Marco},
	doi = {10.1103/PhysRevD.58.043504},
	eprint = {astro-ph/9707278},
	journal = {Phys. Rev. D},
	pages = {043504},
	reportnumber = {SISSA-83-97-A},
	title = {{Second order perturbations of the Einstein-de Sitter universe}},
	volume = {58},
	year = {1998}}

@article{Carbone:2004iv,
	archiveprefix = {arXiv},
	author = {Carbone, Carmelita and Matarrese, Sabino},
	doi = {10.1103/PhysRevD.71.043508},
	eprint = {astro-ph/0407611},
	journal = {Phys. Rev. D},
	pages = {043508},
	reportnumber = {DFPD-04-A-18},
	title = {{A Unified treatment of cosmological perturbations from super-horizon to small scales}},
	volume = {71},
	year = {2005}}

@article{Nakama:2016gzw,
	archiveprefix = {arXiv},
	author = {Nakama, Tomohiro and Silk, Joseph and Kamionkowski, Marc},
	doi = {10.1103/PhysRevD.95.043511},
	eprint = {1612.06264},
	journal = {Phys. Rev. D},
	number = {4},
	pages = {043511},
	primaryclass = {astro-ph.CO},
	title = {{Stochastic gravitational waves associated with the formation of primordial black holes}},
	volume = {95},
	year = {2017}}

@article{Garcia-Bellido:2017aan,
	archiveprefix = {arXiv},
	author = {Garcia-Bellido, Juan and Peloso, Marco and Unal, Caner},
	doi = {10.1088/1475-7516/2017/09/013},
	eprint = {1707.02441},
	journal = {JCAP},
	pages = {013},
	primaryclass = {astro-ph.CO},
	reportnumber = {IFT-UAM-CSIC-17-056, UMN-TH-3630-17, IFT--UAM-CSIC-17-056},
	title = {{Gravitational Wave signatures of inflationary models from Primordial Black Hole Dark Matter}},
	volume = {09},
	year = {2017}}

@article{Cai:2019amo,
	archiveprefix = {arXiv},
	author = {Cai, Rong-Gen and Pi, Shi and Wang, Shao-Jiang and Yang, Xing-Yu},
	doi = {10.1088/1475-7516/2019/05/013},
	eprint = {1901.10152},
	journal = {JCAP},
	pages = {013},
	primaryclass = {astro-ph.CO},
	title = {{Resonant multiple peaks in the induced gravitational waves}},
	volume = {05},
	year = {2019}}

@article{Ragavendra:2020sop,
	archiveprefix = {arXiv},
	author = {Ragavendra, H. V. and Saha, Pankaj and Sriramkumar, L. and Silk, Joseph},
	doi = {10.1103/PhysRevD.103.083510},
	eprint = {2008.12202},
	journal = {Phys. Rev. D},
	number = {8},
	pages = {083510},
	primaryclass = {astro-ph.CO},
	title = {{Primordial black holes and secondary gravitational waves from ultraslow roll and punctuated inflation}},
	volume = {103},
	year = {2021}}

@article{Yuan:2020iwf,
	archiveprefix = {arXiv},
	author = {Yuan, Chen and Huang, Qing-Guo},
	doi = {10.1016/j.physletb.2021.136606},
	eprint = {2007.10686},
	journal = {Phys. Lett. B},
	pages = {136606},
	primaryclass = {astro-ph.CO},
	title = {{Gravitational waves induced by the local-type non-Gaussian curvature perturbations}},
	volume = {821},
	year = {2021}}

@article{Davies:2021loj,
	archiveprefix = {arXiv},
	author = {Davies, Matthew W. and Carrilho, Pedro and Mulryne, David J.},
	doi = {10.1088/1475-7516/2022/06/019},
	eprint = {2110.08189},
	journal = {JCAP},
	number = {06},
	pages = {019},
	primaryclass = {astro-ph.CO},
	title = {{Non-Gaussianity in inflationary scenarios for primordial black holes}},
	volume = {06},
	year = {2022}}

@article{Abe:2022xur,
	archiveprefix = {arXiv},
	author = {Abe, Katsuya T. and Inui, Ryoto and Tada, Yuichiro and Yokoyama, Shuichiro},
	doi = {10.1088/1475-7516/2023/05/044},
	eprint = {2209.13891},
	journal = {JCAP},
	pages = {044},
	primaryclass = {astro-ph.CO},
	title = {{Primordial black holes and gravitational waves induced by exponential-tailed perturbations}},
	volume = {05},
	year = {2023}}

@article{Garcia-Saenz:2022tzu,
	archiveprefix = {arXiv},
	author = {Garcia-Saenz, Sebastian and Pinol, Lucas and Renaux-Petel, S\'ebastien and Werth, Denis},
	doi = {10.1088/1475-7516/2023/03/057},
	eprint = {2207.14267},
	journal = {JCAP},
	pages = {057},
	primaryclass = {astro-ph.CO},
	title = {{No-go theorem for scalar-trispectrum-induced gravitational waves}},
	volume = {03},
	year = {2023}}

@article{Garcia-Saenz:2023zue,
	archiveprefix = {arXiv},
	author = {Garcia-Saenz, Sebastian and Lu, Yizhou and Shuai, Zhiming},
	doi = {10.1103/PhysRevD.108.123507},
	eprint = {2306.09052},
	journal = {Phys. Rev. D},
	number = {12},
	pages = {123507},
	primaryclass = {gr-qc},
	title = {{Scalar-induced gravitational waves from ghost inflation and parity violation}},
	volume = {108},
	year = {2023}}

@article{Li:2023xtl,
	archiveprefix = {arXiv},
	author = {Li, Jun-Peng and Wang, Sai and Zhao, Zhi-Chao and Kohri, Kazunori},
	doi = {10.1088/1475-7516/2024/06/039},
	eprint = {2309.07792},
	journal = {JCAP},
	pages = {039},
	primaryclass = {astro-ph.CO},
	reportnumber = {KEK-Cosmo-0326, KEK-TH-2556, KEK-QUP-2023-0024},
	title = {{Complete analysis of the background and anisotropies of scalar-induced gravitational waves: primordial non-Gaussianity f $_{NL}$ and g $_{NL}$ considered}},
	volume = {06},
	year = {2024}}

@article{Yuan:2023ofl,
	archiveprefix = {arXiv},
	author = {Yuan, Chen and Meng, De-Shuang and Huang, Qing-Guo},
	doi = {10.1088/1475-7516/2023/12/036},
	eprint = {2308.07155},
	journal = {JCAP},
	pages = {036},
	primaryclass = {astro-ph.CO},
	title = {{Full analysis of the scalar-induced gravitational waves for the curvature perturbation with local-type non-Gaussianities}},
	volume = {12},
	year = {2023}}

@article{Ellis:2023oxs,
	archiveprefix = {arXiv},
	author = {Ellis, John and Fairbairn, Malcolm and Franciolini, Gabriele and H\"utsi, Gert and Iovino, Antonio and Lewicki, Marek and Raidal, Martti and Urrutia, Juan and Vaskonen, Ville and Veerm\"ae, Hardi},
	doi = {10.1103/PhysRevD.109.023522},
	eprint = {2308.08546},
	journal = {Phys. Rev. D},
	number = {2},
	pages = {023522},
	primaryclass = {astro-ph.CO},
	reportnumber = {KCL-PH-TH/2023-43, CERN-TH-2023-153, AION-REPORT/2023-08},
	title = {{What is the source of the PTA GW signal?}},
	volume = {109},
	year = {2024}}

@article{LISACosmologyWorkingGroup:2025vdz,
	archiveprefix = {arXiv},
	author = {Gammal, Jonas El and others},
	collaboration = {LISA Cosmology Working Group},
	eprint = {2501.11320},
	month = {1},
	primaryclass = {astro-ph.CO},
	reportnumber = {CERN-TH-2024-217},
	title = {{Reconstructing Primordial Curvature Perturbations via Scalar-Induced Gravitational Waves with LISA}},
	year = {2025}}

@article{Perna:2024ehx,
	archiveprefix = {arXiv},
	author = {Perna, Gabriele and Testini, Chiara and Ricciardone, Angelo and Matarrese, Sabino},
	doi = {10.1088/1475-7516/2024/05/086},
	eprint = {2403.06962},
	journal = {JCAP},
	pages = {086},
	primaryclass = {astro-ph.CO},
	title = {{Fully non-Gaussian Scalar-Induced Gravitational Waves}},
	volume = {05},
	year = {2024}}

@article{Maggiore:1999vm,
	archiveprefix = {arXiv},
	author = {Maggiore, Michele},
	doi = {10.1016/S0370-1573(99)00102-7},
	eprint = {gr-qc/9909001},
	journal = {Phys. Rept.},
	pages = {283--367},
	reportnumber = {IFUP-TH-20-99},
	title = {{Gravitational wave experiments and early universe cosmology}},
	volume = {331},
	year = {2000}}

@article{Zeldovich:1967lct,
	author = {Zel'dovich, Ya. B. and Novikov, I. D.},
	journal = {Sov. Astron.},
	pages = 602,
	title = {{The Hypothesis of Cores Retarded during Expansion and the Hot Cosmological Model}},
	volume = 10,
	year = 1967}

@article{Hawking:1971ei,
	author = {Hawking, Stephen},
	doi = {10.1093/mnras/152.1.75},
	journal = {Mon. Not. Roy. Astron. Soc.},
	pages = 75,
	title = {{Gravitationally collapsed objects of very low mass}},
	volume = 152,
	year = 1971}

@article{Carr:1974nx,
	author = {Carr, Bernard J. and Hawking, S. W.},
	doi = {10.1093/mnras/168.2.399},
	journal = {Mon. Not. Roy. Astron. Soc.},
	pages = {399--415},
	title = {{Black holes in the early Universe}},
	volume = 168,
	year = 1974}

@article{Carr:1975qj,
	author = {Carr, Bernard J.},
	doi = {10.1086/153853},
	journal = {Astrophys. J.},
	pages = {1--19},
	title = {{The Primordial black hole mass spectrum}},
	volume = 201,
	year = 1975}

@article{Chapline:1975ojl,
	author = {Chapline, George F.},
	doi = {10.1038/253251a0},
	journal = {Nature},
	number = 5489,
	pages = {251--252},
	title = {{Cosmological effects of primordial black holes}},
	volume = 253,
	year = 1975}

@article{Escriva:2021aeh,
	archiveprefix = {arXiv},
	author = {Escriv\`a, Albert},
	doi = {10.3390/universe8020066},
	eprint = {2111.12693},
	journal = {Universe},
	number = 2,
	pages = 66,
	primaryclass = {gr-qc},
	title = {{PBH Formation from Spherically Symmetric Hydrodynamical Perturbations: A Review}},
	volume = 8,
	year = 2022}

@article{Shibata:1999zs,
	archiveprefix = {arXiv},
	author = {Shibata, Masaru and Sasaki, Misao},
	doi = {10.1103/PhysRevD.60.084002},
	eprint = {gr-qc/9905064},
	journal = {Phys. Rev. D},
	pages = {084002},
	reportnumber = {OU-TAP-93},
	title = {{Black hole formation in the Friedmann universe: Formulation and computation in numerical relativity}},
	volume = 60,
	year = 1999}

@article{Harada:2023ffo,
	archiveprefix = {arXiv},
	author = {Harada, Tomohiro and Yoo, Chul-Moon and Koga, Yasutaka},
	doi = {10.1103/PhysRevD.108.043515},
	eprint = {2304.13284},
	journal = {Phys. Rev. D},
	number = 4,
	pages = {043515},
	primaryclass = {gr-qc},
	reportnumber = {RUP-23-10},
	title = {{Revisiting compaction functions for primordial black hole formation}},
	volume = 108,
	year = 2023}

@article{Harada:2015yda,
	archiveprefix = {arXiv},
	author = {Harada, Tomohiro and Yoo, Chul-Moon and Nakama, Tomohiro and Koga, Yasutaka},
	doi = {10.1103/PhysRevD.91.084057},
	eprint = {1503.03934},
	journal = {Phys. Rev. D},
	number = 8,
	pages = {084057},
	primaryclass = {gr-qc},
	reportnumber = {RUP-15-5, RESCEU-4-15},
	title = {{Cosmological long-wavelength solutions and primordial black hole formation}},
	volume = 91,
	year = 2015}

@article{DeLuca:2019qsy,
	archiveprefix = {arXiv},
	author = {De Luca, V. and Franciolini, G. and Kehagias, A. and Peloso, M. and Riotto, A. and \"Unal, C.},
	doi = {10.1088/1475-7516/2019/07/048},
	eprint = {1904.00970},
	journal = {JCAP},
	pages = {048},
	primaryclass = {astro-ph.CO},
	title = {{The Ineludible non-Gaussianity of the Primordial Black Hole Abundance}},
	volume = {07},
	year = 2019}

@article{Young:2019yug,
	archiveprefix = {arXiv},
	author = {Young, Sam and Musco, Ilia and Byrnes, Christian T.},
	doi = {10.1088/1475-7516/2019/11/012},
	eprint = {1904.00984},
	journal = {JCAP},
	pages = {012},
	primaryclass = {astro-ph.CO},
	title = {{Primordial black hole formation and abundance: contribution from the non-linear relation between the density and curvature perturbation}},
	volume = 11,
	year = 2019}

@article{Musco:2020jjb,
	archiveprefix = {arXiv},
	author = {Musco, Ilia and De Luca, Valerio and Franciolini, Gabriele and Riotto, Antonio},
	doi = {10.1103/PhysRevD.103.063538},
	eprint = {2011.03014},
	journal = {Phys. Rev. D},
	number = 6,
	pages = {063538},
	primaryclass = {astro-ph.CO},
	title = {{Threshold for primordial black holes. II. A simple analytic prescription}},
	volume = 103,
	year = 2021}

@article{Musco:2012au,
	archiveprefix = {arXiv},
	author = {Musco, Ilia and Miller, John C.},
	doi = {10.1088/0264-9381/30/14/145009},
	eprint = {1201.2379},
	journal = {Class. Quant. Grav.},
	pages = 145009,
	primaryclass = {gr-qc},
	title = {{Primordial black hole formation in the early universe: critical behaviour and self-similarity}},
	volume = 30,
	year = 2013}

@article{Musco:2018rwt,
	archiveprefix = {arXiv},
	author = {Musco, Ilia},
	doi = {10.1103/PhysRevD.100.123524},
	eprint = {1809.02127},
	journal = {Phys. Rev. D},
	number = 12,
	pages = 123524,
	primaryclass = {gr-qc},
	title = {{Threshold for primordial black holes: Dependence on the shape of the cosmological perturbations}},
	volume = 100,
	year = 2019}

@article{Escriva:2019phb,
	archiveprefix = {arXiv},
	author = {Escriv\`a, Albert and Germani, Cristiano and Sheth, Ravi K.},
	doi = {10.1103/PhysRevD.101.044022},
	eprint = {1907.13311},
	journal = {Phys. Rev. D},
	number = 4,
	pages = {044022},
	primaryclass = {gr-qc},
	reportnumber = {ICC-19-013},
	title = {{Universal threshold for primordial black hole formation}},
	volume = 101,
	year = 2020}

@misc{SuppMat,
  note = {See Supplemental Material at [URL will be inserted by publisher] for further details and additional references~\cite{Bardeen:1983qw,DeLuca:2019ufz,Domenech:2020xin,Inomata:2019yww,Li:2023qua,Lyth:2004gb,Ragavendra:2021qdu,Yuan:2019fwv,Zhu:2023faa}.}
}

@article{Luo:1992er,
	author = {Luo, Xiao-chun and Schramm, David N.},
	doi = {10.1086/172567},
	journal = {Astrophys. J.},
	pages = {33--42},
	reportnumber = {FERMILAB-PUB-92-214-A},
	title = {{Kurtosis, skewness, and nonGaussian cosmological density perturbations}},
	volume = 408,
	year = 1993}

@article{Kodama:1982sf,
	author = {Kodama, Hideo and Sasaki, Misao and Sato, Katsuhiko},
	doi = {10.1143/PTP.68.1979},
	journal = {Prog. Theor. Phys.},
	pages = 1979,
	reportnumber = {KUNS 642},
	title = {{Abundance of Primordial Holes Produced by Cosmological First Order Phase Transition}},
	volume = 68,
	year = 1982}

@article{Tomberg:2023kli,
    author = "Tomberg, Eemeli",
    title = "{Stochastic constant-roll inflation and primordial black holes}",
    eprint = "2304.10903",
    archivePrefix = "arXiv",
    primaryClass = "astro-ph.CO",
    doi = "10.1103/PhysRevD.108.043502",
    journal = "Phys. Rev. D",
    volume = "108",
    number = "4",
    pages = "043502",
    year = "2023"
}

@article{Hashino:2025fse,
	archiveprefix = {arXiv},
	author = {Hashino, Katsuya and Kanemura, Shinya and Takahashi, Tomo and Tanaka, Masanori and Yoo, Chul-Moon},
	eprint = {2501.11040},
	month = 1,
	primaryclass = {hep-ph},
	reportnumber = {OU-HET-1260},
	title = {{Super-critical primordial black hole formation via delayed first-order electroweak phase transition}},
	year = 2025}

@article{Gouttenoire:2025jxe,
    author = "Gouttenoire, Yann and Trifinopoulos, Sokratis and Vanvlasselaer, Miguel",
    title = "{Implications for Pulsar Timing Arrays of Sub-solar Black Hole Detections: From LVK to Einstein Telescope and Cosmic Explorer}",
    eprint = "2508.19328",
    archivePrefix = "arXiv",
    primaryClass = "astro-ph.CO",
    reportNumber = "CERN-TH-2025-169, MIT-CTP/5904, MITP-25-056",
    month = "8",
    year = "2025"
}

@article{Flores:2024lng,
	archiveprefix = {arXiv},
	author = {Flores, Marcos M. and Kusenko, Alexander and Sasaki, Misao},
	doi = {10.1103/PhysRevD.110.015005},
	eprint = {2402.13341},
	journal = {Phys. Rev. D},
	number = 1,
	pages = {015005},
	primaryclass = {hep-ph},
	reportnumber = {IPMU23-0053, YITP-23-169},
	title = {{Revisiting formation of primordial black holes in a supercooled first-order phase transition}},
	volume = 110,
	year = 2024}

@article{Hsu:1990fg,
	author = {Hsu, Stephen D. H.},
	doi = {10.1016/0370-2693(90)90717-K},
	journal = {Phys. Lett. B},
	pages = {343--348},
	reportnumber = {UCB-PTH-90/20, LBL-28996},
	title = {{Black Holes From Extended Inflation}},
	volume = 251,
	year = 1990}

@article{Liu:2021svg,
	archiveprefix = {arXiv},
	author = {Liu, Jing and Bian, Ligong and Cai, Rong-Gen and Guo, Zong-Kuan and Wang, Shao-Jiang},
	doi = {10.1103/PhysRevD.105.L021303},
	eprint = {2106.05637},
	journal = {Phys. Rev. D},
	number = 2,
	pages = {L021303},
	primaryclass = {astro-ph.CO},
	title = {{Primordial black hole production during first-order phase transitions}},
	volume = 105,
	year = 2022}

@article{Hashino:2021qoq,
	archiveprefix = {arXiv},
	author = {Hashino, Katsuya and Kanemura, Shinya and Takahashi, Tomo},
	doi = {10.1016/j.physletb.2022.137261},
	eprint = {2111.13099},
	journal = {Phys. Lett. B},
	pages = 137261,
	primaryclass = {hep-ph},
	reportnumber = {OU-HET-1123},
	title = {{Primordial black holes as a probe of strongly first-order electroweak phase transition}},
	volume = 833,
	year = 2022}

@article{Kawana:2022olo,
	archiveprefix = {arXiv},
	author = {Kawana, Kiyoharu and Kim, TaeHun and Lu, Philip},
	doi = {10.1103/PhysRevD.108.103531},
	eprint = {2212.14037},
	journal = {Phys. Rev. D},
	number = 10,
	pages = 103531,
	primaryclass = {astro-ph.CO},
	title = {{PBH formation from overdensities in delayed vacuum transitions}},
	volume = 108,
	year = 2023}

@article{Lewicki:2023ioy,
	archiveprefix = {arXiv},
	author = {Lewicki, Marek and Toczek, Piotr and Vaskonen, Ville},
	doi = {10.1007/JHEP09(2023)092},
	eprint = {2305.04924},
	journal = {JHEP},
	pages = {092},
	primaryclass = {astro-ph.CO},
	title = {{Primordial black holes from strong first-order phase transitions}},
	volume = {09},
	year = 2023}

@article{Gouttenoire:2023naa,
	archiveprefix = {arXiv},
	author = {Gouttenoire, Yann and Volansky, Tomer},
	doi = {10.1103/PhysRevD.110.043514},
	eprint = {2305.04942},
	journal = {Phys. Rev. D},
	number = 4,
	pages = {043514},
	primaryclass = {hep-ph},
	title = {{Primordial black holes from supercooled phase transitions}},
	volume = 110,
	year = 2024}

@article{Baldes:2023rqv,
	archiveprefix = {arXiv},
	author = {Baldes, Iason and Olea-Romacho, Mar\'\i{}a Olalla},
	doi = {10.1007/JHEP01(2024)133},
	eprint = {2307.11639},
	journal = {JHEP},
	pages = 133,
	primaryclass = {hep-ph},
	title = {{Primordial black holes as dark matter: interferometric tests of phase transition origin}},
	volume = {01},
	year = 2024}

@article{Gouttenoire:2023bqy,
	archiveprefix = {arXiv},
	author = {Gouttenoire, Yann},
	doi = {10.1103/PhysRevLett.131.171404},
	eprint = {2307.04239},
	journal = {Phys. Rev. Lett.},
	number = 17,
	pages = 171404,
	primaryclass = {hep-ph},
	title = {{First-Order Phase Transition Interpretation of Pulsar Timing Array Signal Is Consistent with Solar-Mass Black Holes}},
	volume = 131,
	year = 2023}

@article{Salvio:2023ynn,
	archiveprefix = {arXiv},
	author = {Salvio, Alberto},
	doi = {10.1088/1475-7516/2023/12/046},
	eprint = {2307.04694},
	journal = {JCAP},
	pages = {046},
	primaryclass = {hep-ph},
	title = {{Supercooling in radiative symmetry breaking: theory extensions, gravitational wave detection and primordial black holes}},
	volume = 12,
	year = 2023}

@article{Gouttenoire:2023pxh,
	archiveprefix = {arXiv},
	author = {Gouttenoire, Yann},
	doi = {10.1016/j.physletb.2024.138800},
	eprint = {2311.13640},
	journal = {Phys. Lett. B},
	pages = 138800,
	primaryclass = {hep-ph},
	title = {{Primordial black holes from conformal Higgs}},
	volume = 855,
	year = 2024}

@article{Jinno:2023vnr,
	archiveprefix = {arXiv},
	author = {Jinno, Ryusuke and Kume, Jun'ya and Yamada, Masaki},
	doi = {10.1016/j.physletb.2024.138465},
	eprint = {2310.06901},
	journal = {Phys. Lett. B},
	pages = 138465,
	primaryclass = {hep-ph},
	reportnumber = {TU-1209, RESCEU-18/23},
	title = {{Super-slow phase transition catalyzed by BHs and the birth of baby BHs}},
	volume = 849,
	year = 2024}

@article{Lewicki:2024ghw,
	archiveprefix = {arXiv},
	author = {Lewicki, Marek and Toczek, Piotr and Vaskonen, Ville},
	doi = {10.1103/PhysRevLett.133.221003},
	eprint = {2402.04158},
	journal = {Phys. Rev. Lett.},
	number = 22,
	pages = 221003,
	primaryclass = {astro-ph.CO},
	title = {{Black Holes and Gravitational Waves from Slow First-Order Phase Transitions}},
	volume = 133,
	year = 2024}

@article{Lewicki:2024sfw,
	archiveprefix = {arXiv},
	author = {Lewicki, Marek and Toczek, Piotr and Vaskonen, Ville},
	eprint = {2412.10366},
	month = 12,
	primaryclass = {astro-ph.CO},
	title = {{Black holes and gravitational waves from phase transitions in realistic models}},
	year = 2024}

@article{Ai:2024cka,
	archiveprefix = {arXiv},
	author = {Ai, Wen-Yuan and Heurtier, Lucien and Jung, Tae Hyun},
	eprint = {2409.02175},
	month = 9,
	primaryclass = {astro-ph.CO},
	reportnumber = {KCL-PH-TH-2024-46, CTPU-PTC-24-28},
	title = {{Primordial black holes from an interrupted phase transition}},
	year = 2024}

@article{Liu:2022lvz,
	archiveprefix = {arXiv},
	author = {Liu, Jing and Bian, Ligong and Cai, Rong-Gen and Guo, Zong-Kuan and Wang, Shao-Jiang},
	doi = {10.1103/PhysRevLett.130.051001},
	eprint = {2208.14086},
	journal = {Phys. Rev. Lett.},
	number = 5,
	pages = {051001},
	primaryclass = {astro-ph.CO},
	title = {{Constraining First-Order Phase Transitions with Curvature Perturbations}},
	volume = 130,
	year = 2023}

@article{Elor:2023xbz,
	archiveprefix = {arXiv},
	author = {Elor, Gilly and Jinno, Ryusuke and Kumar, Soubhik and McGehee, Robert and Tsai, Yuhsin},
	doi = {10.1103/PhysRevLett.133.211003},
	eprint = {2311.16222},
	journal = {Phys. Rev. Lett.},
	number = 21,
	pages = 211003,
	primaryclass = {hep-ph},
	reportnumber = {FTPI-MINN-23-20, UTWI-39-2023},
	title = {{Finite Bubble Statistics Constrain Late Cosmological Phase Transitions}},
	volume = 133,
	year = 2024}

@book{Byrnes:2025tji,
    editor = "Byrnes, Christian and Franciolini, Gabriele and Harada, Tomohiro and Pani, Paolo and Sasaki, Misao",
    title = "{Primordial Black Holes}",
    isbn = "978-981--978886-6, 978-981--978889-7, 978-981--978887-3",
    publisher = "Springer",
    series = "Springer Series in Astrophysics and Cosmology",
    month = "3",
    year = "2025"
}

@book{Gouttenoire:2022gwi,
	address = {Cham},
	archiveprefix = {arXiv},
	author = {Gouttenoire, Yann},
	doi = {10.1007/978-3-031-11862-3},
	eprint = {2207.01633},
	isbn = {978-3-031-11862-3, 978-3-031-11861-6},
	primaryclass = {hep-ph},
	publisher = {Springer},
	series = {Springer Theses},
	title = {{Beyond the Standard Model Cocktail}},
	year = {2022}}

@article{Kirzhnits:1972ut,
	author = {Kirzhnits, D. A. and Linde, Andrei D.},
	doi = {10.1016/0370-2693(72)90109-8},
	journal = {Phys. Lett. B},
	pages = {471--474},
	title = {{Macroscopic Consequences of the Weinberg Model}},
	volume = {42},
	year = {1972}}

@article{Sasaki:1982fi,
	author = {Sasaki, Misao and Kodama, Hideo and Sato, Katsuhiko},
	doi = {10.1143/PTP.68.1561},
	journal = {Prog. Theor. Phys.},
	pages = {1561--1573},
	title = {{GENERATION OF COSMOLOGICAL PERTURBATIONS BY A FIRST ORDER PHASE TRANSITION}},
	volume = {68},
	year = {1982}}

@article{Ghoshal:2025dmi,
	archiveprefix = {arXiv},
	author = {Ghoshal, Anish and Megias, Eugenio and Nardini, Germano and Quiros, Mariano},
	eprint = {2502.03588},
	month = {2},
	primaryclass = {hep-ph},
	title = {{Complementary Probes of Warped Extra Dimension: Colliders, Gravitational Waves and Primordial Black Holes from Phase Transitions}},
	year = {2025}}

@article{Zou:2025sow,
	archiveprefix = {arXiv},
	author = {Zou, Jintao and Zhu, Zhiqing and Zhao, Zizhuo and Bian, Ligong},
	eprint = {2502.20166},
	month = {2},
	primaryclass = {hep-ph},
	title = {{Numerical simulations of density perturbation and gravitational wave production from cosmological first-order phase transition}},
	year = {2025}}

@article{Murai:2025hse,
	archiveprefix = {arXiv},
	author = {Murai, Kai and Sakurai, Kodai and Takahashi, Fuminobu},
	eprint = {2502.02291},
	month = {2},
	primaryclass = {astro-ph.CO},
	reportnumber = {TU-1254},
	title = {{Primordial Black Hole Formation via Inverted Bubble Collapse}},
	year = {2025}}

@article{Banerjee:2024cwv,
	archiveprefix = {arXiv},
	author = {Banerjee, Indra Kumar and Rescigno, Francesco and Salvio, Alberto},
	eprint = {2412.06889},
	month = {12},
	primaryclass = {hep-ph},
	title = {{Primordial Black Holes (as Dark Matter) from the Supercooled Phase Transitions with Radiative Symmetry Breaking}},
	year = {2024}}

@article{Kanemura:2024pae,
	archiveprefix = {arXiv},
	author = {Kanemura, Shinya and Tanaka, Masanori and Xie, Ke-Pan},
	doi = {10.1007/JHEP06(2024)036},
	eprint = {2404.00646},
	journal = {JHEP},
	pages = {036},
	primaryclass = {hep-ph},
	title = {{Primordial black holes from slow phase transitions: a model-building perspective}},
	volume = {06},
	year = {2024}}

@article{Conaci:2024tlc,
	archiveprefix = {arXiv},
	author = {Conaci, Angela and Delle Rose, Luigi and Dev, P. S. Bhupal and Ghoshal, Anish},
	doi = {10.1007/JHEP12(2024)196},
	eprint = {2401.09411},
	journal = {JHEP},
	pages = {196},
	primaryclass = {astro-ph.CO},
	title = {{Slaying axion-like particles via gravitational waves and primordial black holes from supercooled phase transition}},
	volume = {12},
	year = {2024}}

@article{Banerjee:2023qya,
	archiveprefix = {arXiv},
	author = {Banerjee, Indra Kumar and Dey, Ujjal Kumar},
	doi = {10.1007/JHEP07(2024)006},
	eprint = {2311.03406},
	journal = {JHEP},
	note = {[Erratum: JHEP 08, 054 (2024)]},
	pages = {006},
	primaryclass = {gr-qc},
	title = {{Spinning primordial black holes from first order phase transition}},
	volume = {07},
	year = {2024}}

@article{Jung:2021mku,
	archiveprefix = {arXiv},
	author = {Jung, Tae Hyun and Okui, Takemichi},
	doi = {10.1103/PhysRevD.110.115014},
	eprint = {2110.04271},
	journal = {Phys. Rev. D},
	number = {11},
	pages = {115014},
	primaryclass = {hep-ph},
	reportnumber = {KEK-TH-2350},
	title = {{Primordial black holes from bubble collisions during a first-order phase transition}},
	volume = {110},
	year = {2024}}

@article{Giombi:2023jqq,
	archiveprefix = {arXiv},
	author = {Giombi, Lorenzo and Hindmarsh, Mark},
	doi = {10.1088/1475-7516/2024/03/059},
	eprint = {2307.12080},
	journal = {JCAP},
	pages = {059},
	primaryclass = {astro-ph.CO},
	reportnumber = {HIP-2023-12/TH},
	title = {{General relativistic bubble growth in cosmological phase transitions}},
	volume = {03},
	year = {2024}}

@article{Caprini:2015zlo,
	archiveprefix = {arXiv},
	author = {Caprini, Chiara and others},
	doi = {10.1088/1475-7516/2016/04/001},
	eprint = {1512.06239},
	journal = {JCAP},
	pages = {001},
	primaryclass = {astro-ph.CO},
	reportnumber = {DESY-15-246},
	title = {{Science with the space-based interferometer eLISA. II: Gravitational waves from cosmological phase transitions}},
	volume = {04},
	year = {2016}}

@article{Caprini:2019egz,
	archiveprefix = {arXiv},
	author = {Caprini, Chiara and others},
	doi = {10.1088/1475-7516/2020/03/024},
	eprint = {1910.13125},
	journal = {JCAP},
	pages = {024},
	primaryclass = {astro-ph.CO},
	reportnumber = {DESY-19-159, IPPP/19/27, HIP-2019-14/TH, MITP/19-066, IFT-UAM/CSIC-19-139},
	title = {{Detecting gravitational waves from cosmological phase transitions with LISA: an update}},
	volume = {03},
	year = {2020}}

@article{Li:2023qua,
	archiveprefix = {arXiv},
	author = {Li, Jun-Peng and Wang, Sai and Zhao, Zhi-Chao and Kohri, Kazunori},
	doi = {10.1088/1475-7516/2023/10/056},
	eprint = {2305.19950},
	journal = {JCAP},
	pages = {056},
	primaryclass = {astro-ph.CO},
	reportnumber = {KEK-Cosmo-0315, KEK-TH-2531, KEK-QUP-2023-0012},
	title = {{Primordial non-Gaussianity f $_{NL}$ and anisotropies in scalar-induced gravitational waves}},
	volume = {10},
	year = {2023}}

@article{Zhu:2023faa,
	archiveprefix = {arXiv},
	author = {Wang, Sai and Zhao, Zhi-Chao and Zhu, Qing-Hua},
	doi = {10.1103/PhysRevResearch.6.013207},
	eprint = {2307.03095},
	journal = {Phys. Rev. Res.},
	number = {1},
	pages = {013207},
	primaryclass = {astro-ph.CO},
	title = {{Constraints on scalar-induced gravitational waves up to third order from a joint analysis of BBN, CMB, and PTA data}},
	volume = {6},
	year = {2024}}

@article{Ragavendra:2021qdu,
	archiveprefix = {arXiv},
	author = {Ragavendra, H. V.},
	doi = {10.1103/PhysRevD.105.063533},
	eprint = {2108.04193},
	journal = {Phys. Rev. D},
	number = {6},
	pages = {063533},
	primaryclass = {astro-ph.CO},
	title = {{Accounting for scalar non-Gaussianity in secondary gravitational waves}},
	volume = {105},
	year = {2022}}

@article{Espinosa:2018eve,
	archiveprefix = {arXiv},
	author = {Espinosa, Jos\'e Ram\'on and Racco, Davide and Riotto, Antonio},
	doi = {10.1088/1475-7516/2018/09/012},
	eprint = {1804.07732},
	journal = {JCAP},
	pages = {012},
	primaryclass = {hep-ph},
	title = {{A Cosmological Signature of the SM Higgs Instability: Gravitational Waves}},
	volume = {09},
	year = {2018}}

@article{Inomata:2019yww,
	archiveprefix = {arXiv},
	author = {Inomata, Keisuke and Terada, Takahiro},
	doi = {10.1103/PhysRevD.101.023523},
	eprint = {1912.00785},
	journal = {Phys. Rev. D},
	number = {2},
	pages = {023523},
	primaryclass = {gr-qc},
	reportnumber = {IPMU 19-0178, CTPU-PTC-19-36},
	title = {{Gauge Independence of Induced Gravitational Waves}},
	volume = {101},
	year = {2020}}

@article{DeLuca:2019ufz,
	archiveprefix = {arXiv},
	author = {De Luca, V. and Franciolini, G. and Kehagias, A. and Riotto, A.},
	doi = {10.1088/1475-7516/2020/03/014},
	eprint = {1911.09689},
	journal = {JCAP},
	pages = {014},
	primaryclass = {gr-qc},
	title = {{On the Gauge Invariance of Cosmological Gravitational Waves}},
	volume = {03},
	year = {2020}}

@article{Yuan:2019fwv,
	archiveprefix = {arXiv},
	author = {Yuan, Chen and Chen, Zu-Cheng and Huang, Qing-Guo},
	doi = {10.1103/PhysRevD.101.063018},
	eprint = {1912.00885},
	journal = {Phys. Rev. D},
	number = {6},
	pages = {063018},
	primaryclass = {astro-ph.CO},
	title = {{Scalar induced gravitational waves in different gauges}},
	volume = {101},
	year = {2020}}

@article{Domenech:2020xin,
	archiveprefix = {arXiv},
	author = {Dom\`enech, Guillem and Sasaki, Misao},
	doi = {10.1103/PhysRevD.103.063531},
	eprint = {2012.14016},
	journal = {Phys. Rev. D},
	number = {6},
	pages = {063531},
	primaryclass = {gr-qc},
	reportnumber = {YITP-20-163},
	title = {{Approximate gauge independence of the induced gravitational wave spectrum}},
	volume = {103},
	year = {2021}}

@article{Buckley:2024nen,
	archiveprefix = {arXiv},
	author = {Buckley, Matthew R. and Du, Peizhi and Fernandez, Nicolas and Weikert, Mitchell J.},
	doi = {10.1088/1475-7516/2024/07/031},
	eprint = {2402.13309},
	journal = {JCAP},
	pages = {031},
	primaryclass = {hep-ph},
	title = {{Dark radiation isocurvature from cosmological phase transitions}},
	volume = {07},
	year = 2024}

@article{Bardeen:1980kt,
	author = {Bardeen, James M.},
	doi = {10.1103/PhysRevD.22.1882},
	journal = {Phys. Rev. D},
	pages = {1882--1905},
	title = {{Gauge Invariant Cosmological Perturbations}},
	volume = 22,
	year = 1980}

@article{Kodama:1984ziu,
	author = {Kodama, Hideo and Sasaki, Misao},
	doi = {10.1143/PTPS.78.1},
	journal = {Prog. Theor. Phys. Suppl.},
	pages = {1--166},
	title = {{Cosmological Perturbation Theory}},
	volume = 78,
	year = 1984}

@article{Mukhanov:1990me,
	author = {Mukhanov, Viatcheslav F. and Feldman, H. A. and Brandenberger, Robert H.},
	doi = {10.1016/0370-1573(92)90044-Z},
	journal = {Phys. Rept.},
	pages = {203--333},
	reportnumber = {BROWN-HET-796, BROWN-HET-800, BROWN-HET-780},
	title = {{Theory of cosmological perturbations. Part 1. Classical perturbations. Part 2. Quantum theory of perturbations. Part 3. Extensions}},
	volume = 215,
	year = 1992}

@book{Mukhanov:2005sc,
	address = {Oxford},
	author = {Mukhanov, V.},
	doi = {10.1017/CBO9780511790553},
	isbn = {978-0-521-56398-7},
	publisher = {Cambridge University Press},
	title = {{Physical Foundations of Cosmology}},
	year = 2005}

@book{Baumann:2022mni,
	author = {Baumann, Daniel},
	doi = {10.1017/9781108937092},
	isbn = {978-1-108-93709-2, 978-1-108-83807-8},
	month = 7,
	publisher = {Cambridge University Press},
	title = {{Cosmology}},
	year = 2022}

@book{Peter:2013avv,
	author = {Peter, Patrick and Uzan, Jean-Philippe},
	isbn = {978-0-19-966515-0, 978-0-19-920991-0},
	month = 2,
	publisher = {Oxford University Press},
	series = {Oxford Graduate Texts},
	title = {{Primordial Cosmology}},
	year = 2013}

@article{Komatsu:2001rj,
	archiveprefix = {arXiv},
	author = {Komatsu, Eiichiro and Spergel, David N.},
	doi = {10.1103/PhysRevD.63.063002},
	eprint = {astro-ph/0005036},
	journal = {Phys. Rev. D},
	pages = {063002},
	title = {{Acoustic signatures in the primary microwave background bispectrum}},
	volume = 63,
	year = 2001}

@article{Cai:2018dig,
	archiveprefix = {arXiv},
	author = {Cai, Rong-gen and Pi, Shi and Sasaki, Misao},
	doi = {10.1103/PhysRevLett.122.201101},
	eprint = {1810.11000},
	journal = {Phys. Rev. Lett.},
	number = 20,
	pages = 201101,
	primaryclass = {astro-ph.CO},
	reportnumber = {IPMU18-0172, YITP-18-114},
	title = {{Gravitational Waves Induced by non-Gaussian Scalar Perturbations}},
	volume = 122,
	year = 2019}

@article{Coleman:1980aw,
	author = {Coleman, Sidney R. and De Luccia, Frank},
	doi = {10.1103/PhysRevD.21.3305},
	journal = {Phys. Rev. D},
	pages = 3305,
	reportnumber = {SLAC-PUB-2463},
	title = {{Gravitational Effects on and of Vacuum Decay}},
	volume = 21,
	year = 1980}

@article{Choptuik:1992jv,
	author = {Choptuik, Matthew W.},
	doi = {10.1103/PhysRevLett.70.9},
	journal = {Phys. Rev. Lett.},
	pages = {9--12},
	reportnumber = {FPRINT-92-33},
	title = {{Universality and scaling in gravitational collapse of a massless scalar field}},
	volume = 70,
	year = 1993}

@article{Evans:1994pj,
	archiveprefix = {arXiv},
	author = {Evans, Charles R. and Coleman, Jason S.},
	doi = {10.1103/PhysRevLett.72.1782},
	eprint = {gr-qc/9402041},
	journal = {Phys. Rev. Lett.},
	pages = {1782--1785},
	reportnumber = {TAR-039-UNC},
	title = {{Observation of critical phenomena and selfsimilarity in the gravitational collapse of radiation fluid}},
	volume = 72,
	year = 1994}

@article{Niemeyer:1997mt,
	archiveprefix = {arXiv},
	author = {Niemeyer, Jens C. and Jedamzik, K.},
	doi = {10.1103/PhysRevLett.80.5481},
	eprint = {astro-ph/9709072},
	journal = {Phys. Rev. Lett.},
	pages = {5481--5484},
	title = {{Near-critical gravitational collapse and the initial mass function of primordial black holes}},
	volume = 80,
	year = 1998}

@article{Jinno:2017fby,
	archiveprefix = {arXiv},
	author = {Jinno, Ryusuke and Takimoto, Masahiro},
	doi = {10.1088/1475-7516/2019/01/060},
	eprint = {1707.03111},
	journal = {JCAP},
	pages = {060},
	primaryclass = {hep-ph},
	reportnumber = {CTPU-17-26, KEK-TH-1986},
	title = {{Gravitational waves from bubble dynamics: Beyond the Envelope}},
	volume = {01},
	year = 2019}

@article{Konstandin:2017sat,
	archiveprefix = {arXiv},
	author = {Konstandin, Thomas},
	doi = {10.1088/1475-7516/2018/03/047},
	eprint = {1712.06869},
	journal = {JCAP},
	pages = {047},
	primaryclass = {astro-ph.CO},
	reportnumber = {DESY-17-227},
	title = {{Gravitational radiation from a bulk flow model}},
	volume = {03},
	year = 2018}

@article{Baldes:2024wuz,
	archiveprefix = {arXiv},
	author = {Baldes, Iason and Dichtl, Maximilian and Gouttenoire, Yann and Sala, Filippo},
	doi = {10.1007/JHEP07(2024)231},
	eprint = {2403.05615},
	journal = {JHEP},
	pages = 231,
	primaryclass = {hep-ph},
	title = {{Particle shells from relativistic bubble walls}},
	volume = {07},
	year = 2024}

@article{Bardeen:1983qw,
	author = {Bardeen, James M. and Steinhardt, Paul J. and Turner, Michael S.},
	doi = {10.1103/PhysRevD.28.679},
	journal = {Phys. Rev. D},
	pages = 679,
	reportnumber = {UPR-0202T, EFI-83-13-CHICAGO},
	title = {{Spontaneous Creation of Almost Scale - Free Density Perturbations in an Inflationary Universe}},
	volume = 28,
	year = 1983}

@article{Wands:2000dp,
	archiveprefix = {arXiv},
	author = {Wands, David and Malik, Karim A. and Lyth, David H. and Liddle, Andrew R.},
	doi = {10.1103/PhysRevD.62.043527},
	eprint = {astro-ph/0003278},
	journal = {Phys. Rev. D},
	pages = {043527},
	title = {{A New approach to the evolution of cosmological perturbations on large scales}},
	volume = 62,
	year = 2000}

@article{Lyth:2004gb,
	archiveprefix = {arXiv},
	author = {Lyth, David H. and Malik, Karim A. and Sasaki, Misao},
	doi = {10.1088/1475-7516/2005/05/004},
	eprint = {astro-ph/0411220},
	journal = {JCAP},
	pages = {004},
	reportnumber = {YITP-04-67},
	title = {{A General proof of the conservation of the curvature perturbation}},
	volume = {05},
	year = 2005}

\end{document}